\documentclass[pdflatex,sn-vancouver-num]{sn-jnl}

\usepackage[utf8]{inputenc}
\usepackage{graphicx}%
\usepackage{multirow}%
\usepackage{amsmath,amssymb,amsfonts}%
\usepackage{amsthm}%
\usepackage{mathrsfs}%
\usepackage[title]{appendix}%
\usepackage{xcolor}%
\usepackage{textcomp}%
\usepackage{manyfoot}%
\usepackage{booktabs}%
\usepackage{algorithm}%
\usepackage{algorithmicx}%
\usepackage{algpseudocode}%
\usepackage{listings}%
\usepackage{subcaption}
\usepackage{longtable}
\usepackage{chngcntr}
\usepackage{arydshln}
\usepackage{hyperref}
\usepackage[capitalize]{cleveref}
\usepackage{nameref}

\usepackage{caption}

\DeclareCaptionLabelFormat{figurextop}{Figure~#2~Top}

\DeclareCaptionStyle{special}{%
  labelformat=figurextop,
  labelsep=space,
  justification=centering  
}

\usepackage{caption}

\DeclareCaptionLabelFormat{figurextop}{Figure~#2~Top}

\DeclareCaptionStyle{special}{%
  labelformat=figurextop,
  labelsep=space,
  justification=centering  
}

\usepackage{titlesec}

\titleformat{\section}
  {\normalfont\Large\centering}
  {\thesection} 
  {1em}
  {}

\titleformat{\subsection}
  {\normalfont\itshape\large\centering}
  {\thesubsection}
  {1em}
  {}

\titleformat{\subsubsection}
  {\normalfont\itshape\normalsize}
  {\thesubsubsection}
  {1em}
  {}

\titlespacing*{\section}{0pt}{*2}{*0}
\titlespacing*{\subsection}{0pt}{*2}{*0}
\titlespacing*{\subsubsection}{0pt}{*1.5}{*0}

\usepackage{tikz}
\usetikzlibrary{bayesnet}
\usetikzlibrary{arrows.meta}

\usepackage{amsmath}

\usepackage{times}

\fontsize{12pt}{14pt}\selectfont
\usepackage[acronyms,nonumberlist,nopostdot,nomain,nogroupskip,acronymlists={hidden}]{glossaries} 
\newglossary[algh]{hidden}{acrh}{acnh}{Hidden Acronyms}

\theoremstyle{thmstyleone}%
\theoremstyle{thmstyletwo}%

\theoremstyle{thmstylethree}%

\newacronym{qq}{Q-Q}{Quantile-Quantile}
\newacronym{asd}{ASD}{Autism Spectrum Disorder}
\newacronym{mcmc}{MCMC}{Monte-Carlo Markov-Chain}

\newacronym{aic}{AIC}{Autism Inpatient Collection}
\newacronym{irb}{IRB}{Institutional Review Board}

\newacronym{hpp}{HPP}{Homogeneous Poisson Process}
\newacronym{nhpp}{NHPP}{Non-Homogeneous Poisson Process}
\newacronym{hwkpp}{HawkesPP}{Hawkes Point Process}
\newacronym{hwkexp}{HawkesExpPP}{Hawkes Point Process with Exponential Kernel}
\newacronym{hwk2exp}{Hawkes2ExpPP}{Hawkes Point Process with Two Exponential Kernels}
\newacronym{hwkpl}{HawkesPLPP}{Hawkes Point Process with Power Law Kernel}

\newacronym{hdi}{HDI}{Highest Density Interval}
\newacronym{iwmm}{IWMM}{Importance Weighted Moment Matching}
\newacronym{psis-loo}{PSIS-LOO}{Pareto Smoothed Importance Sampling - Leave One Out}
\newacronym{psis}{PSIS}{Pareto Smoothed Importance Sampling}
\newacronym{elpd}{ELPD}{Expected Log Predictive Density}
\newacronym{hmc}{HMC}{Hamiltonian Monte Carlo}
\newacronym{cdf}{CDF}{Cumulative Density Function}
\newacronym{ecdf}{ECDF}{Empirical Cumulative Density Function}
\newacronym{ppc}{PPC}{Postetior Predictive Check}
\newacronym{rtct}{RTCT}{Random Time Change Theorem}
\newacronym{bic}{BIC}{Bayesian Information Criterion}
\newacronym{waic}{WAIC}{Widely Applicable Information Criterion}
\newacronym{nuts}{NUTS}{No U-Turn Sampler}
\newacronym{pdf}{PDF}{Probability Density Function}
\newacronym{tpp}{TPP}{Temporal Point Process}
\newacronym{iid}{IID}{independent and identically distributed}
\newacronym{is}{IS}{Importance Sampling}
\newacronym{ai}{AI}{Artificial Intelligence}
\newacronym{ml}{ML}{Machine Learning}
\newacronym{dl}{DL}{Deep Learning}
\newacronym{cnn}{CNN}{Convolutional Neural Network}
\newacronym{pl}{PL}{Power Law}
\newacronym{mh}{MH}{Metropolis Hastings}
\newacronym{ess}{ESS}{Effective Sample Size}
\newacronym{mcse}{MCSE}{Monte-Carlo Standard Error}
\newacronym{kde}{KDE}{Kernel Density Estimators}
\newacronym{mape}{MAPE}{Mean Absolute Percent Error}

\newacronym{sib}{SIB}{Self-Injurious Behavior}
\newacronym{ed}{ED}{Emotion Dysregulation}
\newacronym{ato}{ATO}{Aggression Towards Others}

\newacronym{mle}{MLE}{Maximum Likelihood Estimation}
\newacronym{eda}{EDA}{Electrodermal Activity}
\newacronym{lstm}{LSTM}{Long Short Term Memory}
\newacronym{lr}{LR}{Logistic Regression}
\newacronym{svm}{SVM}{Support Vector Machine}
\newacronym{nn}{NN}{Neural Network}
\newacronym{wd}{WD}{Wassertein Distance}
\newacronym{mc}{MC}{Monte Carlo}
\newacronym{gof}{GOF}{Goodness-of-Fit}

\newacronym{rocauc}{ROC-AUC}{Area Under the Receiver Operating Characteristic Curve}

\newacronym{hipaa}{HIPAA}{Health Information Portability and Accountability Act}
\makeglossaries

\begin{document}

\title[Hierarchical Temporal Point Process Modeling of Aggressive Behavior Onset in Psychiatric Inpatient Youth with Autism for Branching Factor Estimation]{Hierarchical Temporal Point Process Modeling of Aggressive Behavior Onset in Psychiatric Inpatient Youth with Autism for Branching Factor Estimation}


\author*[1]{\fnm{Michael} \sur{Potter}}\email{potter.mi@northeastern.edu}

\author[1]{\fnm{Michael} \sur{Everett}}\email{m.everett@northeastern.edu}

\author[1]{\fnm{Deniz} \sur{Erdo\u{g}mu\c{s}}}\email{D.Erdogmus@northeastern.edu}

\author[2]{\fnm{Yuna} \sur{Watanabe}}\email{watanabe.y@northeastern.edu}

\author[3]{\fnm{Tales} \sur{Imbiriba}}\email{tales.imbiriba@umb.edu}
\equalcont{These authors contributed equally to this work.}

\author[2]{\fnm{Matthew S.} \sur{Goodwin}}\email{m.goodwin@northeastern.edu}
\equalcont{These authors contributed equally to this work.}

\affil*[1]{\orgdiv{Electrical and Computer Engineering}, \orgname{Northeastern University}, \orgaddress{\street{360 Huntington}, \city{Boston}, \postcode{02145}, \state{MA}, \country{USA}}}

\affil[2]{\orgdiv{ Bouvé College of Health Sciences and Khoury College of Computer Sciences}, \orgname{Northeastern University}, \orgaddress{\street{360 Huntington}, \city{Boston}, \postcode{02145}, \state{MA}, \country{USA}}}

\affil[3]{\orgdiv{Computer Science}, \orgname{University of Massachusetts}, \orgaddress{\street{Street}, \city{Boston}, \postcode{02145}, \state{MA}, \country{USA}}}

\maketitle

\section*{Abstract}

\noindent \textbf{Background:}
Aggressive behavior in minimally verbal psychiatric inpatient youths with autism often occurs in irregular, temporally clustered bursts, making it challenging to distinguish external versus internal aggression triggers. The sample population branching factor—the expected number of new aggressive behavior onsets triggered by a given event—serves as a summary statistic for understanding the degree of self-excitation in behavior dynamics. Prior approaches using pooled models to estimate this quantity fail to account for between- and within-person variability, leading to potentially inflated and biased estimates. 

\noindent \textbf{Methods:} To address these drawbacks, we model aggressive behavior onsets using a hierarchical Hawkes process with an exponential kernel and edge-effect correction. This approach partially pools data across persons to reduce bias from high-frequency individuals while stabilizing estimates for those with sparse data. We perform Bayesian inference via the \gls{nuts} and evaluate model fit using convergence diagnostics, posterior sensitivity analysis via power-scaling prior distribution and likelihood, and multiple \gls{gof} measures, including \gls{psis-loo}, the Lewis test with Durbin’s modification, and \gls{rtct} residual analysis.

\noindent \textbf{Results:}
The partially pooled (hierarchical) model produced a statistically significantly lower sample population branching factor estimate (mean 0.742 ± 0.026) than the pooled model (mean 0.899 ± 0.015), with narrower credible intervals than the unpooled model (mean 0.717 ± 0.139), indicating more precise and less biased inference. This result translates to a threefold smaller expected number of descendant events from a single parent onset compared to the pooled model. Sensitivity analyses revealed that the partially pooled model is robust to prior and likelihood perturbations, unlike the unpooled model, which showed high sensitivity for individuals with limited data. Goodness-of-fit assessments consistently favored the hierarchical model, confirming superior predictive accuracy and statistical reliability performance.

\noindent \textbf{Conclusions:}
Hierarchical Hawkes process modeling with edge-effect correction enables robust estimation of the sample population branching factor for aggressive behavior onset in autistic inpatient youth by explicitly modeling both within- and between-individual variability. Reliable estimation of the branching factor may help distinguish endogenous from exogenous events, link onsets to physiological signals for clinical insight, and guide early warnings, therapy, and resource allocation based on individual escalation risk.



\section{Background}\label{sec1}

Autism affects approximately 1 in 36 children \cite{maenner2023prevalence}, with up to 80\% displaying challenging behaviors including \gls{sib}, \gls{ed}, and \gls{ato} \cite{hattier2011occurrence,kanne2011aggression,matson2014assessing}. These behaviors are primary reasons for behavioral health referrals and major healthcare cost drivers \cite{arnold2003parent,croen2006comparison}. Many autistic individuals struggle with emotion regulation and communicating internal states, with 30-40\% having minimal verbal abilities and others facing emotional awareness difficulties \cite{mazefsky2013emotion,tager2017conducting}. The unpredictable nature of these behaviors creates barriers to community access, therapy, education, and clinical services. Families experience increased stress, social isolation, and financial strain due to concerns about behavioral incidents across settings \cite{davis2008parenting,hodgetts2013home}, while support staff face higher injury compensation, absenteeism, and turnover rates \cite{kiely1998violence}. These cumulative challenges can demoralize caregivers and clinicians, disrupt care continuity, and in severe cases, require residential placement, reducing quality of life while increasing costs.

Aggressive behavior in inpatient youths with autism exhibits irregular timing and temporal clustering, with bursts of aggressive episodes occurring within short intervals \cite{imbiriba2023wearable}. Several putative mechanisms have been proposed to account for this observation, including individual differences in sensory processing, communication challenges, cognitive impairments, executive function issues, difficulties with emotion regulation, and co-occurring medical and mental health conditions \cite{mazefsky2013emotion,nevin2011behavioral}. Additionally, within-individual differences in behavioral momentum (e.g., reinforcement history, contextual factors) \cite{pritchard2014treatment} and ability to cope with stressors have been posited as explanatory factors \cite{cohen2011assessing}. 

Self-exciting \glspl{tpp}, such as Hawkes Processes, can model irregular inter-onset intervals and support causal inference and knowledge discovery \cite{potter2025temporal}. The current study focuses on reliably estimating the sample population branching factor to describe the expected number of direct offspring events (i.e., subsequent aggression) triggered by an initial parent event (i.e., prior aggression) \cite{laub2024hawkes}. In other words, the branching factor captures the extent to which an aggressive behavior onset increases the likelihood of subsequent aggression onsets. Understanding the branching factor can inform aggression onset simulation, estimate the probability of cascading episodes, and help differentiate between exogenous and endogenous drivers of aggression—where exogenous events are triggered by external stimuli (e.g., environmental disruptions, staffing changes, or medication timing), and endogenous events arise from prior aggression, such as residual distress or ongoing dysregulation \cite{ito2024exogenous}.

To the best of our knowledge, only a single study has explored Hawkes Processes to infer the aggressive behavior onset branching factor in psychiatric inpatient youths with autism \cite{potter2025temporal}. \citet{potter2025temporal} employed a pooled model, which assumes all individuals share identical parameters for their underlying generative process of aggressive behavior events. However, the sample population is not homogeneous—there is non-negligible variability in aggression frequency and patterns across individuals with autism \cite{lydon2016systematic}. As a result, the pooled modeling approach leads to inflated branching factor estimates: a small subset of highly aggressive individuals contributes a disproportionate share of aggressive episodes, causing their behavioral dynamics to dominate the pooled model and biasing the population-level inference of the sample population branching factor upward. This can result in aggressive behavior onsets being misclassified as endogenous rather than exogenous events, which may misguide resource allocation and intervention planning, and does not allow for individual-specific modeling needed for personalized medicine. 

To address the inflation of the sample population branching factor caused by pooling, this paper proposes a partially pooled (hierarchical) Hawkes process model with edge-effect correction. This approach allows for individual-specific parameters while sharing information across persons, thereby reducing bias from high-frequency individuals and stabilizing estimates for those with sparse data. We demonstrate that the partially pooled model produces sample population branching factor estimates statistically significantly lower than the pooled model, while also achieving superior \gls{gof} metrics. 

The main contributions of this paper are as follows:
\begin{itemize}
    \item We propose a partially pooled Hawkes Process model with edge-effect correction for estimating the sample-population branching factor of aggressive behavior onsets in minimally verbal psychiatric inpatient youths with autism.
    \item We show that partial pooling yields statistically significantly lower, less biased branching factor estimates than pooled models, with narrower credible intervals indicating reduced uncertainty compared to unpooled models.
    \item We analyze sensitivity of branching factor estimates to prior and likelihood perturbations via power-scaling, highlighting limitations of unpooled models.
\end{itemize}

\section{Methods}\label{subsec:method}

\subsection{Participants}\label{subsec:participants}
This prognostic study was designed to estimate the branching factor of the sample population and follows the Transparent Reporting of a Multivariable Prediction Model for Individual Prognosis or Diagnosis (TRIPOD) reporting guidelines. It is a secondary analysis of data from \cite{imbiriba2023wearable,potter2025temporal}, acquired from psychiatric inpatients at four clinical sites (Bradley Hospital, Providence, RI; Cincinnati Children's Hospital, Cincinnati, OH; Western Psychiatric Hospital, Pittsburgh, PA; and Spring Harbor Hospital, Portland, ME) participating in the \gls{aic}.  The \glspl{irb} approved the AIC and aggressive behavior prediction protocols of participating study sites. \gls{irb} approval of the AIC extended to this study with an amendment. Guardians of all study participants provided informed consent and were remunerated. The IRB approved this retrospective study in compliance with the \gls{hipaa}. All methods were performed in accordance with relevant guidelines and regulations following the Declaration of Helsinki.

Of 86 enrolled inpatients, 70 were included in the final analysis. Inclusion criteria required autism confirmation via research-reliable ADOS-2 administration and documented physical aggression or self-injurious behavior. Sixteen participants were excluded due to inability to wear the biosensor ($N=8$) or early discharge ($N=8$). Research staff conducted observational coding while participants wore the Empatica E4 biosensor on their non-dominant wrist, collecting physiological signals including electrodermal activity, blood volume pulse, and wrist acceleration. The current study focuses exclusively on annotations of operationally defined aggressive behavior episodes (\gls{sib}, \gls{ed}, \gls{ato}) with start and stop times recorded via a custom mobile application.

The 70 participants were aged 5--19 years ($M=11.85$, $SD=3.5$), predominantly male (88\%), white (90\%), and non-Hispanic (92\%). Nearly half (46\%) were minimally verbal, and 57\% had intellectual disability (Leiter-3 global IQ $M=72.96$, $SD=26.12$). Hospital stays ranged from 8--201 days ($M=37.28$, $SD=33.95$). Data collection occurred from March 2019 to March 2020, yielding 429 observation sessions (median 5 per participant) totaling 497 hours (median 4.4 hours per session).

A total of 6,665 aggressive behaviors were observed: 3,983 \gls{sib} episodes (60\%), 2,063 \gls{ed} episodes (31\%), and 619 \gls{ato} episodes (9\%). Inter-rater reliability was high across all annotated behaviors (kappa values: 0.93, 0.95, and 0.86, respectively). Data preprocessing for this respective study may be found in \cite{potter2025temporal}, and a table outlining the count data for aggressive behavior onset is detailed in \cref{app:data}.

\subsection{Modeling Aggressive Behavior Onsets}
We aim to model the onsets of aggressive behavior as a \gls{tpp} to reliably estimate the sample population branching factor. For a reliable estimate of the branching factor, our method must address several challenges outlined in the seminal work of Filimonov and Sornette (2014), particularly:
\begin{enumerate}
    \item Sensitivity of the self-excitation kernel to outlier inter-arrival times
    \item Bias introduced by edge effects at the boundaries of observation sessions
    \item Presence of multiple likelihood extrema leading to suboptimal parameter estimation
    \item Heterogeneity versus homogeneity in the sample population and its impact on inference
\end{enumerate}

\subsubsection{Temporal Point Process}\label{subsubsec:tpp}
A \gls{tpp} is a statistical method that models the inter-arrival times of event sequences over continuous time. In this study, we leverage TPPs to model the timing of aggressive behavior onsets in minimally verbal psychiatric inpatient autistic youth. Thus, our \gls{tpp} is a stochastic process representing the occurrence of discrete events (aggressive behavior onsets) over a fixed time window (observation session). An observation session, or realization of a TPP, is an ordered sequence of aggressive behavior onset times, denoted as $S_i = \{ t_1, t_2, \ldots, t_{J_i} \}$ where $J_i$ is the number of onsets in session $S_i$ and $t_j$ is the time elapsed since the observation session's start when the $j$-th aggressive behavior onset occurred. We further specify $H_{t^-} = \{ t_j \in S_i,\, t_j < t \}$ as the history of past onsets up to time $t$ in an observation session.

A \gls{tpp} is characterized by the conditional intensity function:
\begin{equation}
    \lambda^*(t) = \lambda(t \mid H_{t}) = \lim_{\Delta t \to 0} \frac{P[t_{n+1} \in [t, t+\Delta t) \mid H_{t^-}]}{\Delta t}
\end{equation}
Heuristically, $\lambda^*(t)\Delta t$ for small $\Delta t$ is proportional to the probability of the next event occurring within a small observation window $[t,t+\Delta t]$, given the history of previous event times up to (but not including) time $t$.

The cumulative intensity function is the integral over the conditional intensity function, which is the expected number of aggressive behavior onsets within a time window $[0, t]$:
\begin{equation}
    \Lambda^*(t) = \int_0^t \lambda(s \mid H_s)\, ds
\end{equation}

The likelihood for an observation session $S_i$ on an observation interval $[0, T)$ is then given by
\begin{equation}
    L(\theta \mid S_i) = \exp\left(-\Lambda^*_\theta(T)\right) \prod_{j=1}^{J_i} \lambda^*_\theta(t_j)
\end{equation}
where $\theta$ denotes the parameters of the conditional intensity function, as we assume a parametric model.

For the entire dataset, With $I$ observation sessions $\mathcal{S} = \{ S_1, S_2, \ldots, S_I \}$, the joint likelihood is $L(\theta \mid \mathcal{S}) = \prod_{i=1}^I L(\theta \mid S_i)$, assuming independence between observation sessions due to the large time gaps separating them. We refer readers to \cite{rasmussen2018lecture} for more details on TPPs.

\subsubsection{Hawkes Point Process with Exponential Kernel}\label{subsubsec:hawkes}
The Hawkes Point Process is a \gls{tpp} where event occurrence increases the probability of future events, a phenomenon known as self-excitation. The conditional intensity function is divided into two components: baseline intensity and excitation trigger:
\begin{equation}
    \lambda^*(t) = \mu + \sum_{t_j < t} \phi(t - t_j)
\end{equation}
The baseline intensity $\mu$ captures events that occur even without prior onsets, while the summation term captures the self-exciting nature of the process. Another way to interpret this is that $\mu$ models exogenous onset occurrences, while the triggering kernel $\phi(t - t_j)$, which is nonnegative and causal, models endogenous onset occurrences. This formulation captures the cascading effect of aggressive behavior, wherein each onset increases subsequent onset likelihood, while the influence of the previous onsets gradually decays over time.

We choose the exponential kernel, $\phi(t) = \alpha \beta \exp(-\beta t)$, as \citet{filimonov2015apparent} demonstrated that this triggering kernel produces branching factor estimates robust to inter-arrival outliers (inter-arrival times that are twice the size of the 99th quantile of inter-arrival times). The branching factor $\alpha$ denotes self-excitation strength, while the time-scale parameter $\beta$ denotes how quickly prior events’ influence on future events weakens over time.

Edge effects arise when the observation session captures only a finite window of the underlying point process, omitting earlier unobserved events that can still influence observed activity, leading to biased estimates of the sample population branching factor. The conditional intensity function is adjusted for edge-effects by enabling the conditional intensity to begin at a specified value $\mu_0$ for the observation period start \cite{laub2021elements}:
\begin{equation}
    \lambda_\theta^*(t) = \mu + (\mu_0 - \mu)\beta \exp(-\beta t) + \alpha \beta \sum_{t_j < t} \exp(-\beta (t - t_j)) \label{eqn:hawkes_conditional_intensity}
\end{equation}

\noindent where $\Lambda^*_\theta(t)=\int_0^t \lambda^*_\theta(s) ds$.

In this formulation, $t=0$ denotes the beginning of the observation session rather than the typical definition of the data-generating process start time.

\subsubsection{Partially Pooled Modeling}\label{subsubsec:pooled}
We compare three modeling approaches: pooled, unpooled, and partially pooled (hierarchical) models. The pooled model assumes that all individuals share identical parameters, effectively treating the entire cohort as generated from a single homogeneous process. While capturing population-level trends, the pooled model ignores individual variation between persons. Previous work has used fully pooled \glspl{tpp} to estimate the sample population-level branching factor \cite{potter2025temporal}. However, because such models neglect non-negligent individual variability in autism \cite{hoemann2020context}, they can produce biased estimates dominated by high-frequency aggression individuals—in our cohort: one individual accounts for 25\% (1,287/4,871) of all onsets. Pooling in general leads to maximal underfitting \cite{gelman2006multilevel}.

In contrast, the unpooled model fits separate parameters for each individual, assuming complete independence and no information sharing across persons. This approach captures within-individual variation but fails to estimate population-level effects. Thus, although the unpooled \glspl{tpp} captures individual differences, it suffers from severe small-sample issues: some individuals have enough data for stable estimation, while others contribute too few events, leading to overfitting and high uncertainty in parameter posteriors. This generally results in maximal overfitting for individuals with few aggressive behavior onsets \cite{gelman2006multilevel}. Notably, 25\% (17/70) of individuals in our corpus did not exhibit any aggressive onsets, making it impossible to estimate individual branching factors and further biasing any population-level summary.

The partially pooled (hierarchical) model introduces a population-level structure by regularizing person-specific parameters through shared hyperpriors. This allows the model to learn the degree of information sharing across persons from the data, balancing individual heterogeneity with population-level trends \cite{gelman2006multilevel,paun2018comparing}. By borrowing strength across persons, the model stabilizes estimates when data is sparse while preserving individual variation. The hierarchical structure enables reliable inference of posterior distributions for the sample-population branching factor, with reduced bias from high- or low-frequency individuals and more realistic credible intervals that reflect uncertainty at individual and population levels \cite{gelman2006multilevel}.

We apply a partially pooled Hawkes Process model with exponential kernel to our data, where the parameters $\mu$, $\alpha$, and $\beta$ are person-specific but regularized by population-level hyperpriors. $\mu_0$ is observation session specific. We choose not to include $\mu_0$ as a person-specific parameter because it is not a parameter of the underlying generative process, but rather a correction for edge effects. Furthermore, when an observation has no aggressive onsets, the model reaches identifiability issues when estimating $\mu_0$ for these observation sessions. Thus, it is not appropriate to include it, which is equivalent to setting $\mu_0 = \mu$ for those observations. The generative process is (\cref{fig:partialpooled_model}):
\begin{figure}[h!]
\centering
\begin{tikzpicture}
\tikzset{
  myarrow/.style={-{Triangle[length=5pt, width=5pt]}},
}


\node[latent,thick] (mu_mu) at (-2.5, 4.2) {$\mu_\mu$};
\node[latent,thick] (sigma_mu) at (-1.2, 4.2) {$\sigma_\mu$};
\node[latent,thick] (mu_beta) at (0.5, 4.2) {$\mu_\beta$};
\node[latent,thick] (sigma_beta) at (1.8, 4.2) {$\sigma_\beta$};
\node[latent,thick] (mu_alpha) at (3.5, 4.2) {$\mu_\alpha$};
\node[latent,thick] (sigma_alpha) at (4.8, 4.2) {$\sigma_\alpha$};

\node[latent,thick] (mu_n) at (-1.85, 2.5) {$\mu_n$};
\node[latent,thick] (beta_n) at (1.15, 2.5) {$\beta_n$};
\node[latent,thick] (alpha_n) at (4.15, 2.5) {$\alpha_n$};

\node[latent,thick] (delta_mu_in) at (-1.8, 0.3) {$\Delta\mu_{in}$};
\node[obs,thick] (s_in) at (1.2, 0.3) {$S_{in}$};

\edge[thick,myarrow] {mu_mu} {mu_n};
\edge[thick,myarrow] {sigma_mu} {mu_n};
\edge[thick,myarrow] {mu_beta} {beta_n};
\edge[thick,myarrow] {sigma_beta} {beta_n};
\edge[thick,myarrow] {mu_alpha} {alpha_n};
\edge[thick,myarrow] {sigma_alpha} {alpha_n};

\edge[thick,myarrow] {mu_n} {s_in};
\edge[thick,dashed,myarrow] {beta_n} {s_in};
\edge[thick,dashed,myarrow] {alpha_n} {s_in};
\edge[thick,dashed,myarrow] {delta_mu_in} {s_in};

\node[font=\small,align=left, anchor=west] at (-5.8, 4.2) {Population Level};
\node[font=\small,align=left, anchor=west] at (-5.8, 2.5) {Person Level};
\node[font=\small,align=left, anchor=west] at (-5.8, 0.3) {Data Level};

\node[font=\footnotesize] at (2.05, 1.8) {if session};
\node[font=\footnotesize] at (2.05, 1.6) {has events};
\node[font=\footnotesize] at (-0.40, 0.65) {if session};
\node[font=\footnotesize] at (-0.40, 0.45) {has events};

\plate[thick,inner sep=0.2cm,yshift=0.1cm,xshift=0.1cm] {plate_i} {(delta_mu_in)(s_in)} {Onset $i=1\!:\!|\mathcal{S}_n|$};
\plate[thick,inner xsep=0.5cm , yshift=0.2cm,inner ysep=0.3cm] {plate_n} {(mu_n)(beta_n)(alpha_n)(plate_i)(s_in)} {Patient $n=1\!:\!N$};

\end{tikzpicture}
\caption{Hierarchical model structure for partially pooled Hawkes process.}
\label{fig:partialpooled_model}
\end{figure}
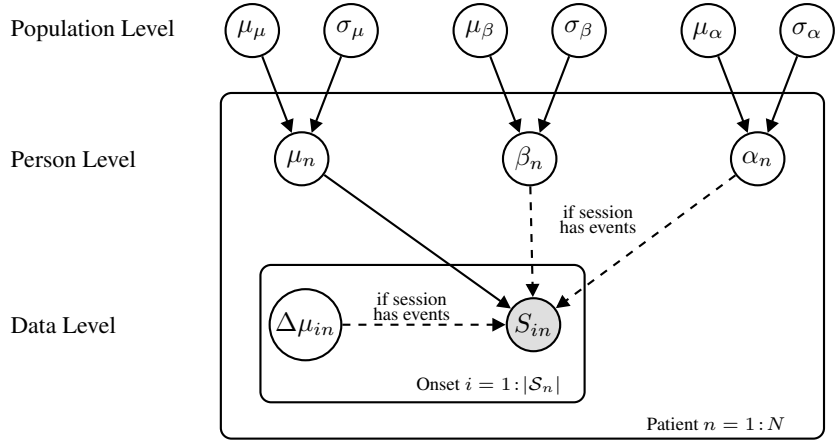

\vspace{-2em}
\noindent
\begin{align*}
    &\textbf{Level 0: Population-level} 
    &&\textbf{Level 1: Person-level} 
    &&\textbf{Level 2: Data-level} \\
    &\quad \mu_\mu \sim \mathrm{Half\text{-}Normal}(0.1) 
    &&\quad \mu_n \sim \mathrm{LogNormal}(\mu'_\mu, \sigma_\mu) 
    &&\quad \Delta\mu_{in} \sim \mathrm{Half\text{-}Cauchy}(0.1) \\
    &\quad \mu_\alpha \sim \mathrm{Gamma}(2.5,0.4) 
    &&\quad \alpha_n \sim \mathrm{LogNormal}(\mu'_\alpha, \sigma_\alpha) 
    &&\quad \mu_{0,in} = \mu_n + \Delta\mu_{in} \\
    &\quad \mu_\beta \sim \mathrm{Half\text{-}Cauchy}(1.5) 
    &&\quad \beta_n \sim \mathrm{LogNormal}(\mu'_\beta, \sigma_\beta) 
    &&\quad S_{in} \sim \mathrm{HawkesProcess}(\mu_n, \alpha_n, \beta_n, \mu_{0,in}) \\
    &\quad \sigma_\mu \sim \mathrm{Half\text{-}Cauchy}(1.5) 
    &&\quad \mu'_\mu = \log(\mu_\mu) - \sigma_\mu^2 / 2 \\
    &\quad \sigma_\alpha \sim \mathrm{Half\text{-}Cauchy}(1.0)
    &&\quad \mu'_\alpha = \log(\mu_\alpha) - \sigma_\alpha^2 / 2 \\
    &\quad \sigma_\beta \sim \mathrm{Half\text{-}Cauchy}(1.5)
    &&\quad \mu'_\beta = \log(\mu_\beta) - \sigma_\beta^2 / 2
\end{align*}

\noindent The generative process of the unpooled and the pooled model formulations is in \cref{app:bayesmodels}.

Following \cite{gelman2006prior}, we choose heavy-tailed weakly informative priors for the population-level variances and two population-level means. The population-level parameter $\mu_\alpha$ is the sample population branching factor, which is the expected number of direct children onsets triggered by an initial parent onset \footnote{A parent event is an event that generates or influences one or more subsequent events. It is considered the source of the triggering. A child event is n event that is generated or triggered by a parent event. It would not have occurred (or would be less likely) without the parent event. An exogenous event is an event not caused by any previous event (they are "born from the base intensity"). An endogenous event is triggered by previous events. Each triggering event becomes a parent, and each event it triggers is a child. }. For the unpooled model, we estimate the population branching factor as the mean of the individual branching factors, $\mu_\alpha = \frac{1}{N} \sum_{n=1}^N \mathbb{E}_{p(\alpha_n | \mathcal{D}_n)} [\alpha_n]$, where $N$ is the number of individuals and $\alpha_n$ is individual $n$ branching factor random variable.

\subsection{Parameter Inference}\label{subsec:inference}

We perform Bayesian inference using \gls{mcmc} to estimate posterior distributions of \gls{tpp} parameters. Specifically, we employ the \gls{nuts}, an adaptive variant of Hamiltonian Monte Carlo that uses gradient information to efficiently explore posterior distributions with strong correlations and nonlinear dependencies \cite{hoffman2014no}. By proposing distant, high-probability model parameter states, \gls{nuts} improves convergence and sampling efficiency compared to traditional random-walk methods.

We assess the ability of the \gls{nuts} to sample the posterior distribution using standard \gls{mcmc} diagnostics: the \gls{ess} to evaluate sampling efficiency (with a target of at least 100 effective samples), the Gelman–Rubin statistic (R-hat) to assess chain convergence (targeting values $<$ 1.05), and divergence checks to identify potential model misspecification \cite{gelman1995bayesian}. We also use rank plots to visually evaluate mixing across chains, where uniform rank distributions indicate good convergence \cite{vehtari2021rank}.

For inference in the pooled, partially pooled, and unpooled models, we ran four \gls{mcmc} chains with 1,000 warm-up iterations and 1,000 sampling iterations each, using a target acceptance probability of 0.95. The \gls{mcmc} demonstrated robust convergence, and the samples from the posterior distribution were of high quality: R-hat values were approximately 1.00, no divergent transitions occurred, \gls{ess} exceeded 3,000 for all key parameters, Monte Carlo standard errors were low, and rank plots showed uniformity across chains. These diagnostics collectively confirm the reliability of the posterior estimates. More details are found in \cref{apptab:mcmc_diagnostics_partial_pooled,apptab:mcmc_diagnostics_pooled,apptab:mcmc_diagnostics_unpooled}.

Bayesian inference offers several advantages over frequentist methods. First, it quantifies uncertainty in parameter estimates through credible intervals, providing more informative summaries than point estimates alone \cite{gelman1995bayesian}. Second, it accommodates complex models with hierarchical structures and latent variables that can be difficult to fit using frequentist techniques such as restricted maximum likelihood with mixed-effect models. Third, Bayesian inference handles multimodal likelihoods by representing uncertainty through the full posterior distribution, rather than converging to a single local optimum. Finally, it supports principled model comparison and selection via techniques such as \gls{psis-loo} cross-validation.

\subsection{Exogenous Versus Endogenous}
In a Hawkes point process, the branching structure refers to the latent, forest-like relationships between events, where each event is either an \emph{exogenous} (parent) or an \emph{endogenous} (child) event triggered by a previous event. Exogenous events are generated spontaneously by the baseline intensity, while earlier events trigger endogenous events according to the self-excitation kernel. This structure forms a branching process: exogenous events can trigger one or more endogenous events, which in turn may trigger further events, resulting in cascades or clusters of activity \cite{laub2021elements}.

For example, in the context of aggressive behavior onsets among psychiatric inpatient youths with autism:
\begin{itemize}
    \item \emph{Exogenous} events might correspond to aggressive behavior onsets triggered by external factors such as a sudden change in environment, a new staff member, medication changes, or an unexpected schedule disruption.
    \item \emph{Endogenous events} could represent subsequent aggressive behavior onsets triggered by distress or \gls{ed} following an initial aggressive incident, such as a chain of outbursts or \gls{sib} that cluster in time after the first event.
\end{itemize}

Although the true branching structure is unobserved, probabilistic inference enables us to estimate, for each event, the likelihood that it was exogenous or endogenous. After generating posterior samples via \gls{mcmc}, we can estimate the probability that each aggressive behavior onset is exogenous or endogenous using the following expressions for each event $i$ \cite{zhang2018efficient}:

\begin{itemize}
    \item Exogenous (parent) event:
    \[
        p_{i0} = \frac{\mu}{\lambda^*(t_i)}
    \]
    \item Endogenous (child) event triggered by the initial edge-effect:
    \[
        p_{i1} = \frac{(\mu_0 - \mu)\beta e^{-\beta (t_i - 0)}}{\lambda^*(t_i)}
    \]
    \item Endogenous event triggered by a previous event $j$ ($j \geq 2$):
    \[
        p_{ij} = \frac{\alpha \beta e^{-\beta (t_i - t_j)}}{\lambda^*(t_i)}
    \]
\end{itemize}
where the event index is $j-2$ because the first two events are the initial edge-effect and the first endogenous event, respectively. The sum of these probabilities, $\sum_{j=0}^N p_{ij}$, is 1, and they can be interpreted as the probability that event $i$ is exogenous, triggered by the initial edge-effect, or triggered by a previous event.

\subsection{Goodness of Fit}\label{subsec:goodness_of_fit}
We assess the \gls{gof} of the Hawkes Point Process with Exponential Kernel for the aggressive behavior onset point data by leveraging the \gls{rtct}. The \gls{rtct} states that given a realization of point data $S = \{ t_1, t_2, \ldots, t_J \}$ over time $[0, T]$ from a point process with a conditional intensity function $\lambda^*(\cdot)$, then the transformed point data $\{t^*_1,t^*_2, \ldots, t^*_{J}\} = \{ \Lambda^*(t_1), \Lambda^*(t_2), \ldots, \Lambda^*(t_J) \}$ come from a Poisson process with unit rate \cite{laub2024hawkes}. The \gls{rtct} interarrival-times $\{\tau^*_1, \tau^*_2, \ldots , \tau^*_J \} = \{\Lambda^*(t_1) - 0, \Lambda^*(t_2) - \Lambda^*(t_1), \ldots , \Lambda^*(t_{J_i}) - \Lambda^*(t_{J_i - 1}) \}$ follow an independent and identically distributed exponential distribution with unit rate. Therefore, the quality of model specification and inference can be assessed using statistical hypothesis tests that evaluate whether the transformed point data are consistent with a unit-rate Poisson process. Since the true intensity function $\lambda^*(t)$ and cumulative intensity function $\Lambda^*(t)$ are unknown, we follow standard practice and use an estimate $\Lambda^*(t) \approx \Lambda^*_{\bar{\theta}}(t)$, where $\bar{\theta} = \mathbb{E}_{p(\theta|\mathcal{S})}[\theta]$. 

\subsubsection{Lewis Test with Durbin's Modification}\label{subsubsec:lewis_test}
We use the Lewis test with Durbin’s modification, as it has been shown to be more powerful than tests based on conditional uniformity \cite{kim2015power}. For each observation session with at least five aggressive behavior onsets, we apply the \gls{rtct} to transform the point data and then conduct the Lewis test with Durbin's modification. The null hypothesis states that the transformed data follow a unit-rate Poisson process, while the alternative hypothesis allows for any deviation from this, i.e., that the data are drawn from a different distribution. Additional hypothesis tests—including the Kolmogorov–Smirnov test for assessing the marginal distribution and autocorrelation tests for detecting serial correlation—along with visualizations such as \gls{ecdf} plots and scatter plots, are presented in \cref{app:hypothesistestingextra}. For further analysis, we compute \glspl{ppc} of the Lewis test to account for uncertainty in parameter inference, yielding a distribution over the p-value for more nuanced interpretation, as shown in \cref{appsec:ppc_rtc}.

\subsubsection{Pareto Smoothed Importance Sampling - Leave One Out Cross-Validation}\label{subsubsec:psislocv}
Since we have posterior distributions for the model parameters, we use \gls{psis-loo} cross-validation when computing the \gls{elpd} to compare predictive performance of pooled, unpooled, and partially pooled models. While \gls{psis-loo} and \gls{waic} are asymptotically equivalent, \gls{psis-loo} is more robust in realistic, non-asymptotic settings, and less biased in the presence of influential observations or weak prior information \cite{vehtari2017practical}. Additionally, \gls{psis-loo} is fully Bayesian, unlike the Bayesian Information Criterion (BIC) or Akaike Information Criterion (AIC), which are frequentist approximations \cite{luo2017performances}. \gls{psis-loo} also provides diagnostic measures, such as the Pareto $\hat{k}$ diagnostic, to assess when its approximation may be unreliable. Moreover, \gls{psis-loo} is computationally efficient, avoiding refitting the model for each data point by using importance sampling. Higher \gls{elpd} values indicate better predictive performance. To improve the robustness of our \gls{elpd} estimates, we apply \gls{iwmm} \cite{paananen2021implicitly} for adaptive importance sampling when computing \gls{psis-loo} cross-validation \gls{elpd}.

\subsection{Sensitivty Analysis}\label{subsec:sensitivity_analysis}
Sensitivity analysis in Bayesian workflows assesses the extent to which posterior distributions are driven by prior data \cite{depaoli2020importance}, with sensitivity to prior or likelihood perturbations indicating potential prior-data conflict or likelihood noninformativity.
We evaluate the sensitivity of the posterior over Hawkes process parameters—particularly the branching factor—using power-scaling \cite{kallioinen2024detecting}, which perturbs the prior or likelihood by raising it to a power \( \delta > 0 \). Specifically, the prior becomes \( p_{\delta_{\text{pr}}}(\theta) = p(\theta)^\delta \), and the likelihood \( L_{\delta_{\text{lik}}}(\theta \mid \mathcal{S}) = L(\theta \mid \mathcal{S})^\delta \). Increasing \( \delta \) strengthens the component’s influence via sharpening; decreasing it weakens it via spreading.

Rather than rerunning \gls{mcmc} to sample from the perturbed posterior before computing a distance metric between the base and perturbed posterior, we use importance sampling to estimate the distance from the base to the perturbed posterior directly, avoiding repeated inference. Given samples \( \theta^{(m)} \sim g(\theta) \) from the base posterior and a target perturbed posterior \( f(\theta) \), we approximate:

\begin{equation}
\mathbb{E}_f[h(\theta)] \approx \frac{\sum_{m} h(\theta^{(m)}) w^{(m)}}{\sum_{m} w^{(m)}}, \quad w^{(m)} = \frac{f(\theta^{(m)})}{g(\theta^{(m)})}.
\end{equation}

Here, \( h(\theta) \) is a distance metric, such as the symmetrized square root of the Cumulative Jensen–Shannon (CJS) distance \cite{kallioinen2024detecting} that measures divergence between the empirical CDFs of the base and perturbed posteriors. For power-scaling, importance weights reduce to:
\begin{align}
w^{(m)}_{\delta_{\text{pr}}} = p(\theta^{(m)})^{\delta - 1}, \quad 
w^{(m)}_{\delta_{\text{lik}}} = L(\theta^{(m)} \mid \mathcal{S})^{\delta - 1}.
\end{align}

This enables efficient posterior comparison without re-estimating the perturbed posterior via \gls{mcmc}. For unpooled model formulations, we find that individuals with few or no aggressive behavior onsets exhibit high sensitivity to power-scaling of the prior and likelihood, reflecting high uncertainty in parameter inference when data is limited.

\section{Results}\label{sec:results}

\subsection{Branching Factor Estimates}\label{subsec:branching_factor_estimates}
The partially pooled model yielded a statistically significantly lower mean sample population branching factor than the pooled model and lower uncertainty than the unpooled model, due to its ability to capture both within- and between-person variability. As a result, the expected number of descendants per parent event is three times smaller than in the pooled model, leading to markedly different clinical implications for intervention planning and resource allocation.

\begin{figure}[h!]
\centering
\includegraphics[width=0.6\textwidth]{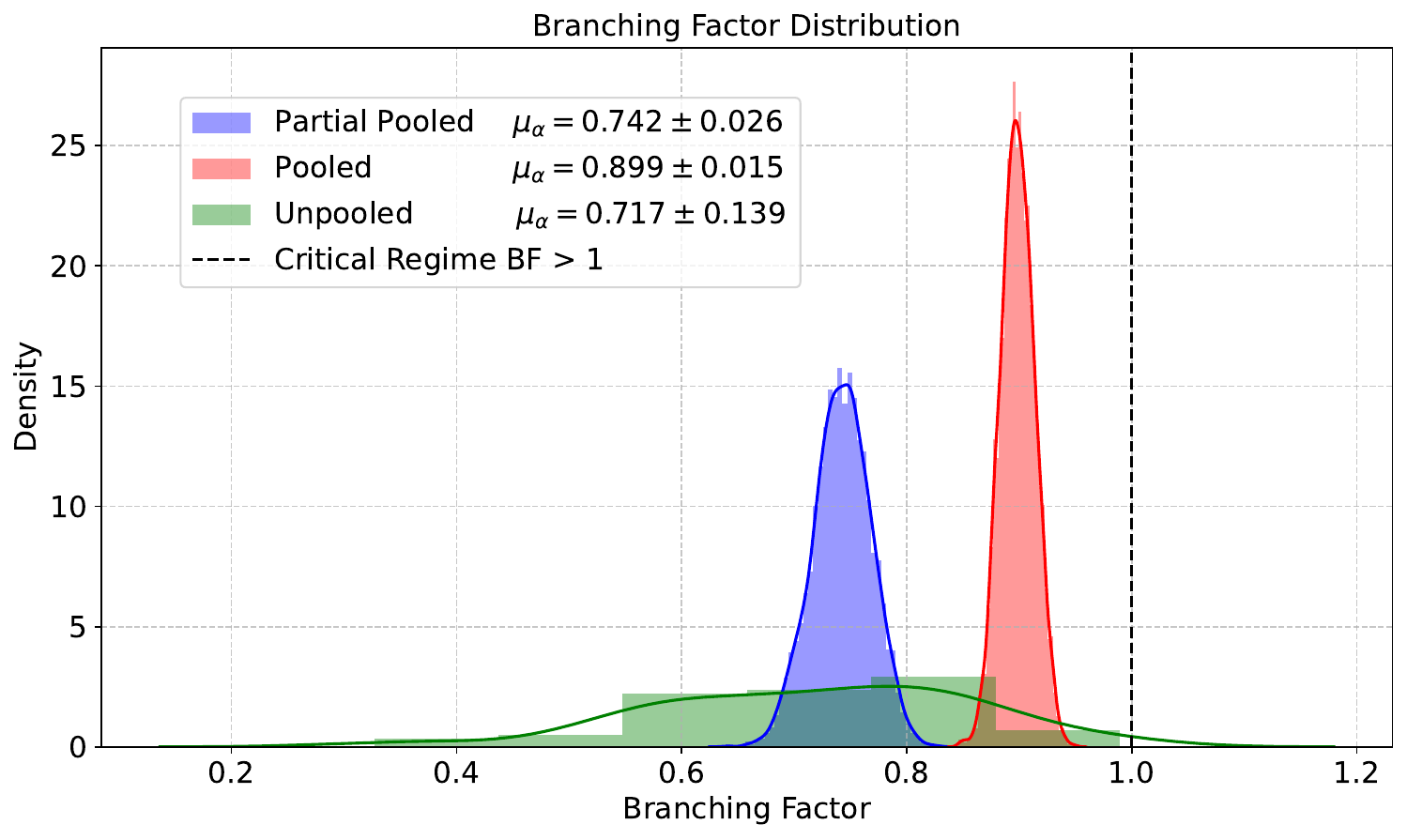}
\caption{Branching factor estimates for pooled, unpooled, and partially pooled models. The partially pooled model has a statistically significantly lower mean than the pooled model, with a substantially lower uncertainty than the unpooled model.}
\label{fig:branching_factor_estimates}
\end{figure}

Posterior distributions are shown in \cref{fig:branching_factor_estimates}. The pooled model yielded a mean branching factor of $0.899 \pm 0.015$, while the partially pooled model produced a statistically significantly lower mean of $0.742 \pm 0.026$ (Welch's $t$-test, $p=10^{-5}$, $n=4000$). The unpooled model had a mean of $0.717 \pm 0.139$, not statistically significantly different from the partially pooled model (Welch's $t$-test, $p=0.191$), but with a much wider credible interval, reflecting greater uncertainty. Thus, partial pooling reduces bias from high-frequency individuals and yields more certain parameter estimates of the sample population branching factor.

\begin{figure}[h!]
    \centering
    \includegraphics[width=0.6\linewidth]{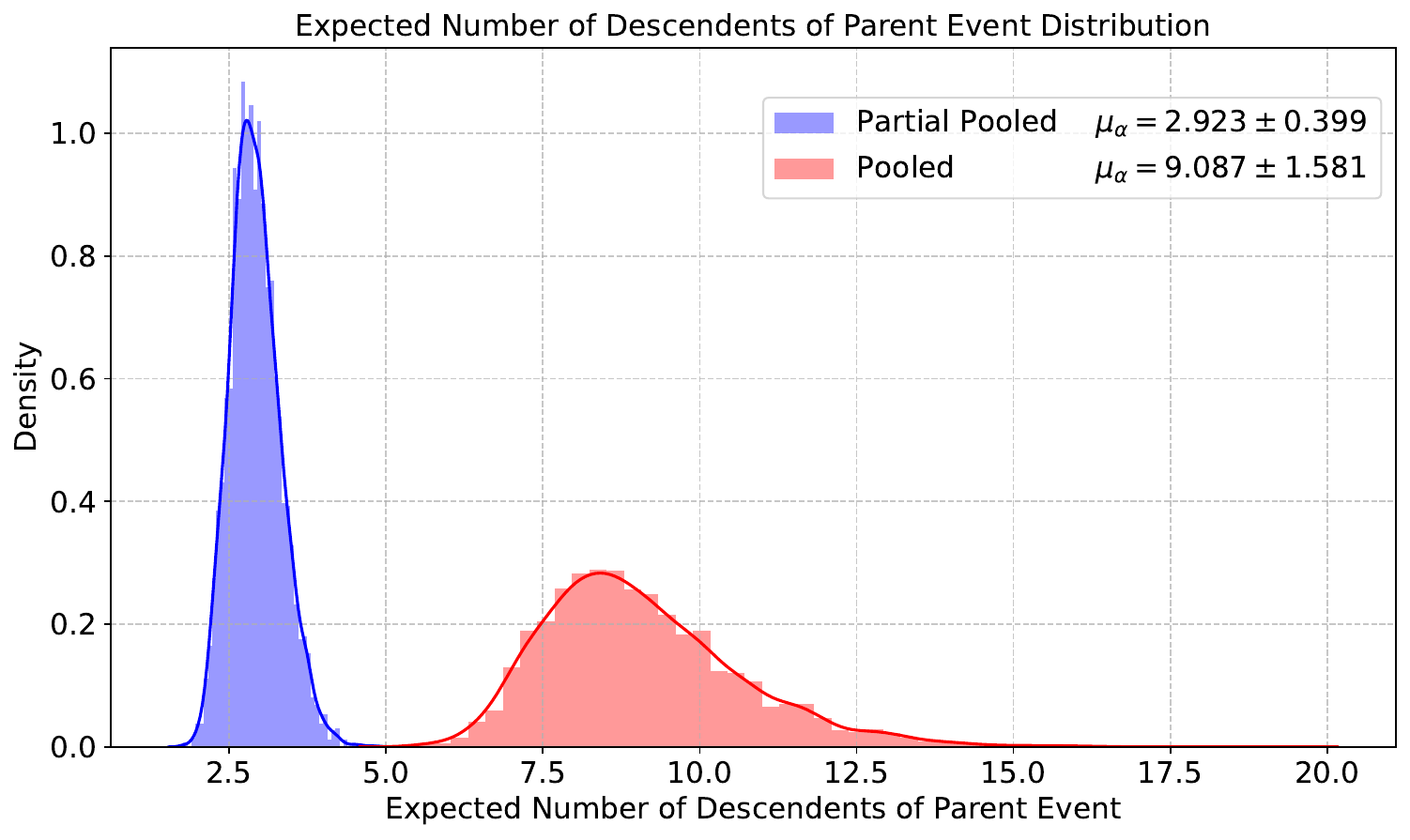}
    \caption{Expected total number of subsequent aggressive behavior onsets (descendants) triggered by a single parent onset, comparing pooled and partially pooled models.}
    \label{fig:expected_descendent}
\end{figure}

Although the sample population branching factor in the unpooled model is close to that of the partially pooled model, this similarity is largely driven by the choice of prior distribution. Specifically, the prior distribution for the sample population branching factor has an expectation of 1.0 and a median of approximately 0.87. Thus, when there is sparse data, or a non-informative likelihood, the posterior distribution becomes the prior distribution.

The unpooled model's posterior distribution is susceptible to the prior specification, particularly the scale, as shown in \cref{subsec:sensitivity_analysis_results}. In contrast, the partially pooled model's posterior distribution is more robust to power-scaling of the prior distribution (and the specification of the prior distribution), as it leverages data across individuals.

The expected total number of subsequent aggressive behavior onsets (descendants) triggered by a single parent aggressive behavior onset is given by $\mu_\alpha / (1-\mu_\alpha)$ when the branching factor $\mu_\alpha < 1$ \cite{laub2021elements}. For the partially pooled model, this expectation is $2.92 \pm 0.40$, while for the pooled model it is $9.09 \pm 1.58$. The difference is statistically significant (Welch's $t$-test, $p < 10^{-5}$, $n=4000$), indicating that the pooled model estaimtes a substantially higher expected cascade size compared to the partially pooled model. 

\subsection{Sensitivity of the Unpooled Model to Prior and Likelihood Power Scaling}
\phantomsection
\label{subsec:sensitivity_analysis_results}

We performed power-scaling sensitivity analysis of the posterior distribution for the sample population branching factor, following \cite{kallioinen2024detecting}. In the unpooled model, individuals with sparse or absent aggressive onsets exhibited high posterior sensitivity to prior and likelihood power-scaling, indicating strong prior influence and non-informative likelihoods. In contrast, the partially pooled model demonstrated robustness, with posterior estimates remaining stable under prior and likelihood perturbations.

A subset of the sensitivity analysis results for the unpooled model is shown in \cref{tab:sensitivity_analysis_unpooled}. The table summarizes how the posterior distributions of unpooled Hawkes Point Process parameters respond to perturbations in the prior and likelihood using the power-scaling approach described in \cite{kallioinen2024detecting}. The posterior is highly sensitive to these perturbations for individuals with few or no aggressive behavior onsets, indicating a lack of robustness due to prior-data conflict and likelihood non-informativity. In contrast, individuals with more aggressive behavior onsets show less sensitivity, reflecting greater robustness. Notably, even individuals with many observation sessions but no aggressive onsets tend to have non-informative likelihoods, resulting in strong prior influence and potential bias in the sample population branching factor estimate.

\begin{figure}[h!]
\centering
\begin{minipage}[t]{0.55\textwidth}
\centering
\captionof{table}{Prior/Likelihood Power Scaling Sensitivity Analysis Results for the Unpooled Model. The table shows the sensitivity of the posterior distribution over Hawkes Point Process parameters to perturbations in the prior and likelihood using the power-scaling approach. The subscript $i$ indicates the participant. The ``prior'' and ``likelihood'' columns show the cumulative Jensen-Shannon distance between the base and perturbed posterior distribution under power-scaling of the prior or likelihood, respectively. The ``diagnosis'' column indicates whether the parameter is robust (\checkmark) or shows potential sensitivity to the prior or likelihood.}
\begin{tabular}{lcccc}
\toprule
$\theta$ & Prior & Likelihood & Diagnosis \\
\midrule
$\alpha_0$ & 0.03 & 0.01 & \checkmark \\
$\beta_0$ & 0.05 & 0.01 & \checkmark \\
$\mu_0$ & 0.04 & 0.01 & \checkmark \\
\cdashline{1-4}[0.5pt/2pt]
$\alpha_{46}$ & 0.74 & 0.06 & Strong prior / weak likelihood \\
$\beta_{46}$ & 0.48 & 0.04 & Prior-data conflict \\
$\mu_{46}$ & 0.17 & 0.01 & Strong prior / weak likelihood \\
\cdashline{1-4}[0.5pt/2pt]
$\alpha_{49}$ & 0.40 & 0.05 & Strong prior / weak likelihood \\
$\beta_{49}$ & 2.41 & 0.15 & Prior-data conflict \\
$\mu_{49}$ & 0.06 & 0.02 & \checkmark \\
\cdashline{1-4}[0.5pt/2pt]
$\alpha_{52}$ & 0.39 & 0.05 & Strong prior / weak likelihood \\
$\beta_{52}$ & 0.73 & 0.24 & Prior-data conflict \\
$\mu_{52}$ & 0.06 & 0.02 & \checkmark \\
\bottomrule
\end{tabular}
\label{tab:sensitivity_analysis_unpooled}
\end{minipage}
\hspace{0.04\textwidth}
\begin{minipage}[t]{0.38\textwidth}
\centering
\captionof{table}{Prior/Likelihood Power Scaling Sensitivity Analysis Results for the Partial Pooled Model. The table shows the sensitivity of the posterior distribution over population-level Hawkes Point Process parameters to perturbations in the prior and likelihood using the power-scaling approach. The ``prior'' and ``likelihood'' columns show the cumulative Jensen-Shannon distance between the baseline and perturbed posterior distribution under power-scaling of the prior or likelihood, respectively. The ``diagnosis'' column indicates whether the parameter is robust (\checkmark) or shows potential sensitivity to the prior or likelihood.}
\begin{tabular}{lccc}
\toprule
$\theta$ & Prior & Likelihood & Diagnosis \\
\midrule
$\mu_\alpha$ & 0.00 & 0.38 & \checkmark \\
$\mu_\beta$ & 0.03 & 0.48 & \checkmark \\
$\mu_\mu$ & 0.04 & 0.25 & \checkmark \\
$\sigma_\beta$ & 0.03 & 0.51 & \checkmark \\
$\sigma_\mu$ & 0.03 & 0.35 & \checkmark \\
$\sigma_\alpha$ & 0.00 & 0.46 & \checkmark \\
$\alpha_{1:70}$ & NA & NA & \checkmark \\
$\beta_{1:70}$ & NA & NA & \checkmark \\
$\mu_{1:70}$ & NA & NA & \checkmark \\
\bottomrule
\end{tabular}
\label{tab:sensitivity_analysis_partialpooled}
\end{minipage}
\end{figure}

We include several representative individuals in \cref{tab:sensitivity_analysis_unpooled} for the discussion section, with additional examples provided in \cref{apptab:power_scale_analysis_unpooled}. Participant 0 has 1287 aggressive behavior onsets recorded over 2000 minutes of observation across 24 sessions. In contrast, participant 46 has only 3 onsets within 22 minutes from a single session; participant 49 has 2 onsets across 129 minutes over 2 sessions; and participant 52 has just 1 onset in 104 minutes, also across 2 sessions. Although participants 49 and 52 each have approximately two hours of observation, the sparse number of onsets per session and the average observation per session is around an hour or less, resulting in high sensitivity to prior and likelihood perturbations. Conversely, participant 0’s large number of onsets yields a more stable and robust posterior distribution. This may be explained by the low baseline intensity at the population level—and even lower values for specific individuals. For example, if a person has a baseline intensity of 0.01, the likelihood of observing no aggressive behavior onsets in a 50-minute session is given by $L(\theta \mid S_i) = \exp(-0.01 \cdot 50) = 0.6065$. This relatively high likelihood implies that the absence of observed onsets is not unexpected under the model, suggesting that such individuals should be observed for longer durations to obtain more informative data.

However, for the partial pooled model, the sensitivity analysis results indicate that the posterior distributions are robust to power-scale perturbations in the prior (hyperprior, as suggested by \cite{kallioinen2024detecting}) and likelihood \cref{tab:sensitivity_analysis_partialpooled,apptab:power_scale_analysis_partialpooled}.

\begin{figure}[h!]
    \centering
    \includegraphics[width=0.6\textwidth]{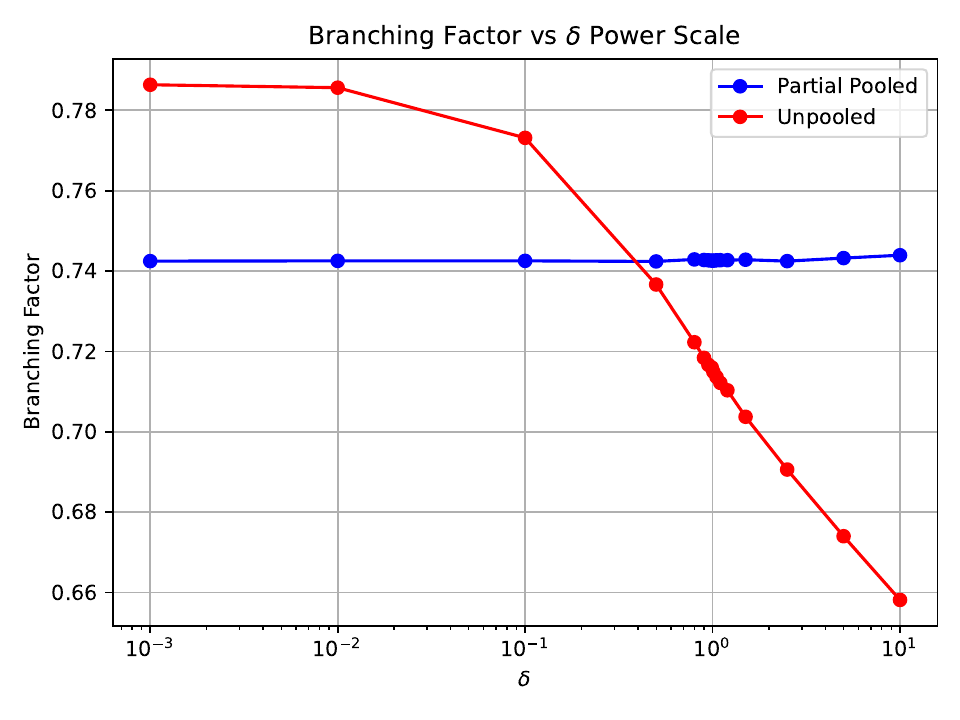}
    \caption{Branching factor estimates for the unpooled and partially pooled models after various power-scale perturbations in the prior of the branching factor. The unpooled model's branching factor estimates are highly sensitive to prior perturbations, while the partially pooled model's estimates remain stable across power-scale perturbations.}
    \label{fig:bf_sensitivity}
\end{figure}

To underscore the importance of sensitivity analysis, we apply power-scaling to the sample population branching factor prior in both the unpooled and partially pooled models (\cref{fig:bf_sensitivity}).

The unpooled model exhibits substantial variation in sample population branching factor estimates under branching factor prior perturbations via power-scaling, indicating high sensitivity. In contrast, the partially pooled model yields stable estimates of the population branching factor across power-scaled priors, demonstrating greater robustness. Furthermore, when we rerun posterior distribution inference via \gls{mcmc}—rather than relying on importance sampling—we observe even larger shifts in the unpooled model's branching factor estimates when altering the prior distribution (e.g., changing from a half-Cauchy to a gamma prior or adjusting the scale parameters). This further highlights the unpooled model's sensitivity to prior specification. 

\subsection{Branching Cascades}

Visualizing inferred branching structures and exogenous event probabilities alongside physiological signals (e.g., \gls{eda} and accelerometer energy defined as the $\ell_2$ norm of x-y-z acceleration) reveals promising qualitative alignment between physiological changes and behavioral events. These preliminary findings highlight the potential of Hawkes models to support cognitive-behavioral assessment by identifying physiological precursors to aggression, motivating future quantitative validation and further modeling improvement. 

\begin{figure}[h!]
    \centering
    \begin{minipage}[t]{0.9\textwidth}
        \centering
        \includegraphics[width=\textwidth]{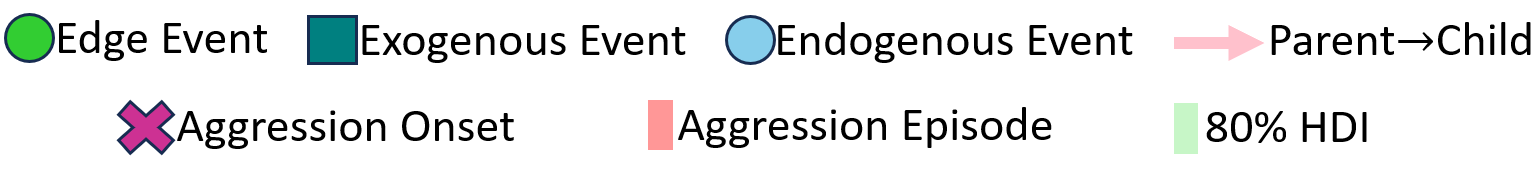}
    \end{minipage}
    \vspace{0.5em}
    \begin{minipage}[t]{0.465\textwidth}
        \centering
        \includegraphics[width=\textwidth]{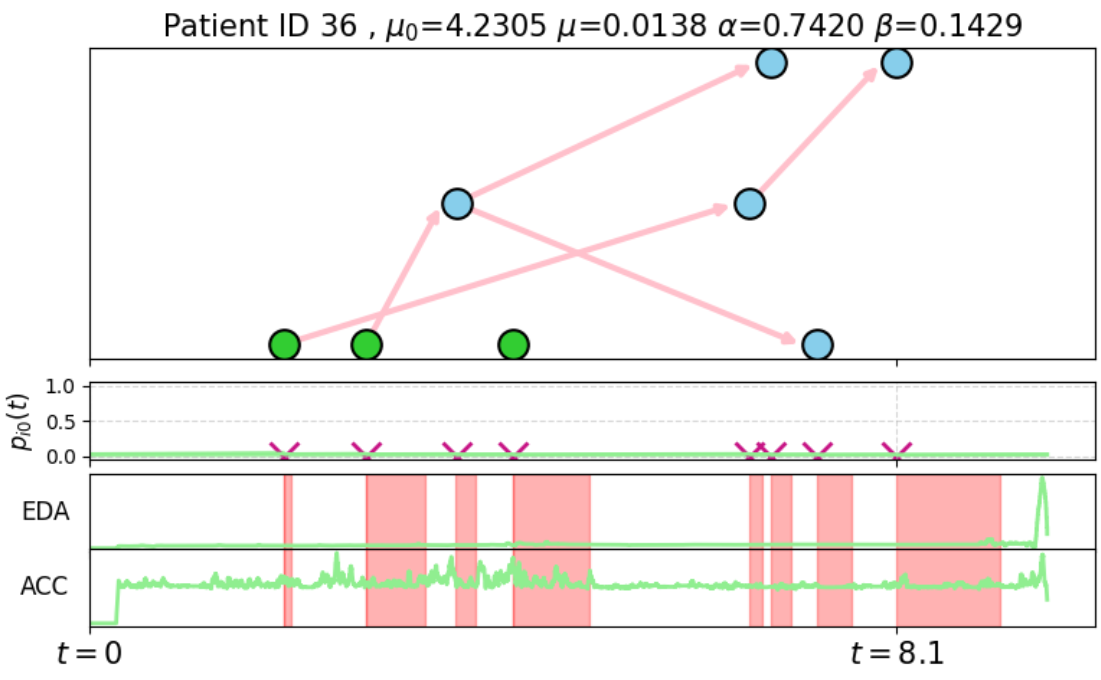}
        \subcaption{Session 131 Branching Structure}
        \label{fig:branching_structure_a}
    \end{minipage}\hfill
    \begin{minipage}[t]{0.48\textwidth}
        \centering
        \includegraphics[width=\textwidth]{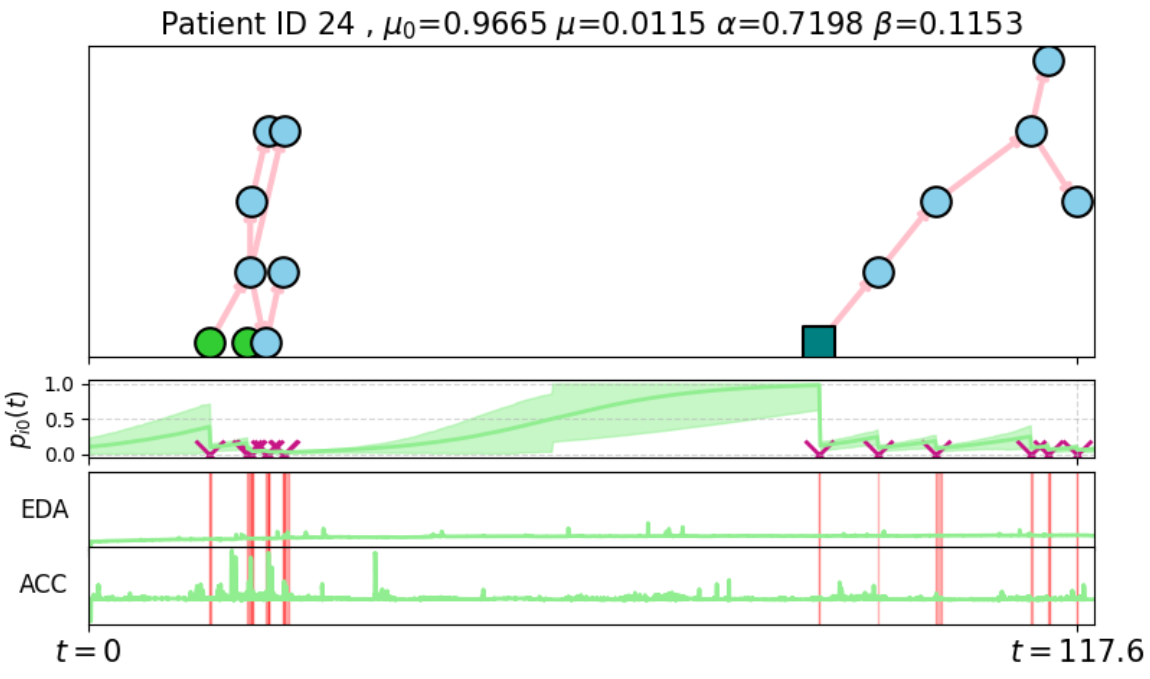}
        \subcaption{Session 77 Branching Structure}
        \label{fig:branching_structure_b}
    \end{minipage}
    \vspace{0.5em}
    \begin{minipage}[t]{0.48\textwidth}
        \centering
        \includegraphics[width=\textwidth]{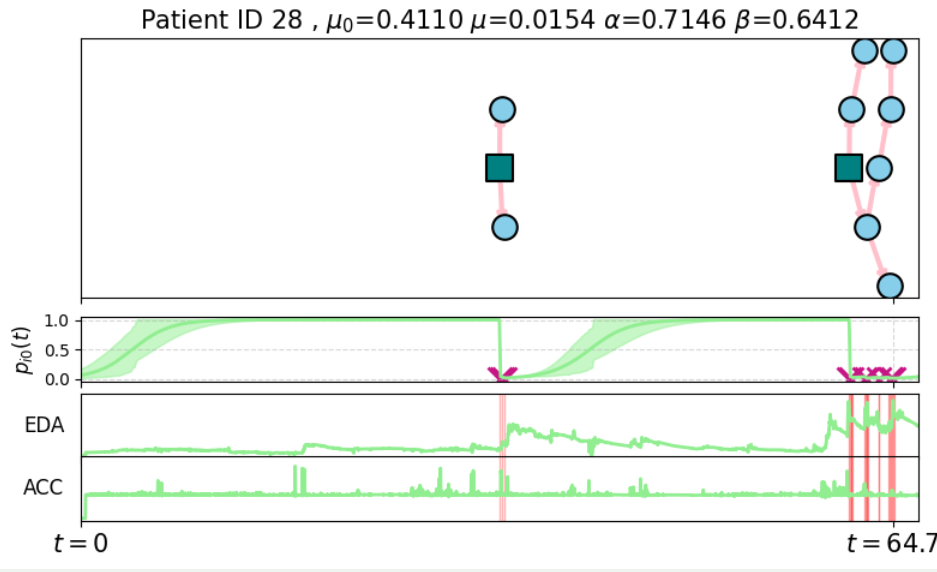}
        \subcaption{Session 95 Branching Structure}
        \label{fig:branching_structure_c}
    \end{minipage}\hfill
    \begin{minipage}[t]{0.48\textwidth}
        \centering
        \includegraphics[width=\textwidth]{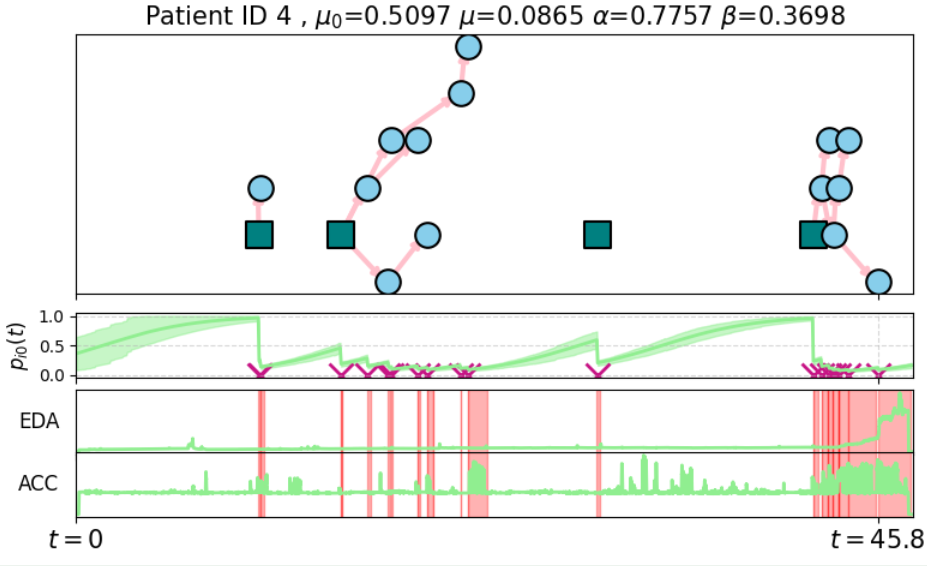}
        \subcaption{Session 62 Branching Structure}
        \label{fig:branching_structure_d}
    \end{minipage}
    \vspace{0.5em}
    \begin{minipage}[t]{0.48\textwidth}
        \centering
        \includegraphics[width=\textwidth]{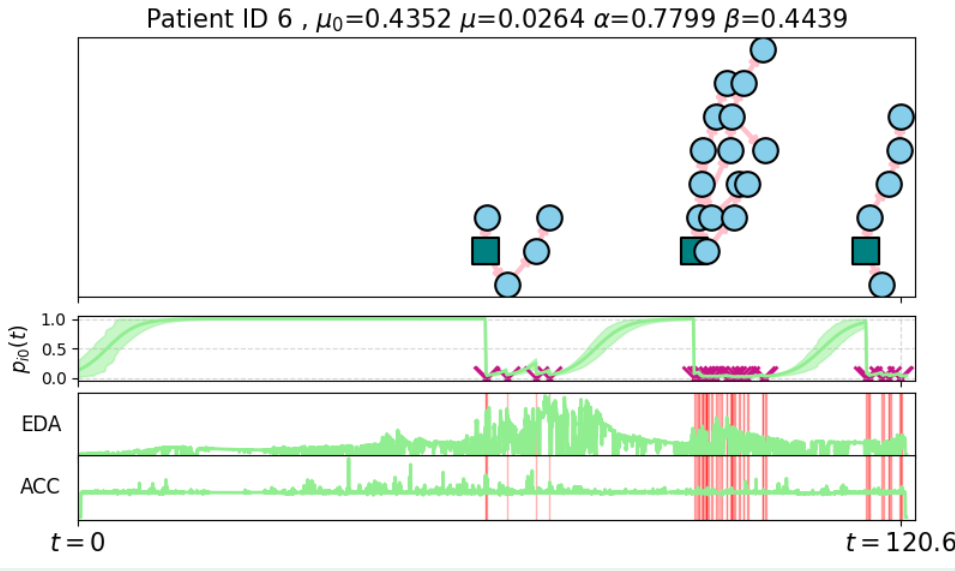}
        \subcaption{Session 40 Branching Structure}
        \label{fig:branching_structure_e}
    \end{minipage}\hfill
    \begin{minipage}[t]{0.48\textwidth}
        \centering
        \includegraphics[width=\textwidth]{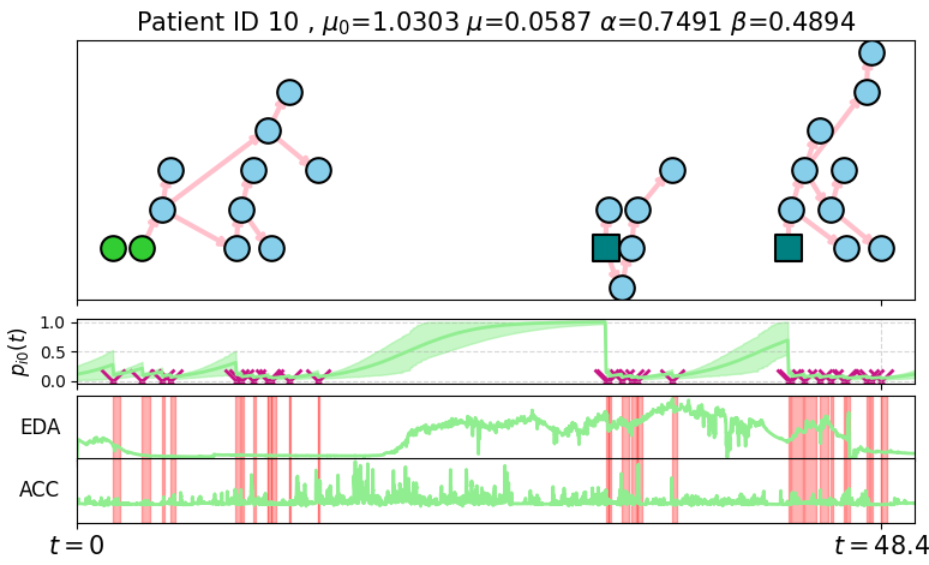}
        \subcaption{Session 41 Branching Structure}
        \label{fig:branching_structure_f}
    \end{minipage}
    \vspace{0.5em}
    \caption{Examples of inferred branching structures (first row of each subfigure) and probability of exogenous event (second row of each subfigure) for six observation sessions. (a) Session 131, (b) Session 77, (c) Session 95, (d) Session 62, (e) Session 40, and (g) Session 41. Green circles indicate endogenous events triggered by the edge effect, light blue circles represent child (offspring) events from subsequent generations, and green squares denote parent (exogenous) events. Magenta x's mark observed aggressive behavior onset times (point data). The shaded area in the probability of onset being exogenous if occurring at time $t$ is the highest density interval from the posterior distribution. In the physiological signals, the shaded red line indicates aggressive behavior (not just the onset, but the continued behavior). The timeline is in minutes.}
    \label{fig:branching_structure_examples}
    \label{fig:branching_structure_examples}
\end{figure}

We present examples of branching structures sampled from the inferred posterior distribution of the partially pooled model in \cref{fig:branching_structure_examples}. In each subfigure (\cref{fig:branching_structure_a,fig:branching_structure_b,fig:branching_structure_c,fig:branching_structure_d,fig:branching_structure_e,fig:branching_structure_f}), the top row shows the branching cascade for a single observation session: endogenous events are depicted as light blue circles, exogenous events as green squares, and initial edge-effect events as green circles (child events with unobserved parents). The middle row displays the probability (with 80\% \gls{hdi}) that an event is exogenous at time $t$, with magenta x's marking observed aggressive behavior onsets. The bottom row shows peripheral physiological signals, with lightly shaded red regions indicating aggressive episodes. Note that an aggressive behavior episode may consist of a continuous sequence of aggressive events, but only the first event in each episode is considered the onset. The red shading indicates the entire duration of the aggressive episode, while the onset corresponds to the initial event that triggers the episode. These visualizations illustrate how onsets may be triggered by preceding events, forming a forest-like structure of causally related events.

In participant 36 (\cref{fig:branching_structure_a}), multiple endogenous events are attributed to a parent event that occurred before the start of the observation session, highlighting the role of edge effects. Without an initial intensity term $\mu_0$, the first event in each session would necessarily be assigned to an exogenous source. Participant 28 (\cref{fig:branching_structure_c}) demonstrates two exogenous events spaced far apart in time, with multiple generations of child events occurring between them. This illustrates the temporal clustering of aggressive behavior onsets, where sparse exogenous events spawn bursts of endogenous activity. For participant 24 (\cref{fig:branching_structure_b}), the first tree of events is attributed to edge effects—consistent with its proximity to the beginning of the session—while a second tree is triggered by an exogenous event occurring later in the session. 

A particularly illustrative example of how Hawkes models can relate physiological signals to behavioral observations is seen in participant 28 (\cref{fig:branching_structure_c}). In this session, bursts of aggressive behavior consistently coincide with sharp increases in \gls{eda}, which are often preceded by elevated acceleration. After each aggressive episode, \gls{eda} levels gradually decrease. This pattern suggests that changes in physiological signals may serve as precursors or correlates to aggressive behavior onsets, highlighting the potential utility of Hawkes models for linking physiological dynamics to behavioral events. Similar patterns are observed in a session for participant 6, where elevated or variable \gls{eda} and high acceleration occur before or during aggressive onsets (\cref{fig:branching_structure_e}). 

These visualizations suggest Hawkes point process models may help align physiological changes with behavioral events, supporting cognitive-behavioral assessment and identifying prodromal or precursor physiological markers. However, these are preliminary observations; rigorous quantitative analyses—such as statistical correlation of physiological change points, trends, or peaks with behavioral onsets—are needed to confirm these relationships and guide future data collection. We include this selective analysis as motivation for the importance of linking exogenous events with physiological signals.

\subsection{Goodness of Fit}\label{subsec:goodness_of_fit_results
}

We demonstrate that the partially pooled (hierarchical) model consistently achieves superior or comparable \gls{gof} across multiple metrics—including \gls{psis-loo} cross-validation, the Lewis test with Durbin’s modification, and analyses of \gls{rtct} inter-arrival times—relative to both pooled and unpooled model formulations. These results indicate that the hierarchical approach provides more reliable and robust modeling of aggressive behavior onset dynamics in minimally verbal psychiatric inpatient youth with autism.

The \gls{psis-loo} cross-validation estimate of \gls{elpd} results for the pooled, unpooled, and partially pooled models are shown in \cref{tab:goodness_of_fit_results}. The effective number of parameters ($p$) is highest for the unpooled model, indicating that it has the most flexibility to fit the data, but has a slightly lower \gls{elpd} compared to the partially pooled model. The partially pooled model has the lowest \gls{elpd} value, indicating that it is the best model for predicting imminent aggressive behavior onsets in this sample population. While there is a statistically significant difference between the pooled and partially pooled / unpooled models (based on a two-sample Welch t-test, $p=0$), the difference between the unpooled and partially pooled models is not statistically significant ($p=0.88$). The weight for the partially pooled model is 0.50, indicating that it is the preferred model for predicting aggressive behavior onsets in this sample population.

\begin{table}[h!]
\centering
\caption{Goodness of Fit Results. The table shows results of the Pareto-Smoothed Importance Sampling - Leave One Out Cross-Validation (PSIS-LOO-CV) Expected Log Density Prediction (ELPD) for the pooled, unpooled, and partially pooled models. The effective number of parameters (p) is also shown. The weight is calculated using stacking to estimate model preference for prediction. Estimated from 429 observation sessions.}

\begin{tabular}{lrrrrrrr}
\toprule
 & rank & elpd & p & elpd\_diff & weight & se & dse \\
\midrule
Partial Pooled & 0 & -7379.82 & 78.9 & 0.00 & 0.51 & 458.71 & 0.00 \\
Unpooled & 1 & -7384.62 & 89.78 & 4.81 & 0.41 & 457.54 & 17.22 \\
Pooled & 2 & -7640.04 & 39.93 & 260.22 & 0.08 & 474.27 & 39.47 \\
\bottomrule
\end{tabular}
\label{tab:goodness_of_fit_results}
\end{table}

The p-values for the Lewis test with Durbin’s modification are presented in \cref{tab:lewis_test_results}. At the session level, the partially pooled model yields the highest proportion of sessions that do not reject the null hypothesis, followed by the unpooled and pooled models. At the individual level, the partially pooled and unpooled models perform similarly. The pooled model consistently exhibits the lowest proportion of non-rejections in both session- and person-level analyses, suggesting the poorest fit among the three modeling approaches.

\begin{table}[h!]
\centering
\caption{Lewis test with Durbin's modification: rejection rates at various significance levels for each model. Results are reported for both session- and person-level analyses. At the session level, we report the average proportion that does not reject the null hypothesis. At the person-level, we report the average proportion where the null hypothesis was rejected. Only sessions with more than 5 events were included (157 out of 429 sessions, corresponding to 39 out of 70 participants).}
\begin{tabular}{lccccccc}
\toprule
 & \multicolumn{3}{c}{Session-level} & & \multicolumn{3}{c}{Person-level} \\
\cmidrule{2-4} \cmidrule{6-8}
Significance Level & Partial Pooled & Unpooled & Pooled & & Partial Pooled & Unpooled & Pooled \\
\midrule
0.050 & 0.86 & 0.88 & 0.83 & & 0.86 & 0.90 & 0.86 \\
0.100 & 0.81 & 0.83 & 0.69 & & 0.83 & 0.84 & 0.71 \\
0.150 & 0.75 & 0.73 & 0.64 & & 0.76 & 0.72 & 0.67 \\
\bottomrule
\end{tabular}
\label{tab:lewis_test_results}
\end{table}

As a baseline reference, we fit a Homogeneous Poisson point process model, which assumes a constant aggressive behavior onset rate, and compute the \gls{psis-loo} cross-validation estimate of the \gls{elpd} for the pooled, unpooled, and partially pooled versions. The resulting \gls{elpd} values are $-13727.11 \pm 1004.24$ for the pooled model, $-9913.58 \pm 606.97$ for the unpooled model, and $-9892.86 \pm 608.32$ for the partially pooled model. These values are substantially lower than those obtained in the Hawkes point process with an exponential kernel. This demonstrates that the Hawkes model provides a statistically significantly better fit to the aggressive behavior onset point data. In addition, the Lewis test with Durbin’s modification rejects the null hypothesis (that the data follow a unit-rate Poisson process) far more frequently under the Homogeneous Poisson point process model than under the Hawkes point process model (\cref{tab:poisson_gof_results}), further supporting the superiority \gls{gof} of the latter approach. Similar findings have been reported in previous work \cite{potter2025temporal}.

\begin{table}[htbp]
\centering
\caption{Goodness of fit results for the Poisson Process model. The table shows the proportion of sessions that do not reject the null hypothesis at various significance levels for the partial pooled and pooled models.}
\begin{tabular}{llll}
\toprule
 & Partial Pooled & Unpooled & Pooled \\
Significance Level &  &  &  \\
\midrule
0.050000 & 0.496815 & 0.496815 & 0.496815 \\
0.100000 & 0.420382 & 0.420382 & 0.420382 \\
0.150000 & 0.369427 & 0.369427 & 0.369427 \\
\bottomrule
\end{tabular}
\label{tab:poisson_gof_results}
\end{table}

\section{Discussion}\label{sec:discussion}

Estimating the branching factor is a common problem across finance, epidemiology, and seismology domains. In finance, self-exciting Hawkes processes are applied to high-frequency data to estimate the criticality index, a proxy for market volatility and dynamics \cite{hardiman2013critical}. In epidemiology, the branching factor helps determine whether disease spread will subside or persist indefinitely \cite{jacob2010branching}. In seismology, Epidemic-Type Aftershock Sequences models estimate branching factors that quantify the expected number of triggered aftershocks following a primary earthquake \cite{ogata1988statistical}.

In most of these applications, the datasets either consists of a single observation session with an extremely long time window and a very high event count (on the order of at least 100{,}000 events), or assumes that all observation sessions are generated from the same underlying process—effectively treating the set of observation sessions as homogeneous. Non-parametric estimation techniques for the self-excitation kernel, which directly inform the branching factor, typically require approximately 100{,}000 events per session to yield stable and reliable estimates \cite{bacry2016first}. However, our dataset contains fewer than 5{,}000 events generated from 70 participants, making it unsuitable for such methods. Therefore, we employ parametric modeling for the self-exciting process.

Parametric modeling of Hawkes processes yields closed-form expressions for the branching factor but introduces its own challenges. First, specific excitation kernels used in the literature can introduce significant bias. For instance, power-law kernels often ``exogenize" endogenous events by misclassifying them—particularly when inter-event intervals are much longer than average—leading to inflated branching factor estimates \cite{filimonov2015apparent}. To address this, we selected the exponential kernel, which is more robust to outliers and better preserves the classification of events as endogenous, yielding more reliable branching factor estimates.

A second issue involves edge effects, which occur because the observation session does not start at the true beginning of the underlying point process, and the session is observed over a finite time window rather than the entire process duration. As a result, events that occurred before the start of the observation window—but are unobserved—can still influence events within the session. This unobserved history can bias parameter estimates, for example, by causing the model to misclassify the first few observed events as exogenous (background) events, thereby inflating the estimated baseline intensity. Accounting for edge effects is essential to avoid biased temporal point process modeling inference. To mitigate this, we incorporated an initial intensity term to model latent triggering activity before the observation window \cite{laub2024hawkes}. We conceptualize exogenous events as aggression caused by external stimuli—such as environmental disruptions, staffing changes, or medication timing—while prior aggressive episodes, such as residual distress or ongoing dysregulation, trigger endogenous events. This distinction is critical for understanding individual differences in aggression dynamics and informing whether intervention strategies should, for instance, target upstream environmental controls or post-incident de-escalation.

A third issue involves within- and between-person variability in estimating the sample population branching factor. In the context of aggression dynamics, the branching factor can suggest non-negligible variance across individuals due to differences in underlying neurobiology, environmental factors, and intervention responses. For example, some individuals may have a higher propensity for aggression due to neurobiological predispositions or past trauma. In contrast, others may respond more effectively to interventions involving behavior modification or PRN medications. Our previous work assumed participants were homogeneous, which can lead to biased estimates of the sample population branching factor and potential misinterpretation of aggression dynamics \cite{potter2025temporal}. Thus, in the current study, we employed a partially pooled model to capture both within- and between-person variability in the branching factor. This approach allowed us to obtain more accurate estimates of the sample population branching factor while accounting for individual differences.

To the best of our knowledge, the current work addresses all the aforementioned issues. In summary, we use the exponential kernel to avoid "homogenizing" endogenous events, incorporate an initial intensity term to account for edge effects, and employ a partially pooled model to capture within- and between-person variability in the branching factor. This approach allows us to obtain more accurate estimates of the sample population branching factor and better understand the dynamics of aggression in this sample population.

\subsection{Implications for Clinical Practice}
Separating exogenous versus endogenous aggression behavior onset factors has potential utility in guiding more individualized and thus effective clinical care.  For example, if an individual has a high branching factor, it suggests a higher likelihood of more frequent endogenous aggression behavior onsets, which could be due to residual agitation or ongoing emotional dysregulation. In this case, clinicians may want to focus on interventions that target the individual's underlying neurobiology or physiology, such as medication adjustments or behavioral therapy. On the other hand, if an individual has a low branching factor, they may be more likely to have exogenous aggression behavior onsets due to environmental factors or external stimuli. In this case, clinicians may want to focus on interventions that target the environment, such as reducing noise levels or providing means for augmentative and alternative communication. 

Additionally, the branching factor can quantify the total expected number of subsequent aggressive onsets triggered by a single event. Clinicians can use this information to anticipate escalation risk and tailor intervention intensity: higher branching factors suggest the need for closer monitoring and proactive intervention, while lower values may warrant less intensive approaches.

Furthermore, isolating exogenous events can help clinicians correlate physiological signals, such as heart rate, electrodermal activity, or acceleration data, with aggression behavior onsets. This would enable more precise investigation of whether physiological changes precede exogenous aggression, supporting causal inference about the relationship between physiological arousal and behavior. By distinguishing between aggression triggered by external factors and that arising from internal states, researchers could better identify predictive biomarkers and refine risk assessment models. Additionally, this separation allows for more targeted analysis of antecedents and consequences, facilitating the development of early warning systems that leverage physiological monitoring to anticipate and potentially prevent aggressive episodes.

\subsection{Limitations and Future Work}\label{sec:limitations_future_work}
This study has several limitations that suggest directions for future research. First, the partially pooled model could be extended to include additional hierarchical levels or stratification by hospital, gender, or age, potentially improving branching factor estimation. Incorporating covariates—such as physiological signals or contextual variables—may enhance predictive accuracy, interpretability, and model fit. Exploring alternative excitation kernels with more extended memory, or zero-inflated models to address the substantial proportion of participants / sessions with no observed aggression, could further refine inference. Finally, while bootstrap methods could help correct for the conservativeness of hypothesis tests—such as the Lewis test with Durbin's correction—the computational cost of running bootstrapping in conjunction with \gls{mcmc} is prohibitive. As such, we may consider reverting to \gls{mle}-based inference in future work to enable more practical resampling approaches. However, to compensate this currently we compute \gls{ppc} over the p-value of the Lewis test with Durbin's modification to incorporate the model estimation uncertainty in the p-value, as shown in \cref{appfig:ppc_lewis_combined}.

It is also essential to consider that inpatient hospitalization often coincides with initiation or adjustment of prescription medications, which may influence observed behaviors. As such, the findings reported here may not generalize to the broader population of inpatient autistic youth. Nevertheless, the sizeable subset of autistic individuals at risk for psychiatric hospitalization and severe \gls{sib}, \gls{ato}, or \gls{ed} represents a critical population to characterize, understand, and support better.

\glsresetall
\section{Conclusion}\label{sec:conclusion}
In this work, we developed and rigorously evaluated a Bayesian hierarchical Hawkes point process model with an exponential kernel to estimate the sample population branching factor governing aggressive behavior onsets in a sizeable group of minimally verbal psychiatric inpatient youth with autism. Our approach systematically addressed key challenges in temporal point process modeling of clinical event data, including edge-effect correction, between- and within-person variance, and instability in parameter inference under small-sample regimes. Introducing a partially pooled (hierarchical) structure enabled principled regularization of person-specific parameters through population-level hyperpriors, mitigating overfitting in low-data individuals and bias from high-frequency individuals—issues that limit pooled and unpooled models.

Empirical results demonstrated that the partially pooled model produced statistically significantly lower branching factor estimates than pooled models, with credible intervals that better captured population uncertainty than unpooled models. Sensitivity analyses using power-scaling of priors and likelihoods confirmed the robustness of the hierarchical model, in contrast to the high sensitivity observed in unpooled models due to prior-data conflict and strong prior influence with non-informative likelihoods. \gls{gof} was validated using \gls{psis-loo} cross-validation to estimate the \gls{elpd}, the Lewis test with Durbin’s modification, and analyses of \gls{rtct} inter-arrival times—all supporting the superiority of the partially pooled Hawkes process over both pooled and unpooled alternatives and Poisson process baselines.

Additionally, our probabilistic inference of latent branching structures enabled decomposition of observed events into exogenous (environmentally triggered) and endogenous (self-excited) components, providing a principled basis for causal interpretation of aggression cascades. This distinction has direct clinical implications, potentially informing more targeted intervention strategies and supporting future physiological or contextual covariates integration into the model.

\glsresetall
\section{Software Packages}\label{sec:software_shoutout}
All analyses were conducted in Python, following the \textit{Cookiecutter Data Science} project structure and workflow \cite{drivendata2025}. Bayesian inference was performed using \verb|NumPyro| \cite{phan2019composable}, a probabilistic programming language built on \verb|JAX| \cite{jax2018github} for efficient \gls{nuts} sampling. Model diagnostics (including \gls{psis-loo} cross-validation) and posterior exploratory analysis utilized \verb|ArviZ| \cite{arviz_2019}, a toolkit for exploratory analysis of Bayesian models, and \verb|ArviZ-stats|. Data preprocessing employed \verb|pandas|, \verb|numpy|, and \verb|xarray|. Visualizations were created using \verb|matplotlib|, \verb|seaborn|, and \verb|ArviZ-plots|.

\section{List of abbreviations}
\printglossary[type=\acronymtype]

\pagebreak

\section{Declarations}

\subsection{Ethics approval and consent to participate} 
The \glspl{irb} approved the \gls{aic} and aggressive behavior prediction protocols of participating study sites. \gls{irb} approval of the \gls{aic} extended to this study with an amendment. Guardians of all study participants provided informed consent and were remunerated. The \gls{irb} approved this retrospective study in compliance with the Health Information Portability and Accountability Act. All methods were performed in accordance with relevant guidelines and regulations following the Declaration of Helsinki.

\subsection{Consent for publication}
Not applicable. Data reported is de-identified and the results have not been submitted to or published elsewhere.

\subsection{Availability of data and material}
The datasets generated during and/or analysed during the current study are available from the corresponding author upon reasonable request.
\begin{itemize}
\item Data types: Deidentified participant data, Data dictionary
\item Additional Information: Data access instructions, deidentified data description, data dictionary description, acceptable use policy.
\item How to access documents: Will be made available pre-production. 
\item When available: With publication. 
\item Who can access the data: Researchers whose proposed use of the data has been approved.
\item Types of analyses: The scientific community for non-commercial research purposes.
\item Mechanisms of data availability: Per the Simons Foundation data sharing policy.
\end{itemize}

\subsection{Competing interests}
The author(s) declare no competing interests.

\subsection{Funding}
This study was supported by grant R01LM014191 from the National Institutes of Health, the Simons Foundation for Autism
Research Initiative, and the Nancy Lurie Marks Family Foundation.

\subsection{Authors' contributions}
M.P., D.E., T.I., M.S.G conceived the experiments. M.S.G., Y.W. conducted the data collection. M.P., M.E., T.I. conducted the experiments. M.P., M.E., T.I., M.S.G. analyzed the results. All authors reviewed the manuscript. D.E., M.S.G. obtained funding.

\subsection{Acknowledgements}

This study was supported by grant R01LM014191 from the National Institutes of Health, the Simons Foundation for Autism Research Initiative, and the Nancy Lurie Marks Family Foundation. We are grateful to participating families. Families were offered \$75 for their child’s participation in the study. All project personnel were paid as research assistants for the duration of their involvement.





\bibliography{sn-bibliography}

\pagebreak

\newcounter{savefigure}
\newcounter{savetable}
\setcounter{savefigure}{\value{figure}}
\setcounter{savetable}{\value{table}}

\begin{appendices}

  \renewcommand{\thesection}{Appendix~\Alph{section}}
  \renewcommand{\thesubsection}{\thesection.\arabic{subsection}}
  \renewcommand{\thesubsubsection}{\thesubsection.\arabic{subsubsection}}
\renewcommand{\thefigure}{\arabic{figure}}
\renewcommand{\thetable}{\arabic{table}}
\counterwithout{figure}{section}
\counterwithout{table}{section}
\crefname{figure}{Figure}{Figures}
\crefname{table}{Table}{Tables}

\setcounter{figure}{\value{savefigure}}
\setcounter{table}{\value{savetable}}

\section{Individual Data Statistics}\label{app:data}

\begin{table}[h!]
    \centering
\scalebox{0.92}{
\begin{tabular}{lcccc}
\toprule
Individual ID & \# Of Onsets & Total Observation Time (min) & \# of Sessions & Time (min) / Session \\
\midrule
0 & 1287 & 2000.08 & 24 & 83.34 \\
1 & 741 & 528.88 & 7 & 75.55 \\
2 & 195 & 434.00 & 7 & 62.00 \\
3 & 481 & 629.65 & 7 & 89.95 \\
4 & 407 & 940.49 & 17 & 55.32 \\
5 & 118 & 262.33 & 3 & 87.44 \\
6 & 140 & 931.94 & 9 & 103.55 \\
7 & 65 & 253.31 & 3 & 84.44 \\
8 & 79 & 211.61 & 7 & 30.23 \\
9 & 79 & 339.70 & 5 & 67.94 \\
10 & 54 & 203.55 & 5 & 40.71 \\
11 & 150 & 1052.95 & 16 & 65.81 \\
12 & 55 & 177.53 & 4 & 44.38 \\
13 & 57 & 386.73 & 4 & 96.68 \\
14 & 77 & 2263.22 & 25 & 90.53 \\
15 & 22 & 558.22 & 5 & 111.64 \\
16 & 44 & 446.34 & 5 & 89.27 \\
17 & 94 & 719.15 & 16 & 44.95 \\
18 & 98 & 376.91 & 8 & 47.11 \\
19 & 42 & 377.38 & 7 & 53.91 \\
20 & 19 & 204.75 & 3 & 68.25 \\
21 & 55 & 740.04 & 6 & 123.34 \\
22 & 28 & 164.43 & 4 & 41.11 \\
23 & 64 & 746.63 & 20 & 37.33 \\
24 & 37 & 987.37 & 13 & 75.95 \\
25 & 50 & 1597.28 & 20 & 79.86 \\
26 & 33 & 1088.57 & 9 & 120.95 \\
27 & 21 & 657.87 & 8 & 82.23 \\
28 & 38 & 949.90 & 11 & 86.35 \\
29 & 19 & 296.57 & 4 & 74.14 \\
30 & 12 & 384.54 & 6 & 64.09 \\
31 & 9 & 77.82 & 1 & 77.82 \\
32 & 26 & 792.47 & 13 & 60.96 \\
33 & 21 & 267.56 & 6 & 44.59 \\
34 & 33 & 550.03 & 7 & 78.58 \\
35 & 14 & 772.40 & 7 & 110.34 \\
36 & 22 & 330.09 & 9 & 36.68 \\
37 & 13 & 516.60 & 7 & 73.80 \\
38 & 9 & 278.54 & 6 & 46.42 \\
39 & 6 & 644.56 & 10 & 64.46 \\
40 & 5 & 75.77 & 1 & 75.77 \\
41 & 6 & 252.44 & 6 & 42.07 \\
42 & 8 & 686.19 & 5 & 137.24 \\
43 & 5 & 241.64 & 5 & 48.33 \\
44 & 4 & 71.95 & 1 & 71.95 \\
45 & 10 & 256.91 & 6 & 42.82 \\
46 & 3 & 22.32 & 1 & 22.32 \\
47 & 2 & 196.33 & 3 & 65.44 \\
48 & 9 & 908.44 & 18 & 50.47 \\
49 & 2 & 129.07 & 2 & 64.54 \\
50 & 1 & 762.71 & 6 & 127.12 \\
51 & 1 & 80.94 & 2 & 40.47 \\
52 & 1 & 103.88 & 2 & 51.94 \\
53 & 0 & 71.05 & 1 & 71.05 \\
54 & 0 & 67.88 & 1 & 67.88 \\
55 & 0 & 57.98 & 1 & 57.98 \\
56 & 0 & 21.15 & 1 & 21.15 \\
57 & 0 & 26.78 & 1 & 26.78 \\
58 & 0 & 45.85 & 1 & 45.85 \\
59 & 0 & 46.63 & 1 & 46.63 \\
60 & 0 & 27.18 & 1 & 27.18 \\
61 & 0 & 66.48 & 1 & 66.48 \\
62 & 0 & 40.00 & 1 & 40.00 \\
63 & 0 & 39.83 & 1 & 39.83 \\
64 & 0 & 79.50 & 1 & 79.50 \\
65 & 0 & 112.40 & 1 & 112.40 \\
66 & 0 & 64.95 & 1 & 64.95 \\
67 & 0 & 49.75 & 1 & 49.75 \\
68 & 0 & 4.98 & 1 & 4.98 \\
69 & 0 & 73.85 & 1 & 73.85 \\
\bottomrule
\end{tabular}
}
\end{table}

\section{Bayesian Models}\label{app:bayesmodels}

\subsection{Pooled}
\noindent 
\begin{align*}
    \mu \sim \mathrm{Uniform}(0,3) \quad
    \alpha \sim \mathrm{Uniform}(0,3)
    \quad
    \beta \sim
    \mathrm{Uniform(0,3)}
\end{align*}
\subsection{Unpooled}
\noindent
\begin{align*}
    &\textbf{Level 0: Person-level} 
    &&\textbf{Level 1: Data-level} \\
    &\quad \mu_n \sim \mathrm{Half\text{-}Cauchy}(0.1) 
    &&\quad \Delta\mu_{in} \sim \mathrm{Half\text{-}Cauchy}(0.1) \\
    &\quad \alpha_n \sim \mathrm{Gamma}(2.5,0.4) 
    &&\quad \mu_{0,in} = \mu_n + \Delta\mu_{in} \\
    &\quad \beta_n \sim \mathrm{Half\text{-}Cauchy}(1.5) 
    &&\quad S_{in} \sim \mathrm{HawkesProcess}(\mu_n, \alpha_n, \beta_n, \mu_{0,in}) \\
\end{align*}

\section{Additional Hypothesis Testing}\label{app:hypothesistestingextra}
Another \gls{gof} approach involves two steps. First, we use the Kolmogorov-Smirnov test to assess whether the \gls{rtct} inter-arrival times for each person follow a unit-rate exponential distribution. This is complemented by visual inspection with Q–Q plots, comparing the empirical \gls{rtct} inter-arrival times to the theoretical quantiles of a unit-rate exponential distribution; close alignment along the diagonal indicates a good fit. If the exponentiality assumption is not rejected, we then assess the independence of these \gls{rtct} inter-arrival times by analyzing autocorrelation in the uniform-transformed values $U_k = F(t^*_k - t^*_{k-1})$, where $F$ is the cumulative distribution function of the exponential distribution. We apply the Ljung–Box test for a formal assessment of autocorrelation and use scatter plots of $U_{k+1}$ versus $U_k$ for visual inspection; a random scatter without discernible structure suggests independence.

\begin{figure}[H]
    \centering
    \includegraphics[width=1.0\textwidth]{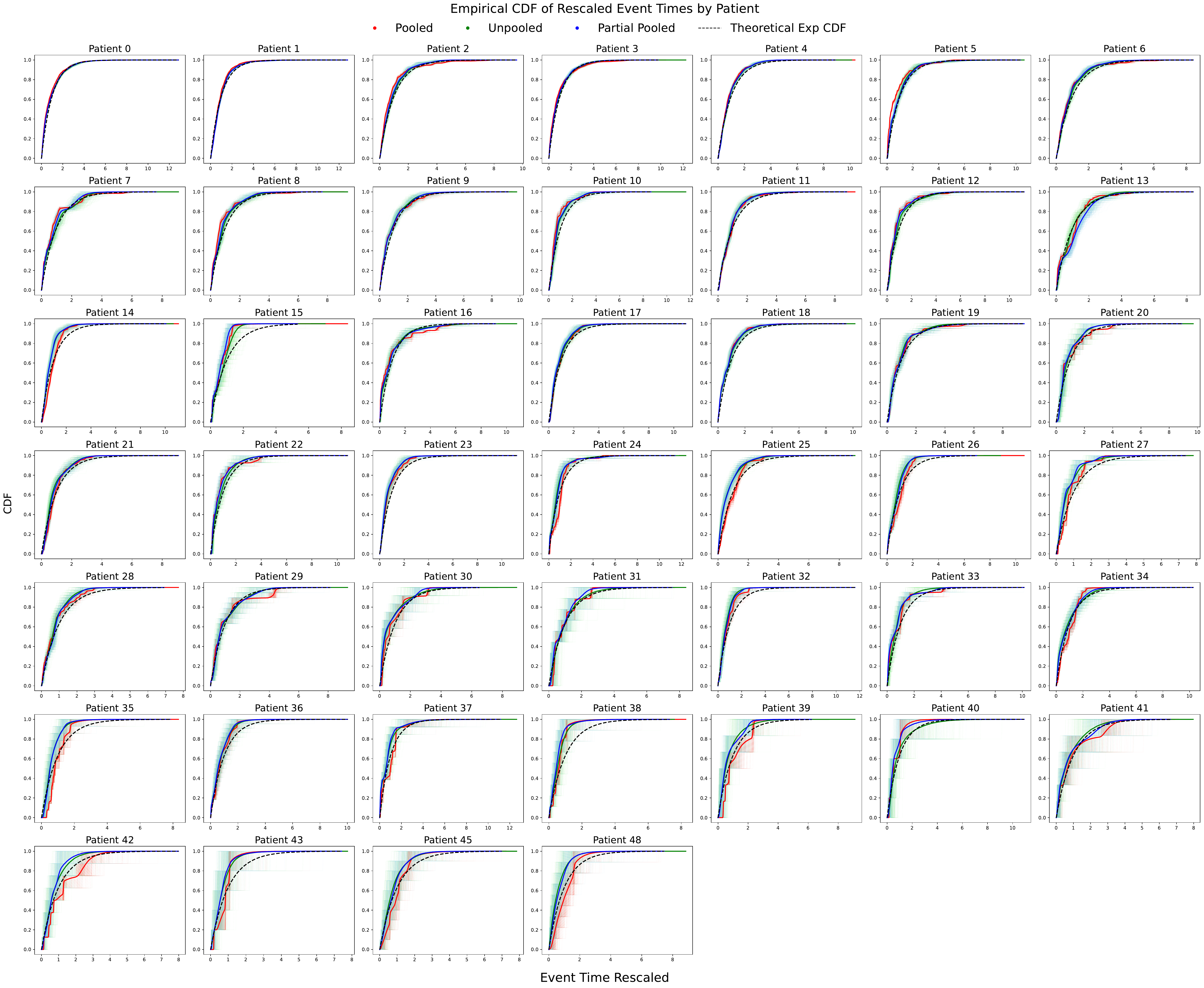}
    \caption{Posterior Predictive Checks for the Empirical Cumulative Density Function plots of the RTCT inter-arrival times of aggressive behavior onsets for a sample participant. The plot compares the empirical quantiles of the inter-arrival times to the theoretical quantiles of the unit-rate exponential distribution. The points closely follow the diagonal line, indicating that the inter-arrival times follow a unit-rate exponential distribution.}
    \label{appfig:qq_plot_inter_arrival_times}
\end{figure}

\begin{figure}[H]
    \centering
        \includegraphics[width=1.0\textwidth, trim=0 0 0 120, clip]{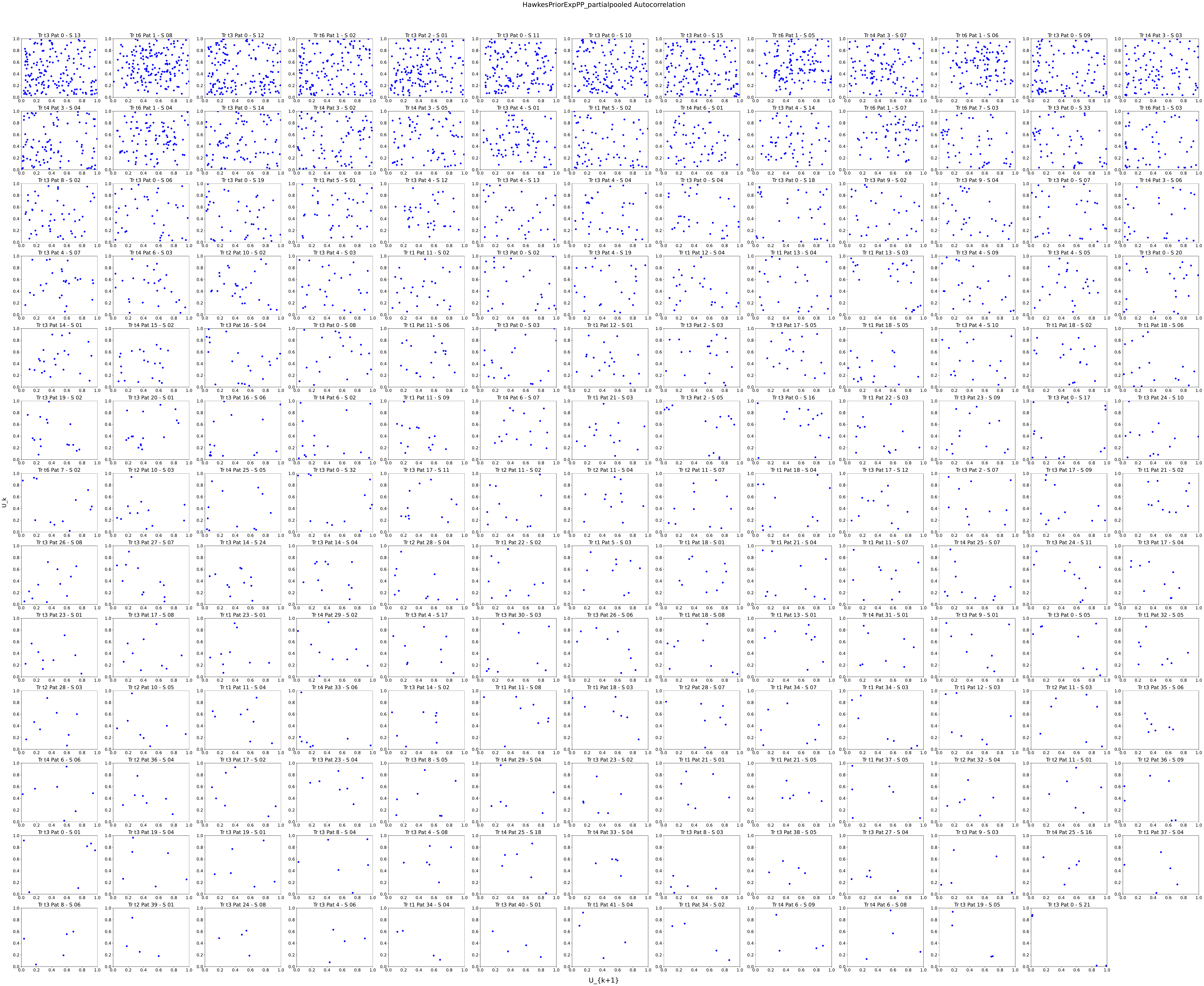}
        \caption{Scatter plot of the uniform-transformed \gls{rtct} inter-arrival times for observation sessions with more than 5 aggressive behavior onsets. The plot shows the relationship between the uniform-transformed inter-arrival times at lag 1 and lag 2. The randomly scattered points indicate that the inter-arrival times are independent and identically distributed.}
        \label{appfig:autocorrelation}
    \end{figure}

While the Lewis test with Durbin’s modification offers greater statistical power, we also report results from tests of independent and identically distributed (i.i.d.) inter-arrival times for completeness, as shown in \cref{apptab:gof_independence_exponential}. The Kolmogorov–Smirnov test indicates that a high percentage of participants have \gls{rtct} inter-arrival times consistent with a unit-rate exponential distribution. However, due to the Kolmogorov-Smirnov test’s sensitivity with larger sample size \cite{mason1983modified}, individuals contributing the majority of the data often reject the null hypothesis of exponential marginal distribution with statistical significance—even when \gls{qq} plots suggest a good visual fit to the unit-rate exponential distribution (\cref{appfig:qq_plot_inter_arrival_times}). The independence test on the uniform transform of the inter-arrival times shows that most observation sessions exhibit no significant autocorrelation at lag 1, suggesting that the \gls{rtct} inter-arrival times are not temporally dependent. Visually, this is confirmed in the scatterplots shown in \cref{appfig:autocorrelation}. 

\begin{table}[h!]
\centering
\caption{Goodness of fit results for the two additional tests: the Independence Test (Ljung–Box test on uniform-transformed inter-arrival times) and the Marginal Exponential Test (Kolmogorov–Smirnov test on transformed inter-arrival times per person). Results are reported as the proportion of sessions out of 157 sessions (Independence Test) and out of 46 individuals (Marginal Exponential Test) that do not reject the null hypothesis at various significance levels for each model. Only sessions with more than 5 events were included in the analyses.}
\begin{tabular}{llccclcccc}
\toprule
 & \multicolumn{3}{c}{Independence Test} & & \multicolumn{3}{c}{Marginal Exponential Test} \\
\cmidrule{2-4} \cmidrule{6-8}
Significance Level & Partial Pooled & Unpooled & Pooled & & Partial Pooled & Unpooled & Pooled \\
\midrule
0.050 & 0.96 & 0.95 & 0.96 & & 0.87 & 0.93 & 0.74 \\
0.100 & 0.89 & 0.90 & 0.89 & & 0.83 & 0.89 & 0.65 \\
0.150 & 0.86 & 0.86 & 0.87 & & 0.76 & 0.85 & 0.54 \\
\bottomrule
\end{tabular}
\label{apptab:gof_independence_exponential}
\end{table}

We note that it is well known that statistical tests such as the Kolmogorov-Smirnov test assume that the parameters of the hypothesized distribution are fully specified. However, in many real-world scenarios—including our case—these parameters are unknown and must be estimated from the sample data. This estimation process effectively ``fits" the hypothesized distribution to the observed data, thereby reducing the discrepancy between the empirical and hypothesized distributions for a specific sample. As a result, the KS test statistic tends to be smaller than it would be if the true, fixed parameters were known, which in turn leads to inflated p-values and a higher risk of Type II error. This provides a partial explanation for why the unpooled model formulation yields slightly better p-values across all hypothesis tests—particularly for the per-individual Kolmogorv-Smirnov test—as there is no regularization induced by hyperpriors in the partial pooled model when fitting a separate model per individual.

\pagebreak

\pagebreak

\section{Monte Carlo Markov Chain (MCMC) Diagnostics}\label{app:mcmc_diagnostics}
We do not include the estimates of the prior basline intensity,$\mu_{in}$, for the sake of clarity and pagelength.
{\footnotesize
\begin{longtable}{lrrrrrrrrrrr}
\caption{MCMC diagnostics for the partial pooled model parameters. Statistics are based on 4000 MCMC samples from the posterior distributions. Reported metrics include Monte Carlo standard errors (mcse), effective sample sizes (ess), and $\hat{R}$ convergence diagnostics for each parameter.} 
\label{apptab:mcmc_diagnostics_partial_pooled} \\
\toprule
label & mcse-mean & mcse-sd & ess-bulk & ess-tail & r-hat & mean & std & median-std & 5\% & median & 95\% \\
\midrule
\endfirsthead

\toprule
label & mcse-mean & mcse-sd & ess-bulk & ess-tail & r-hat & mean & std & median-std & 5\% & median & 95\% \\
\midrule
\endhead

\midrule
\multicolumn{12}{r}{\textit{Continued on next page}} \\
\endfoot

\bottomrule
\endlastfoot
$\mu_\alpha$ & 0.000450 & 0.000292 & 3282 & 3763 & 1.001091 & 0.742489 & 0.025748 & 0.025749 & 0.698528 & 0.742699 & 0.784302 \\
$\mu_\beta$ & 0.001580 & 0.002347 & 3770 & 3930 & 1.001087 & 0.432781 & 0.097057 & 0.098250 & 0.311677 & 0.417514 & 0.607113 \\
$\mu_\mu$ & 0.000228 & 0.000416 & 3141 & 3714 & 1.000167 & 0.036306 & 0.013127 & 0.013425 & 0.022275 & 0.033492 & 0.058177 \\
$\sigma_\alpha$ & 0.000720 & 0.000512 & 2903 & 2567 & 1.001046 & 0.092903 & 0.039981 & 0.040028 & 0.029096 & 0.090962 & 0.161191 \\
$\sigma_\beta$ & 0.003024 & 0.002373 & 2942 & 3574 & 1.002966 & 0.932407 & 0.162999 & 0.163839 & 0.692477 & 0.915846 & 1.221616 \\
$\sigma_\mu$ & 0.003702 & 0.002840 & 2951 & 3468 & 1.000824 & 1.445303 & 0.202659 & 0.203330 & 1.151815 & 1.428806 & 1.797775 \\
$\alpha_0$ & 0.000591 & 0.000392 & 3662 & 3740 & 1.000603 & 0.824332 & 0.035984 & 0.036018 & 0.767887 & 0.822785 & 0.886532 \\
$\alpha_1$ & 0.000953 & 0.000594 & 3142 & 3725 & 1.001092 & 0.696101 & 0.053365 & 0.053417 & 0.606293 & 0.698469 & 0.777280 \\
$\alpha_2$ & 0.000890 & 0.000644 & 3387 & 3724 & 1.000573 & 0.779652 & 0.051842 & 0.051902 & 0.699138 & 0.777167 & 0.870659 \\
$\alpha_3$ & 0.000698 & 0.000496 & 3603 & 3937 & 1.000003 & 0.805769 & 0.041801 & 0.041861 & 0.741403 & 0.803538 & 0.877646 \\
$\alpha_4$ & 0.000808 & 0.000537 & 3046 & 3726 & 1.000902 & 0.776722 & 0.044656 & 0.044668 & 0.704344 & 0.775712 & 0.853599 \\
$\alpha_5$ & 0.001044 & 0.000701 & 3265 & 3784 & 1.000504 & 0.721480 & 0.059861 & 0.059909 & 0.618122 & 0.723873 & 0.813668 \\
$\alpha_6$ & 0.000919 & 0.000741 & 3782 & 3859 & 1.000420 & 0.783809 & 0.056483 & 0.056616 & 0.698749 & 0.779932 & 0.884015 \\
$\alpha_7$ & 0.001038 & 0.000918 & 3756 & 3772 & 0.999966 & 0.775680 & 0.064052 & 0.064157 & 0.675998 & 0.772000 & 0.885807 \\
$\alpha_8$ & 0.001062 & 0.000926 & 3653 & 3684 & 1.000128 & 0.770304 & 0.064039 & 0.064062 & 0.669123 & 0.768618 & 0.878084 \\
$\alpha_9$ & 0.001158 & 0.000801 & 3048 & 3752 & 1.000660 & 0.725571 & 0.064045 & 0.064178 & 0.614816 & 0.729711 & 0.820967 \\
$\alpha_{10}$ & 0.001228 & 0.001074 & 3071 & 3547 & 0.999707 & 0.747473 & 0.068336 & 0.068355 & 0.637389 & 0.749083 & 0.862808 \\
$\alpha_{11}$ & 0.000966 & 0.000655 & 3457 & 3940 & 1.000455 & 0.713127 & 0.056746 & 0.056807 & 0.615980 & 0.715744 & 0.799810 \\
$\alpha_{12}$ & 0.001147 & 0.001051 & 3467 & 3788 & 1.000572 & 0.776421 & 0.069491 & 0.069716 & 0.670481 & 0.770829 & 0.900004 \\
$\alpha_{13}$ & 0.001651 & 0.000987 & 2521 & 3542 & 1.003183 & 0.665904 & 0.083831 & 0.084157 & 0.514884 & 0.673306 & 0.787318 \\
$\alpha_{14}$ & 0.001130 & 0.000870 & 3185 & 3595 & 1.001700 & 0.744367 & 0.063764 & 0.063787 & 0.639016 & 0.746064 & 0.848437 \\
$\alpha_{15}$ & 0.001370 & 0.001210 & 2971 & 4003 & 1.000550 & 0.774643 & 0.075591 & 0.075920 & 0.663106 & 0.767582 & 0.911006 \\
$\alpha_{16}$ & 0.001120 & 0.000855 & 3629 & 3949 & 1.000157 & 0.731221 & 0.067585 & 0.067739 & 0.618179 & 0.735788 & 0.834564 \\
$\alpha_{17}$ & 0.001030 & 0.000816 & 3514 & 3319 & 0.999821 & 0.767258 & 0.061328 & 0.061390 & 0.669436 & 0.764490 & 0.871793 \\
$\alpha_{18}$ & 0.001132 & 0.000828 & 2987 & 3571 & 1.000907 & 0.754760 & 0.062118 & 0.062137 & 0.651296 & 0.756306 & 0.856288 \\
$\alpha_{19}$ & 0.001141 & 0.000891 & 3687 & 3837 & 1.001583 & 0.714741 & 0.068985 & 0.069094 & 0.594741 & 0.718616 & 0.819203 \\
$\alpha_{20}$ & 0.001353 & 0.001063 & 2792 & 3542 & 1.000178 & 0.740427 & 0.071345 & 0.071428 & 0.622526 & 0.743880 & 0.851972 \\
$\alpha_{21}$ & 0.001322 & 0.000838 & 2558 & 3797 & 1.001402 & 0.708643 & 0.066972 & 0.067190 & 0.593217 & 0.714054 & 0.805878 \\
$\alpha_{22}$ & 0.001258 & 0.001174 & 3470 & 3504 & 1.000039 & 0.759309 & 0.074048 & 0.074073 & 0.642674 & 0.757389 & 0.882181 \\
$\alpha_{23}$ & 0.001121 & 0.000928 & 3393 & 3338 & 1.000640 & 0.745776 & 0.064947 & 0.064954 & 0.640386 & 0.746743 & 0.852503 \\
$\alpha_{24}$ & 0.001206 & 0.000960 & 3523 & 3957 & 1.000853 & 0.715757 & 0.072179 & 0.072292 & 0.590286 & 0.719801 & 0.823451 \\
$\alpha_{25}$ & 0.001143 & 0.001021 & 3744 & 3852 & 1.000281 & 0.753240 & 0.069814 & 0.069815 & 0.640849 & 0.753610 & 0.865480 \\
$\alpha_{26}$ & 0.001303 & 0.000924 & 2934 & 3336 & 1.001202 & 0.713553 & 0.070981 & 0.071152 & 0.590960 & 0.718487 & 0.817091 \\
$\alpha_{27}$ & 0.001255 & 0.001070 & 3359 & 3715 & 1.000808 & 0.752489 & 0.072888 & 0.072889 & 0.637691 & 0.752086 & 0.872685 \\
$\alpha_{28}$ & 0.001138 & 0.000906 & 3608 & 3866 & 1.000782 & 0.710363 & 0.068521 & 0.068650 & 0.592157 & 0.714568 & 0.811167 \\
$\alpha_{29}$ & 0.001289 & 0.001046 & 3070 & 3497 & 0.999827 & 0.730289 & 0.072294 & 0.072389 & 0.609096 & 0.733984 & 0.844021 \\
$\alpha_{30}$ & 0.001409 & 0.001134 & 2916 & 3894 & 1.002108 & 0.743983 & 0.075278 & 0.075314 & 0.621364 & 0.746303 & 0.865543 \\
$\alpha_{31}$ & 0.001346 & 0.001120 & 3297 & 3435 & 1.000668 & 0.742512 & 0.076406 & 0.076447 & 0.611211 & 0.745011 & 0.862233 \\
$\alpha_{32}$ & 0.001305 & 0.000887 & 2994 & 3691 & 1.000120 & 0.714594 & 0.071417 & 0.071665 & 0.588908 & 0.720545 & 0.821367 \\
$\alpha_{33}$ & 0.001240 & 0.001050 & 3414 & 3754 & 1.003567 & 0.737623 & 0.072712 & 0.072751 & 0.613043 & 0.739996 & 0.850580 \\
$\alpha_{34}$ & 0.001369 & 0.001092 & 2929 & 3536 & 1.000581 & 0.741183 & 0.076959 & 0.077039 & 0.611285 & 0.744678 & 0.864956 \\
$\alpha_{35}$ & 0.001263 & 0.001040 & 3416 & 3958 & 1.000234 & 0.733364 & 0.072967 & 0.073046 & 0.611727 & 0.736771 & 0.845039 \\
$\alpha_{36}$ & 0.001309 & 0.001242 & 3151 & 3531 & 1.000727 & 0.737456 & 0.074901 & 0.075039 & 0.613903 & 0.742010 & 0.852954 \\
$\alpha_{37}$ & 0.001314 & 0.001384 & 3619 & 3888 & 1.000465 & 0.748834 & 0.078873 & 0.078894 & 0.624550 & 0.750666 & 0.872618 \\
$\alpha_{38}$ & 0.001256 & 0.001076 & 3643 & 3787 & 1.000758 & 0.750154 & 0.076483 & 0.076483 & 0.627076 & 0.750519 & 0.879521 \\
$\alpha_{39}$ & 0.001290 & 0.001163 & 3433 & 3754 & 1.000755 & 0.741933 & 0.077402 & 0.077412 & 0.611674 & 0.743186 & 0.868888 \\
$\alpha_{40}$ & 0.001232 & 0.001054 & 3734 & 3518 & 1.002158 & 0.745482 & 0.075375 & 0.075447 & 0.619574 & 0.748768 & 0.869308 \\
$\alpha_{41}$ & 0.001224 & 0.001084 & 3894 & 3889 & 1.000539 & 0.728630 & 0.076692 & 0.076951 & 0.594748 & 0.734935 & 0.847123 \\
$\alpha_{42}$ & 0.001456 & 0.001100 & 2746 & 3395 & 1.002666 & 0.733506 & 0.076989 & 0.077065 & 0.600543 & 0.736931 & 0.851751 \\
$\alpha_{43}$ & 0.001366 & 0.001161 & 3204 & 3377 & 1.001785 & 0.740503 & 0.076966 & 0.076979 & 0.615625 & 0.741923 & 0.864014 \\
$\alpha_{44}$ & 0.001567 & 0.001045 & 2245 & 3639 & 1.001450 & 0.732091 & 0.075552 & 0.075669 & 0.603638 & 0.736294 & 0.851391 \\
$\alpha_{45}$ & 0.001304 & 0.001072 & 3547 & 3623 & 1.001395 & 0.716122 & 0.077411 & 0.077660 & 0.583522 & 0.722333 & 0.829355 \\
$\alpha_{46}$ & 0.001490 & 0.001236 & 2942 & 3782 & 1.000884 & 0.749204 & 0.081084 & 0.081089 & 0.619344 & 0.750080 & 0.884547 \\
$\alpha_{47}$ & 0.001530 & 0.001330 & 2766 & 3913 & 1.001698 & 0.742219 & 0.078152 & 0.078155 & 0.615738 & 0.742906 & 0.865961 \\
$\alpha_{48}$ & 0.001358 & 0.001188 & 3502 & 3303 & 1.000581 & 0.730595 & 0.078283 & 0.078427 & 0.599226 & 0.735355 & 0.852192 \\
$\alpha_{49}$ & 0.001273 & 0.001146 & 3686 & 3893 & 1.001015 & 0.731430 & 0.077510 & 0.077585 & 0.602731 & 0.734839 & 0.850143 \\
$\alpha_{50}$ & 0.001296 & 0.001157 & 3545 & 3830 & 1.001538 & 0.739146 & 0.077418 & 0.077522 & 0.606362 & 0.743165 & 0.860547 \\
$\alpha_{51}$ & 0.001344 & 0.001348 & 3365 & 3601 & 0.999995 & 0.737140 & 0.078983 & 0.079058 & 0.602950 & 0.740594 & 0.861787 \\
$\alpha_{52}$ & 0.001325 & 0.001069 & 3428 & 3926 & 0.999991 & 0.737407 & 0.077109 & 0.077219 & 0.607405 & 0.741530 & 0.858593 \\
$\beta_0$ & 0.001027 & 0.000685 & 3626 & 3816 & 1.000732 & 0.442579 & 0.061578 & 0.061651 & 0.346879 & 0.439568 & 0.552339 \\
$\beta_1$ & 0.001082 & 0.000804 & 3701 & 3996 & 1.000745 & 0.414489 & 0.065783 & 0.065950 & 0.315076 & 0.409803 & 0.529192 \\
$\beta_2$ & 0.002808 & 0.002039 & 3560 & 3812 & 1.000421 & 0.709799 & 0.167735 & 0.167992 & 0.449356 & 0.700516 & 0.999577 \\
$\beta_3$ & 0.001688 & 0.001277 & 4040 & 4100 & 1.000224 & 0.701694 & 0.107446 & 0.107731 & 0.543093 & 0.693871 & 0.886418 \\
$\beta_4$ & 0.000864 & 0.000675 & 3960 & 3959 & 0.999613 & 0.373348 & 0.054374 & 0.054492 & 0.289103 & 0.369757 & 0.468130 \\
$\beta_5$ & 0.009113 & 0.007845 & 4005 & 4020 & 0.999656 & 1.613279 & 0.578593 & 0.585648 & 0.852762 & 1.522655 & 2.682222 \\
$\beta_6$ & 0.001245 & 0.000930 & 3822 & 3567 & 0.999548 & 0.448605 & 0.077117 & 0.077263 & 0.329800 & 0.443857 & 0.583413 \\
$\beta_7$ & 0.002768 & 0.002724 & 4321 & 4012 & 1.000243 & 0.635957 & 0.178900 & 0.179873 & 0.378902 & 0.617269 & 0.953784 \\
$\beta_8$ & 0.002740 & 0.002833 & 3911 & 3694 & 1.000396 & 0.581265 & 0.175396 & 0.177040 & 0.342001 & 0.557196 & 0.896481 \\
$\beta_9$ & 0.001987 & 0.001701 & 3774 & 3947 & 0.999998 & 0.457752 & 0.122100 & 0.122695 & 0.283802 & 0.445689 & 0.675473 \\
$\beta_{10}$ & 0.003198 & 0.002551 & 3566 & 3537 & 1.000677 & 0.516231 & 0.191683 & 0.193549 & 0.249779 & 0.489419 & 0.872698 \\
$\beta_{11}$ & 0.000892 & 0.000787 & 4051 & 3815 & 0.999584 & 0.286364 & 0.056680 & 0.056870 & 0.204046 & 0.281720 & 0.385825 \\
$\beta_{12}$ & 0.003018 & 0.002838 & 3626 & 3877 & 1.000262 & 0.491245 & 0.181567 & 0.183580 & 0.250377 & 0.464138 & 0.830492 \\
$\beta_{13}$ & 0.009670 & 0.012560 & 3292 & 3854 & 0.999750 & 0.759995 & 0.544407 & 0.561150 & 0.162967 & 0.623944 & 1.821434 \\
$\beta_{14}$ & 0.000562 & 0.000412 & 3454 & 3971 & 1.000657 & 0.118664 & 0.032998 & 0.033118 & 0.070505 & 0.115847 & 0.178625 \\
$\beta_{15}$ & 0.003257 & 0.002924 & 3214 & 3329 & 1.001699 & 0.458104 & 0.190023 & 0.192378 & 0.206138 & 0.428093 & 0.808158 \\
$\beta_{16}$ & 0.004857 & 0.007660 & 3847 & 4130 & 1.000445 & 0.700023 & 0.309372 & 0.314108 & 0.323492 & 0.645684 & 1.254553 \\
$\beta_{17}$ & 0.001005 & 0.000823 & 4033 & 3891 & 1.001019 & 0.253643 & 0.063766 & 0.064319 & 0.163050 & 0.245234 & 0.369926 \\
$\beta_{18}$ & 0.001422 & 0.001253 & 4131 & 3969 & 1.000759 & 0.340443 & 0.091623 & 0.092186 & 0.211256 & 0.330270 & 0.501499 \\
$\beta_{19}$ & 0.002123 & 0.002266 & 4094 & 3691 & 1.001210 & 0.356037 & 0.136680 & 0.138034 & 0.168852 & 0.336752 & 0.595465 \\
$\beta_{20}$ & 0.002710 & 0.003347 & 3910 & 3991 & 0.999606 & 0.410042 & 0.166297 & 0.168549 & 0.196756 & 0.382578 & 0.712321 \\
$\beta_{21}$ & 0.001177 & 0.000946 & 3844 & 3468 & 1.000604 & 0.288103 & 0.073249 & 0.073696 & 0.183583 & 0.280003 & 0.420024 \\
$\beta_{22}$ & 0.003484 & 0.003259 & 3758 & 3648 & 1.000086 & 0.431986 & 0.208789 & 0.213075 & 0.168356 & 0.389465 & 0.823335 \\
$\beta_{23}$ & 0.001256 & 0.001013 & 3463 & 3852 & 1.000121 & 0.273650 & 0.075502 & 0.075920 & 0.163593 & 0.265694 & 0.406393 \\
$\beta_{24}$ & 0.001084 & 0.000843 & 3250 & 3973 & 1.003183 & 0.121238 & 0.061532 & 0.061814 & 0.031479 & 0.115339 & 0.230002 \\
$\beta_{25}$ & 0.000599 & 0.000531 & 2762 & 3562 & 1.000911 & 0.075435 & 0.033756 & 0.034370 & 0.032785 & 0.068967 & 0.139058 \\
$\beta_{26}$ & 0.002085 & 0.002453 & 3772 & 3755 & 1.000321 & 0.228666 & 0.129320 & 0.133189 & 0.081609 & 0.196799 & 0.474085 \\
$\beta_{27}$ & 0.000823 & 0.000843 & 3547 & 3878 & 0.999894 & 0.131971 & 0.049052 & 0.049524 & 0.064367 & 0.125145 & 0.220712 \\
$\beta_{28}$ & 0.003921 & 0.005251 & 3965 & 3929 & 0.999508 & 0.678666 & 0.246202 & 0.249043 & 0.361058 & 0.641159 & 1.127294 \\
$\beta_{29}$ & 0.004281 & 0.004109 & 3855 & 3511 & 0.999812 & 0.595108 & 0.266114 & 0.268306 & 0.234337 & 0.560883 & 1.078392 \\
$\beta_{30}$ & 0.002296 & 0.005128 & 2782 & 3294 & 0.999916 & 0.156088 & 0.126703 & 0.130720 & 0.035237 & 0.123931 & 0.379760 \\
$\beta_{31}$ & 0.002371 & 0.005107 & 3388 & 3891 & 1.000518 & 0.206240 & 0.142441 & 0.146236 & 0.062392 & 0.173138 & 0.455582 \\
$\beta_{32}$ & 0.003391 & 0.003044 & 3574 & 3843 & 0.999677 & 0.492685 & 0.202358 & 0.204625 & 0.221194 & 0.462313 & 0.862288 \\
$\beta_{33}$ & 0.004078 & 0.012557 & 3526 & 3586 & 1.000139 & 0.256864 & 0.246994 & 0.255924 & 0.083448 & 0.189849 & 0.639378 \\
$\beta_{34}$ & 0.000921 & 0.001739 & 3171 & 3684 & 1.000898 & 0.082072 & 0.053934 & 0.055223 & 0.024410 & 0.070211 & 0.175888 \\
$\beta_{35}$ & 0.001192 & 0.001280 & 3735 & 3415 & 0.999977 & 0.145555 & 0.073498 & 0.074811 & 0.054451 & 0.131597 & 0.285611 \\
$\beta_{36}$ & 0.001506 & 0.001886 & 3306 & 3628 & 1.000479 & 0.159635 & 0.089108 & 0.090658 & 0.049546 & 0.142948 & 0.327007 \\
$\beta_{37}$ & 0.001040 & 0.002658 & 3472 & 3645 & 1.001187 & 0.095702 & 0.062965 & 0.064829 & 0.031807 & 0.080267 & 0.211683 \\
$\beta_{38}$ & 0.001914 & 0.002075 & 3428 & 3574 & 1.000168 & 0.195731 & 0.115186 & 0.118111 & 0.061750 & 0.169609 & 0.423634 \\
$\beta_{39}$ & 0.002839 & 0.004075 & 3881 & 3754 & 1.000789 & 0.207185 & 0.183648 & 0.191715 & 0.037165 & 0.152158 & 0.593989 \\
$\beta_{40}$ & 0.004177 & 0.005741 & 4006 & 4012 & 0.999945 & 0.408263 & 0.265361 & 0.271083 & 0.103653 & 0.352859 & 0.905334 \\
$\beta_{41}$ & 0.003029 & 0.005273 & 3604 & 3101 & 1.000967 & 0.254048 & 0.188478 & 0.193392 & 0.055931 & 0.210731 & 0.591894 \\
$\beta_{42}$ & 0.002479 & 0.003420 & 3802 & 3421 & 1.000002 & 0.274710 & 0.147740 & 0.150185 & 0.092960 & 0.247721 & 0.542786 \\
$\beta_{43}$ & 0.003336 & 0.004749 & 2929 & 3455 & 0.999961 & 0.221700 & 0.188527 & 0.196506 & 0.033558 & 0.166270 & 0.586766 \\
$\beta_{44}$ & 0.003359 & 0.007453 & 3736 & 3761 & 1.000374 & 0.314153 & 0.208498 & 0.211947 & 0.064728 & 0.276073 & 0.682982 \\
$\beta_{45}$ & 0.001377 & 0.003250 & 3450 & 3829 & 1.001189 & 0.118027 & 0.083335 & 0.085371 & 0.026410 & 0.099494 & 0.265601 \\
$\beta_{46}$ & 0.006783 & 0.011387 & 3896 & 3952 & 0.999508 & 0.476124 & 0.425107 & 0.441778 & 0.075098 & 0.355910 & 1.271134 \\
$\beta_{47}$ & 0.019648 & 0.059848 & 3266 & 3534 & 1.001498 & 0.905707 & 1.175462 & 1.224277 & 0.094858 & 0.563447 & 2.734785 \\
$\beta_{48}$ & 0.000647 & 0.001199 & 3685 & 3802 & 0.999798 & 0.051497 & 0.039411 & 0.040959 & 0.014761 & 0.040344 & 0.124923 \\
$\beta_{49}$ & 0.009347 & 0.062392 & 2520 & 3666 & 1.000049 & 0.311893 & 0.525008 & 0.546021 & 0.022885 & 0.161876 & 0.999671 \\
$\beta_{50}$ & 0.009311 & 0.060785 & 3402 & 3770 & 1.000784 & 0.412381 & 0.570715 & 0.590421 & 0.045918 & 0.261113 & 1.198323 \\
$\beta_{51}$ & 0.006149 & 0.059103 & 3129 & 3575 & 1.000184 & 0.280543 & 0.378916 & 0.387397 & 0.052361 & 0.199925 & 0.700215 \\
$\beta_{52}$ & 0.007107 & 0.012836 & 3843 & 4056 & 1.001018 & 0.540757 & 0.448452 & 0.464422 & 0.102589 & 0.420013 & 1.419425 \\
$\mu_0$ & 0.000283 & 0.000197 & 3935 & 3888 & 1.000477 & 0.106942 & 0.017740 & 0.017745 & 0.078968 & 0.106497 & 0.136177 \\
$\mu_1$ & 0.001208 & 0.000806 & 3605 & 3801 & 1.000414 & 0.435390 & 0.072555 & 0.072577 & 0.318492 & 0.433579 & 0.558225 \\
$\mu_2$ & 0.000305 & 0.000214 & 3923 & 3744 & 1.000423 & 0.080198 & 0.019175 & 0.019179 & 0.049616 & 0.079806 & 0.113376 \\
$\mu_3$ & 0.000388 & 0.000282 & 3888 & 4006 & 1.000573 & 0.111277 & 0.024267 & 0.024300 & 0.073523 & 0.110020 & 0.152430 \\
$\mu_4$ & 0.000270 & 0.000195 & 3992 & 3693 & 1.001031 & 0.087404 & 0.017027 & 0.017052 & 0.060365 & 0.086489 & 0.115849 \\
$\mu_5$ & 0.000509 & 0.000377 & 4058 & 3882 & 1.000126 & 0.131466 & 0.032419 & 0.032423 & 0.080228 & 0.131013 & 0.186217 \\
$\mu_6$ & 0.000102 & 0.000079 & 4197 & 3928 & 1.000042 & 0.026919 & 0.006567 & 0.006588 & 0.017023 & 0.026391 & 0.038721 \\
$\mu_7$ & 0.000205 & 0.000175 & 4308 & 3823 & 0.999836 & 0.035927 & 0.013469 & 0.013647 & 0.017186 & 0.033730 & 0.061028 \\
$\mu_8$ & 0.000412 & 0.000322 & 3678 & 3756 & 1.000765 & 0.070775 & 0.024769 & 0.024883 & 0.034384 & 0.068395 & 0.115535 \\
$\mu_9$ & 0.000291 & 0.000214 & 3709 & 3970 & 1.000374 & 0.045831 & 0.017846 & 0.017907 & 0.019520 & 0.044356 & 0.077553 \\
$\mu_{10}$ & 0.000392 & 0.000323 & 3755 & 3436 & 1.000116 & 0.061561 & 0.023937 & 0.024102 & 0.026912 & 0.058748 & 0.105130 \\
$\mu_{11}$ & 0.000135 & 0.000102 & 4085 & 3593 & 0.999789 & 0.040045 & 0.008593 & 0.008606 & 0.026856 & 0.039588 & 0.054690 \\
$\mu_{12}$ & 0.000366 & 0.000298 & 3931 & 3928 & 0.999983 & 0.048873 & 0.023138 & 0.023343 & 0.016601 & 0.045785 & 0.093205 \\
$\mu_{13}$ & 0.000326 & 0.000233 & 3759 & 3643 & 1.000145 & 0.076753 & 0.019944 & 0.019944 & 0.044176 & 0.076732 & 0.109972 \\
$\mu_{14}$ & 0.000028 & 0.000023 & 3975 & 3648 & 0.999365 & 0.004744 & 0.001733 & 0.001743 & 0.002250 & 0.004556 & 0.007899 \\
$\mu_{15}$ & 0.000039 & 0.000049 & 3861 & 3896 & 1.002131 & 0.003505 & 0.002429 & 0.002492 & 0.000745 & 0.002949 & 0.008167 \\
$\mu_{16}$ & 0.000151 & 0.000114 & 4055 & 3779 & 1.000439 & 0.029789 & 0.009690 & 0.009743 & 0.015416 & 0.028777 & 0.047098 \\
$\mu_{17}$ & 0.000126 & 0.000102 & 3767 & 3668 & 1.001484 & 0.021947 & 0.007730 & 0.007774 & 0.010447 & 0.021118 & 0.035939 \\
$\mu_{18}$ & 0.000289 & 0.000213 & 4189 & 3624 & 0.999797 & 0.052663 & 0.018814 & 0.018863 & 0.024146 & 0.051311 & 0.086603 \\
$\mu_{19}$ & 0.000199 & 0.000163 & 4010 & 3526 & 1.000794 & 0.037923 & 0.012592 & 0.012660 & 0.019251 & 0.036609 & 0.059949 \\
$\mu_{20}$ & 0.000161 & 0.000189 & 4051 & 3867 & 1.000563 & 0.016729 & 0.010133 & 0.010300 & 0.003991 & 0.014884 & 0.035720 \\
$\mu_{21}$ & 0.000111 & 0.000083 & 3868 & 3851 & 1.000142 & 0.023123 & 0.006939 & 0.006959 & 0.012797 & 0.022595 & 0.035448 \\
$\mu_{22}$ & 0.000325 & 0.000257 & 3248 & 3728 & 1.000241 & 0.030513 & 0.018505 & 0.018745 & 0.006018 & 0.027520 & 0.065272 \\
$\mu_{23}$ & 0.000088 & 0.000078 & 3920 & 3731 & 1.000149 & 0.011986 & 0.005482 & 0.005522 & 0.004330 & 0.011322 & 0.021938 \\
$\mu_{24}$ & 0.000080 & 0.000063 & 3697 & 3523 & 1.003275 & 0.011845 & 0.004913 & 0.004927 & 0.004382 & 0.011477 & 0.020517 \\
$\mu_{25}$ & 0.000040 & 0.000033 & 3931 & 3766 & 1.000640 & 0.004568 & 0.002485 & 0.002511 & 0.001190 & 0.004208 & 0.009180 \\
$\mu_{26}$ & 0.000061 & 0.000045 & 3959 & 3996 & 0.999943 & 0.011173 & 0.003906 & 0.003914 & 0.005367 & 0.010930 & 0.018094 \\
$\mu_{27}$ & 0.000040 & 0.000046 & 3724 & 3582 & 1.000607 & 0.002912 & 0.002387 & 0.002468 & 0.000367 & 0.002285 & 0.007725 \\
$\mu_{28}$ & 0.000071 & 0.000059 & 3894 & 4099 & 1.000186 & 0.015787 & 0.004417 & 0.004434 & 0.009167 & 0.015401 & 0.023629 \\
$\mu_{29}$ & 0.000132 & 0.000125 & 4001 & 3772 & 0.999523 & 0.018605 & 0.008402 & 0.008508 & 0.007331 & 0.017271 & 0.033884 \\
$\mu_{30}$ & 0.000074 & 0.000077 & 3701 & 3811 & 0.999837 & 0.005740 & 0.004538 & 0.004694 & 0.000688 & 0.004540 & 0.014632 \\
$\mu_{31}$ & 0.000405 & 0.000426 & 3538 & 3500 & 1.000128 & 0.027418 & 0.023535 & 0.024359 & 0.002546 & 0.021134 & 0.074985 \\
$\mu_{32}$ & 0.000064 & 0.000056 & 4115 & 3857 & 1.000031 & 0.012049 & 0.004044 & 0.004064 & 0.006207 & 0.011652 & 0.019236 \\
$\mu_{33}$ & 0.000190 & 0.000177 & 3845 & 3746 & 1.000039 & 0.017189 & 0.012020 & 0.012316 & 0.002752 & 0.014504 & 0.040664 \\
$\mu_{34}$ & 0.000141 & 0.000112 & 3879 & 4021 & 1.000114 & 0.022960 & 0.008772 & 0.008811 & 0.010182 & 0.022131 & 0.038517 \\
$\mu_{35}$ & 0.000052 & 0.000045 & 3503 & 3684 & 1.000873 & 0.006055 & 0.003046 & 0.003079 & 0.001890 & 0.005599 & 0.011495 \\
$\mu_{36}$ & 0.000124 & 0.000127 & 3968 & 3560 & 1.001497 & 0.014785 & 0.007801 & 0.007861 & 0.004138 & 0.013814 & 0.028849 \\
$\mu_{37}$ & 0.000055 & 0.000063 & 3769 & 4013 & 1.000986 & 0.004203 & 0.003420 & 0.003523 & 0.000512 & 0.003359 & 0.011193 \\
$\mu_{38}$ & 0.000082 & 0.000094 & 3681 & 3861 & 1.000602 & 0.006123 & 0.004978 & 0.005165 & 0.000744 & 0.004745 & 0.015847 \\
$\mu_{39}$ & 0.000041 & 0.000041 & 4042 & 3848 & 1.000233 & 0.004355 & 0.002654 & 0.002712 & 0.001036 & 0.003801 & 0.009402 \\
$\mu_{40}$ & 0.000239 & 0.000334 & 3448 & 3710 & 1.001402 & 0.017308 & 0.014271 & 0.014735 & 0.002404 & 0.013638 & 0.044869 \\
$\mu_{41}$ & 0.000073 & 0.000085 & 3988 & 3689 & 1.000146 & 0.005893 & 0.004669 & 0.004852 & 0.000847 & 0.004570 & 0.015311 \\
$\mu_{42}$ & 0.000045 & 0.000046 & 4073 & 3823 & 0.999858 & 0.005564 & 0.002844 & 0.002887 & 0.001898 & 0.005069 & 0.010869 \\
$\mu_{43}$ & 0.000093 & 0.000093 & 4096 & 3723 & 1.001056 & 0.008151 & 0.005986 & 0.006132 & 0.001028 & 0.006824 & 0.019966 \\
$\mu_{44}$ & 0.000280 & 0.000349 & 3555 & 3728 & 1.001778 & 0.020074 & 0.016851 & 0.017482 & 0.002613 & 0.015419 & 0.053637 \\
$\mu_{45}$ & 0.000153 & 0.000140 & 3690 & 3712 & 1.000213 & 0.013901 & 0.009344 & 0.009476 & 0.002202 & 0.012329 & 0.031749 \\
$\mu_{46}$ & 0.000605 & 0.001414 & 3916 & 3812 & 0.999804 & 0.037083 & 0.037525 & 0.039208 & 0.003787 & 0.025720 & 0.108487 \\
$\mu_{47}$ & 0.000089 & 0.000113 & 3804 & 3742 & 1.000060 & 0.007288 & 0.005515 & 0.005675 & 0.001282 & 0.005952 & 0.018089 \\
$\mu_{48}$ & 0.000039 & 0.000036 & 4036 & 4040 & 1.000416 & 0.003720 & 0.002420 & 0.002462 & 0.000621 & 0.003269 & 0.008248 \\
$\mu_{49}$ & 0.000156 & 0.000164 & 3627 & 3666 & 1.000931 & 0.013435 & 0.009601 & 0.009826 & 0.002032 & 0.011343 & 0.032341 \\
$\mu_{50}$ & 0.000027 & 0.000028 & 3726 & 3726 & 1.000274 & 0.002524 & 0.001673 & 0.001712 & 0.000552 & 0.002161 & 0.005779 \\
$\mu_{51}$ & 0.000145 & 0.000249 & 3483 & 3848 & 0.999718 & 0.008496 & 0.008752 & 0.009149 & 0.000634 & 0.005831 & 0.025440 \\
$\mu_{52}$ & 0.000091 & 0.000129 & 3997 & 3645 & 0.999845 & 0.005786 & 0.005618 & 0.005890 & 0.000561 & 0.004016 & 0.016848 \\
$\mu_{53}$ & 0.000125 & 0.000211 & 3597 & 3868 & 0.999960 & 0.007015 & 0.007456 & 0.007840 & 0.000604 & 0.004592 & 0.021655 \\
$\mu_{54}$ & 0.000117 & 0.000176 & 4017 & 3754 & 0.999688 & 0.007225 & 0.007342 & 0.007704 & 0.000607 & 0.004890 & 0.022113 \\
$\mu_{55}$ & 0.000151 & 0.000236 & 3469 & 3428 & 1.000353 & 0.007914 & 0.008580 & 0.009018 & 0.000581 & 0.005138 & 0.024698 \\
$\mu_{56}$ & 0.000297 & 0.000914 & 2875 & 3735 & 1.001293 & 0.012772 & 0.017807 & 0.018666 & 0.000746 & 0.007175 & 0.042940 \\
$\mu_{57}$ & 0.000243 & 0.000511 & 3986 & 4002 & 0.999916 & 0.011783 & 0.014988 & 0.015858 & 0.000777 & 0.006606 & 0.040127 \\
$\mu_{58}$ & 0.000169 & 0.000330 & 3528 & 3928 & 1.001571 & 0.009178 & 0.010095 & 0.010610 & 0.000758 & 0.005914 & 0.028453 \\
$\mu_{59}$ & 0.000155 & 0.000254 & 3722 & 3586 & 1.000848 & 0.008965 & 0.009764 & 0.010292 & 0.000676 & 0.005708 & 0.028429 \\
$\mu_{60}$ & 0.000223 & 0.000473 & 4151 & 3982 & 0.999468 & 0.011272 & 0.013704 & 0.014453 & 0.000770 & 0.006680 & 0.038191 \\
$\mu_{61}$ & 0.000121 & 0.000211 & 4261 & 3972 & 0.999650 & 0.007239 & 0.007805 & 0.008219 & 0.000580 & 0.004663 & 0.022694 \\
$\mu_{62}$ & 0.000182 & 0.000353 & 3375 & 3626 & 1.000764 & 0.009554 & 0.011121 & 0.011701 & 0.000639 & 0.005916 & 0.030981 \\
$\mu_{63}$ & 0.000171 & 0.000316 & 3593 & 3761 & 1.000850 & 0.009689 & 0.010807 & 0.011391 & 0.000742 & 0.006088 & 0.032065 \\
$\mu_{64}$ & 0.000104 & 0.000162 & 3722 & 3821 & 1.001126 & 0.006619 & 0.006586 & 0.006905 & 0.000588 & 0.004545 & 0.019419 \\
$\mu_{65}$ & 0.000085 & 0.000152 & 3916 & 3890 & 1.001329 & 0.005310 & 0.005302 & 0.005551 & 0.000491 & 0.003667 & 0.015495 \\
$\mu_{66}$ & 0.000130 & 0.000229 & 3901 & 3390 & 1.000396 & 0.007436 & 0.007846 & 0.008243 & 0.000630 & 0.004906 & 0.023291 \\
$\mu_{67}$ & 0.000175 & 0.000264 & 3288 & 3209 & 1.000092 & 0.008678 & 0.009795 & 0.010346 & 0.000610 & 0.005346 & 0.028232 \\
$\mu_{68}$ & 0.000492 & 0.001248 & 3756 & 3692 & 0.999895 & 0.021100 & 0.031804 & 0.033592 & 0.000938 & 0.010289 & 0.080758 \\
$\mu_{69}$ & 0.000110 & 0.000175 & 3890 & 3749 & 1.000234 & 0.006810 & 0.006940 & 0.007292 & 0.000616 & 0.004572 & 0.020578 \\
\end{longtable}
}

{\footnotesize
\begin{longtable}{lrrrrrrrrrrr}
\caption{MCMC diagnostics for the pooled model parameters. Statistics are based on 4000 MCMC samples from the posterior distributions. Reported metrics include Monte Carlo standard errors (mcse), effective sample sizes (ess), and $\hat{R}$ convergence diagnostics for each parameter.} \label{apptab:mcmc_diagnostics_pooled} \\
\toprule
label & mcse-mean & mcse-sd & ess-bulk & ess-tail & r-hat & mean & std & median-std & 5\% & median & 95\% \\
\midrule
\endfirsthead

\toprule
label & mcse-mean & mcse-sd & ess-bulk & ess-tail & r-hat & mean & std & median-std & 5\% & median & 95\% \\
\midrule
\endhead

\midrule
\multicolumn{12}{r}{\textit{Continued on next page}} \\
\endfoot

\bottomrule
\endlastfoot
$\alpha$ & 0.000238 & 0.000172 & 3894.887234 & 3409.326031 & 1.000263 & 0.898585 & 0.014843 & 0.014845 & 0.874801 & 0.898304 & 0.922887 \\
$\beta$ & 0.000184 & 0.000137 & 4167.728028 & 3879.384501 & 1.000216 & 0.272208 & 0.011869 & 0.011871 & 0.253230 & 0.271985 & 0.291764 \\
$\mu$ & 0.000015 & 0.000011 & 4115.363637 & 3961.730651 & 0.999922 & 0.014709 & 0.000982 & 0.000982 & 0.013122 & 0.014692 & 0.016303 \\
\end{longtable}
}

{\footnotesize
\begin{longtable}{lrrrrrrrrrrr}
\caption{MCMC diagnostics for the unpooled model parameters. Statistics are based on 4000 MCMC samples from the posterior distributions. Reported metrics include Monte Carlo standard errors (mcse), effective sample sizes (ess), and $\hat{R}$ convergence diagnostics for each parameter.} \label{apptab:mcmc_diagnostics_unpooled} \\
\toprule
label & mcse-mean & mcse-sd & ess-bulk & ess-tail & r-hat & mean & std & median-std & 5\% & median & 95\% \\
\midrule
\endfirsthead

\toprule
label & mcse-mean & mcse-sd & ess-bulk & ess-tail & r-hat & mean & std & median-std & 5\% & median & 95\% \\
\midrule
\endhead

\midrule
\multicolumn{12}{r}{\textit{Continued on next page}} \\
\endfoot

\bottomrule
\endlastfoot

$\alpha_0$ & 0.000574 & 0.000412 & 4116 & 4002 & 0.999717 & 0.842833 & 0.036825 & 0.036828 & 0.781704 & 0.842367 & 0.903373 \\
$\alpha_1$ & 0.001061 & 0.000733 & 3813 & 3931 & 1.000356 & 0.630499 & 0.065449 & 0.065456 & 0.525318 & 0.629550 & 0.738096 \\
$\alpha_2$ & 0.001187 & 0.000819 & 3715 & 3890 & 0.999529 & 0.809215 & 0.072259 & 0.072295 & 0.696981 & 0.806916 & 0.932088 \\
$\alpha_3$ & 0.000776 & 0.000565 & 3972 & 3454 & 1.000535 & 0.833217 & 0.048867 & 0.048875 & 0.754563 & 0.832366 & 0.915015 \\
$\alpha_4$ & 0.000928 & 0.000648 & 3912 & 3877 & 1.000259 & 0.793520 & 0.058007 & 0.058007 & 0.696032 & 0.793585 & 0.887924 \\
$\alpha_5$ & 0.001389 & 0.000992 & 4048 & 4128 & 1.000538 & 0.643709 & 0.088308 & 0.088374 & 0.506953 & 0.640296 & 0.798237 \\
$\alpha_6$ & 0.001286 & 0.000941 & 4119 & 4052 & 1.000554 & 0.835543 & 0.082570 & 0.082588 & 0.701662 & 0.833837 & 0.976241 \\
$\alpha_7$ & 0.001917 & 0.001296 & 3807 & 3660 & 1.000356 & 0.854283 & 0.118613 & 0.118746 & 0.666461 & 0.848669 & 1.057624 \\
$\alpha_8$ & 0.001892 & 0.001347 & 3845 & 3725 & 1.000413 & 0.819994 & 0.117536 & 0.117664 & 0.633520 & 0.814498 & 1.022494 \\
$\alpha_9$ & 0.001721 & 0.001294 & 4296 & 3968 & 0.999886 & 0.638928 & 0.113122 & 0.113190 & 0.464364 & 0.634997 & 0.827322 \\
$\alpha_{10}$ & 0.002213 & 0.001702 & 3846 & 3749 & 0.999665 & 0.719180 & 0.137276 & 0.137492 & 0.510801 & 0.711480 & 0.954714 \\
$\alpha_{11}$ & 0.001306 & 0.000924 & 3750 & 3837 & 0.999610 & 0.653333 & 0.079663 & 0.079720 & 0.526536 & 0.650322 & 0.787870 \\
$\alpha_{12}$ & 0.002327 & 0.001839 & 4101 & 4140 & 0.999408 & 0.862268 & 0.149145 & 0.149396 & 0.629408 & 0.853601 & 1.117425 \\
$\alpha_{13}$ & 0.001322 & 0.001105 & 4433 & 4099 & 1.000239 & 0.326736 & 0.087837 & 0.087989 & 0.194506 & 0.321572 & 0.482351 \\
$\alpha_{14}$ & 0.001883 & 0.001334 & 3846 & 3746 & 0.999716 & 0.757756 & 0.116785 & 0.116870 & 0.570123 & 0.753280 & 0.956136 \\
$\alpha_{15}$ & 0.003427 & 0.004832 & 4352 & 4053 & 1.001437 & 0.989429 & 0.225987 & 0.226873 & 0.674355 & 0.969397 & 1.375153 \\
$\alpha_{16}$ & 0.002324 & 0.001675 & 4174 & 3873 & 0.999863 & 0.603663 & 0.149723 & 0.149788 & 0.370732 & 0.599251 & 0.859425 \\
$\alpha_{17}$ & 0.001765 & 0.001262 & 3897 & 3869 & 1.000112 & 0.816884 & 0.110102 & 0.110223 & 0.641909 & 0.811730 & 1.003424 \\
$\alpha_{18}$ & 0.001788 & 0.001294 & 4056 & 4052 & 1.000073 & 0.758486 & 0.113809 & 0.113843 & 0.576341 & 0.755700 & 0.951823 \\
$\alpha_{19}$ & 0.002078 & 0.001595 & 4182 & 3566 & 0.999654 & 0.578030 & 0.134376 & 0.134652 & 0.371307 & 0.569417 & 0.811100 \\
$\alpha_{20}$ & 0.003122 & 0.002523 & 3987 & 4072 & 0.999960 & 0.735030 & 0.196516 & 0.197004 & 0.440796 & 0.721175 & 1.080236 \\
$\alpha_{21}$ & 0.001999 & 0.001439 & 3788 & 3641 & 1.000835 & 0.587661 & 0.122971 & 0.123401 & 0.399469 & 0.577369 & 0.802922 \\
$\alpha_{22}$ & 0.003219 & 0.002508 & 3924 & 3799 & 0.999325 & 0.811567 & 0.201610 & 0.202330 & 0.513033 & 0.794507 & 1.174709 \\
$\alpha_{23}$ & 0.001999 & 0.001552 & 4036 & 3939 & 1.001155 & 0.760526 & 0.127739 & 0.127890 & 0.563571 & 0.754310 & 0.982899 \\
$\alpha_{24}$ & 0.002756 & 0.003208 & 3948 & 3971 & 0.999808 & 0.572129 & 0.172820 & 0.173668 & 0.326493 & 0.554988 & 0.872180 \\
$\alpha_{25}$ & 0.003377 & 0.003362 & 4073 & 3791 & 0.999989 & 0.870482 & 0.216339 & 0.218054 & 0.564606 & 0.843191 & 1.260032 \\
$\alpha_{26}$ & 0.002291 & 0.001760 & 3903 & 3969 & 0.999730 & 0.534847 & 0.143186 & 0.143545 & 0.320446 & 0.524702 & 0.786186 \\
$\alpha_{27}$ & 0.003043 & 0.002615 & 4273 & 3473 & 1.000191 & 0.837370 & 0.198337 & 0.198914 & 0.540968 & 0.822229 & 1.187554 \\
$\alpha_{28}$ & 0.002014 & 0.001561 & 4163 & 3957 & 1.000713 & 0.589712 & 0.129686 & 0.129890 & 0.390589 & 0.582427 & 0.819531 \\
$\alpha_{29}$ & 0.003177 & 0.002359 & 3694 & 3793 & 1.000527 & 0.658228 & 0.192006 & 0.192666 & 0.370001 & 0.642291 & 0.998341 \\
$\alpha_{30}$ & 0.005069 & 0.005362 & 4022 & 4056 & 1.000547 & 0.778044 & 0.326861 & 0.329970 & 0.320869 & 0.732856 & 1.351714 \\
$\alpha_{31}$ & 0.004710 & 0.003806 & 3899 & 4085 & 1.000294 & 0.597273 & 0.291291 & 0.293610 & 0.187229 & 0.560443 & 1.130762 \\
$\alpha_{32}$ & 0.002299 & 0.001718 & 3968 & 3832 & 1.001538 & 0.549661 & 0.145037 & 0.145593 & 0.335037 & 0.536945 & 0.813240 \\
$\alpha_{33}$ & 0.003180 & 0.002584 & 3606 & 3774 & 1.001064 & 0.531480 & 0.189484 & 0.191148 & 0.269389 & 0.506317 & 0.880843 \\
$\alpha_{34}$ & 0.008364 & 0.008380 & 3275 & 3927 & 0.999676 & 0.901331 & 0.487613 & 0.496204 & 0.310595 & 0.809398 & 1.809883 \\
$\alpha_{35}$ & 0.003545 & 0.003151 & 3841 & 3912 & 0.999816 & 0.677798 & 0.221974 & 0.223354 & 0.361906 & 0.653007 & 1.074839 \\
$\alpha_{36}$ & 0.004716 & 0.005504 & 3586 & 3707 & 1.000960 & 0.720982 & 0.285556 & 0.286888 & 0.315146 & 0.693364 & 1.215502 \\
$\alpha_{37}$ & 0.005942 & 0.006497 & 3459 & 3475 & 1.000291 & 0.871847 & 0.365596 & 0.367390 & 0.336214 & 0.835579 & 1.502804 \\
$\alpha_{38}$ & 0.006282 & 0.005539 & 3504 & 3697 & 1.000520 & 0.962226 & 0.369418 & 0.372073 & 0.453472 & 0.917862 & 1.624296 \\
$\alpha_{39}$ & 0.007446 & 0.008321 & 3679 & 3862 & 1.001204 & 0.863818 & 0.446796 & 0.456315 & 0.329838 & 0.771100 & 1.699990 \\
$\alpha_{40}$ & 0.005805 & 0.004743 & 3652 & 3799 & 1.001241 & 0.786867 & 0.354264 & 0.357231 & 0.289355 & 0.740915 & 1.434882 \\
$\alpha_{41}$ & 0.004698 & 0.004737 & 4042 & 3739 & 1.000218 & 0.512872 & 0.292542 & 0.298162 & 0.139243 & 0.455257 & 1.042348 \\
$\alpha_{42}$ & 0.004556 & 0.004420 & 3847 & 3617 & 1.000662 & 0.663392 & 0.282520 & 0.285021 & 0.280351 & 0.625716 & 1.179221 \\
$\alpha_{43}$ & 0.005940 & 0.007650 & 4192 & 3594 & 1.000412 & 0.729523 & 0.381419 & 0.386621 & 0.249470 & 0.666316 & 1.418397 \\
$\alpha_{44}$ & 0.005565 & 0.006397 & 3822 & 3974 & 1.000086 & 0.570019 & 0.347211 & 0.353126 & 0.147357 & 0.505655 & 1.226759 \\
$\alpha_{45}$ & 0.005500 & 0.009050 & 3383 & 3492 & 1.000219 & 0.392500 & 0.308433 & 0.316509 & 0.082995 & 0.321456 & 0.940905 \\
$\alpha_{46}$ & 0.009015 & 0.009264 & 3203 & 3744 & 1.000570 & 0.953397 & 0.522976 & 0.530556 & 0.295200 & 0.864032 & 1.955573 \\
$\alpha_{47}$ & 0.006654 & 0.006475 & 4193 & 3776 & 0.999799 & 0.791390 & 0.430025 & 0.436722 & 0.237742 & 0.715200 & 1.623362 \\
$\alpha_{48}$ & 0.007048 & 0.007849 & 4010 & 3899 & 0.999734 & 0.668447 & 0.444339 & 0.454424 & 0.143931 & 0.573240 & 1.498485 \\
$\alpha_{49}$ & 0.006256 & 0.007114 & 3544 & 3767 & 1.000224 & 0.582443 & 0.381611 & 0.390691 & 0.129412 & 0.498701 & 1.298209 \\
$\alpha_{50}$ & 0.007604 & 0.008132 & 3603 & 3872 & 0.999780 & 0.727243 & 0.466316 & 0.474772 & 0.162354 & 0.638040 & 1.609145 \\
$\alpha_{51}$ & 0.007761 & 0.007560 & 3706 & 3862 & 1.001773 & 0.717764 & 0.461715 & 0.472990 & 0.162217 & 0.615105 & 1.590726 \\
$\alpha_{52}$ & 0.007232 & 0.007647 & 4049 & 3993 & 1.000904 & 0.721864 & 0.460115 & 0.469586 & 0.163409 & 0.628025 & 1.615064 \\
$\beta_0$ & 0.000967 & 0.000703 & 4037 & 4140 & 1.000586 & 0.437691 & 0.061665 & 0.061727 & 0.341443 & 0.434926 & 0.543873 \\
$\beta_1$ & 0.001273 & 0.000996 & 3571 & 3679 & 1.000301 & 0.450662 & 0.076233 & 0.076558 & 0.337286 & 0.443621 & 0.583071 \\
$\beta_2$ & 0.002936 & 0.002364 & 3913 & 3677 & 1.000772 & 0.776649 & 0.183595 & 0.184010 & 0.500514 & 0.764302 & 1.101255 \\
$\beta_3$ & 0.001733 & 0.001384 & 3973 & 3891 & 0.999524 & 0.726047 & 0.109509 & 0.109702 & 0.559285 & 0.719531 & 0.912679 \\
$\beta_4$ & 0.000920 & 0.000733 & 3799 & 3870 & 1.001387 & 0.379256 & 0.056547 & 0.056636 & 0.295094 & 0.376087 & 0.477908 \\
$\beta_5$ & 0.012171 & 0.009816 & 3762 & 3707 & 1.000706 & 2.332840 & 0.744487 & 0.748139 & 1.242505 & 2.259010 & 3.680980 \\
$\beta_6$ & 0.001291 & 0.000942 & 3841 & 3878 & 1.000011 & 0.462278 & 0.079928 & 0.080158 & 0.340167 & 0.456218 & 0.599770 \\
$\beta_7$ & 0.003232 & 0.002737 & 3802 & 3598 & 0.999589 & 0.729135 & 0.197019 & 0.198026 & 0.445317 & 0.709186 & 1.084719 \\
$\beta_8$ & 0.003370 & 0.003023 & 3725 & 3684 & 1.000718 & 0.672481 & 0.205433 & 0.206901 & 0.385671 & 0.647880 & 1.053476 \\
$\beta_9$ & 0.002867 & 0.002923 & 3507 & 3810 & 0.999916 & 0.576113 & 0.169231 & 0.170802 & 0.348071 & 0.553000 & 0.885806 \\
$\beta_{10}$ & 0.003723 & 0.003572 & 4167 & 3898 & 1.000247 & 0.674675 & 0.240050 & 0.241842 & 0.333772 & 0.645292 & 1.108244 \\
$\beta_{11}$ & 0.001015 & 0.000837 & 4140 & 3962 & 0.999822 & 0.313779 & 0.065538 & 0.065784 & 0.217663 & 0.308098 & 0.432549 \\
$\beta_{12}$ & 0.003549 & 0.002985 & 3907 & 4016 & 0.999942 & 0.577759 & 0.222117 & 0.224760 & 0.273655 & 0.543392 & 0.998235 \\
$\beta_{13}$ & 0.020803 & 0.024743 & 3639 & 3690 & 0.999739 & 2.512741 & 1.243850 & 1.262789 & 0.989525 & 2.294854 & 4.808529 \\
$\beta_{14}$ & 0.000542 & 0.000471 & 4130 & 3936 & 1.001502 & 0.117672 & 0.034564 & 0.034762 & 0.066971 & 0.113962 & 0.180927 \\
$\beta_{15}$ & 0.003827 & 0.003274 & 3923 & 3932 & 1.000247 & 0.537361 & 0.240260 & 0.241616 & 0.197111 & 0.511803 & 0.979179 \\
$\beta_{16}$ & 0.027680 & 0.034160 & 4173 & 3972 & 1.000134 & 1.878550 & 1.750354 & 1.883515 & 0.509705 & 1.182930 & 5.946013 \\
$\beta_{17}$ & 0.001103 & 0.000954 & 4015 & 3835 & 1.000598 & 0.268346 & 0.070541 & 0.071005 & 0.166274 & 0.260241 & 0.396700 \\
$\beta_{18}$ & 0.001762 & 0.002638 & 3656 & 3971 & 1.000725 & 0.380531 & 0.106879 & 0.107523 & 0.230406 & 0.368783 & 0.565948 \\
$\beta_{19}$ & 0.003300 & 0.004000 & 3848 & 3717 & 1.000166 & 0.486570 & 0.203963 & 0.207026 & 0.226725 & 0.451090 & 0.866350 \\
$\beta_{20}$ & 0.003905 & 0.004793 & 4102 & 3849 & 1.000156 & 0.553527 & 0.249581 & 0.255493 & 0.257138 & 0.498882 & 1.037742 \\
$\beta_{21}$ & 0.001477 & 0.001287 & 3706 & 3963 & 0.999937 & 0.324734 & 0.089368 & 0.090123 & 0.200289 & 0.313089 & 0.484767 \\
$\beta_{22}$ & 0.004549 & 0.004401 & 4112 & 3900 & 1.000170 & 0.637070 & 0.292195 & 0.294634 & 0.241619 & 0.599234 & 1.175005 \\
$\beta_{23}$ & 0.001459 & 0.001423 & 3703 & 3515 & 1.000240 & 0.307433 & 0.089337 & 0.090117 & 0.185492 & 0.295602 & 0.473916 \\
$\beta_{24}$ & 0.001538 & 0.002385 & 3485 & 3919 & 1.001515 & 0.146847 & 0.089455 & 0.089996 & 0.027448 & 0.136992 & 0.292106 \\
$\beta_{25}$ & 0.000633 & 0.000789 & 3387 & 3728 & 1.000068 & 0.061193 & 0.036686 & 0.037783 & 0.020499 & 0.052152 & 0.130625 \\
$\beta_{26}$ & 0.003398 & 0.003923 & 3261 & 3651 & 1.000424 & 0.355494 & 0.202949 & 0.205892 & 0.103132 & 0.320808 & 0.717746 \\
$\beta_{27}$ & 0.000853 & 0.001064 & 4229 & 3826 & 0.999618 & 0.131653 & 0.055289 & 0.055789 & 0.056551 & 0.124199 & 0.229284 \\
$\beta_{28}$ & 0.005167 & 0.004965 & 4036 & 3841 & 1.000589 & 0.924093 & 0.329099 & 0.333242 & 0.485955 & 0.871705 & 1.538444 \\
$\beta_{29}$ & 0.005378 & 0.004993 & 3967 & 3969 & 0.999972 & 0.821659 & 0.338319 & 0.340930 & 0.342232 & 0.779548 & 1.428293 \\
$\beta_{30}$ & 0.004668 & 0.008893 & 3147 & 3902 & 0.999816 & 0.223819 & 0.276430 & 0.292609 & 0.016721 & 0.127869 & 0.764836 \\
$\beta_{31}$ & 0.039404 & 0.947601 & 3355 & 3321 & 0.999857 & 0.678730 & 2.518477 & 2.535606 & 0.055492 & 0.384507 & 1.745557 \\
$\beta_{32}$ & 0.004474 & 0.004455 & 3707 & 3928 & 1.000335 & 0.672192 & 0.270315 & 0.273717 & 0.314579 & 0.629176 & 1.182217 \\
$\beta_{33}$ & 0.020625 & 0.018670 & 3486 & 3567 & 1.000301 & 1.625399 & 1.233904 & 1.250317 & 0.163306 & 1.423470 & 3.894843 \\
$\beta_{34}$ & 0.003555 & 0.009453 & 2771 & 3685 & 1.000294 & 0.098741 & 0.204110 & 0.213617 & 0.007774 & 0.035722 & 0.430914 \\
$\beta_{35}$ & 0.001749 & 0.002441 & 3731 & 3537 & 1.000010 & 0.164420 & 0.102358 & 0.104911 & 0.044594 & 0.141417 & 0.362593 \\
$\beta_{36}$ & 0.002050 & 0.003555 & 3463 & 3517 & 1.001423 & 0.180188 & 0.125700 & 0.127943 & 0.035962 & 0.156334 & 0.398123 \\
$\beta_{37}$ & 0.008248 & 0.033517 & 3426 & 3473 & 1.001044 & 0.187033 & 0.485283 & 0.500382 & 0.017967 & 0.065039 & 0.875574 \\
$\beta_{38}$ & 0.003025 & 0.006802 & 3144 & 3954 & 1.000828 & 0.237555 & 0.176592 & 0.181780 & 0.053867 & 0.194439 & 0.564825 \\
$\beta_{39}$ & 0.006184 & 0.022799 & 2622 & 3410 & 1.000879 & 0.304840 & 0.348150 & 0.370224 & 0.013842 & 0.178917 & 0.967476 \\
$\beta_{40}$ & 0.011627 & 0.068569 & 3730 & 3405 & 0.999514 & 0.774221 & 0.713928 & 0.730262 & 0.136749 & 0.620634 & 1.877686 \\
$\beta_{41}$ & 0.008121 & 0.027427 & 3963 & 3945 & 1.000038 & 0.498319 & 0.510851 & 0.532269 & 0.070717 & 0.348851 & 1.384917 \\
$\beta_{42}$ & 0.006684 & 0.083372 & 3633 & 3829 & 1.000282 & 0.393274 & 0.417792 & 0.422666 & 0.097362 & 0.329272 & 0.847375 \\
$\beta_{43}$ & 0.010707 & 0.094213 & 2648 & 3533 & 1.001288 & 0.392839 & 0.606558 & 0.617429 & 0.015297 & 0.277491 & 1.113168 \\
$\beta_{44}$ & 0.199341 & 2.225104 & 3540 & 3426 & 1.002084 & 1.403364 & 11.878705 & 11.917869 & 0.067306 & 0.437972 & 2.292227 \\
$\beta_{45}$ & 0.028137 & 0.719950 & 1923 & 2801 & 1.000037 & 0.315662 & 1.752492 & 1.763564 & 0.002737 & 0.118361 & 1.234751 \\
$\beta_{46}$ & 0.052211 & 1.266507 & 3446 & 3584 & 1.000394 & 1.230975 & 3.312102 & 3.331851 & 0.104030 & 0.868749 & 3.084096 \\
$\beta_{47}$ & 0.160878 & 2.703025 & 3279 & 4003 & 1.000652 & 5.901596 & 9.613227 & 9.884609 & 0.648863 & 3.601299 & 18.529569 \\
$\beta_{48}$ & 0.000980 & 0.011974 & 3169 & 3336 & 1.000318 & 0.034674 & 0.059722 & 0.060890 & 0.006767 & 0.022804 & 0.099025 \\
$\beta_{49}$ & 15.489460 & 480.776515 & 1981 & 2477 & 1.000140 & 23.464457 & 980.821086 & 981.076575 & 0.005687 & 1.075984 & 17.261768 \\
$\beta_{50}$ & 13.018856 & 262.931628 & 2929 & 2581 & 0.999911 & 28.534296 & 729.525990 & 730.030467 & 0.062316 & 1.399203 & 20.882902 \\
$\beta_{51}$ & 0.564686 & 8.549065 & 2550 & 3304 & 1.000257 & 5.042335 & 33.824437 & 34.089965 & 0.055500 & 0.795789 & 13.244361 \\
$\beta_{52}$ & 0.144860 & 3.216954 & 3337 & 3491 & 1.001730 & 1.733833 & 9.227895 & 9.252985 & 0.223794 & 1.052890 & 3.472127 \\
$\mu_0$ & 0.000281 & 0.000185 & 3679 & 3616 & 1.000922 & 0.103897 & 0.017131 & 0.017138 & 0.076540 & 0.103397 & 0.132888 \\
$\mu_1$ & 0.001387 & 0.000952 & 3712 & 4014 & 0.999754 & 0.510356 & 0.084479 & 0.084482 & 0.373630 & 0.509717 & 0.648903 \\
$\mu_2$ & 0.000321 & 0.000247 & 4049 & 3480 & 0.999893 & 0.086000 & 0.020438 & 0.020463 & 0.054547 & 0.084992 & 0.121356 \\
$\mu_3$ & 0.000389 & 0.000299 & 3980 & 3647 & 1.000121 & 0.113085 & 0.024456 & 0.024496 & 0.074751 & 0.111691 & 0.155691 \\
$\mu_4$ & 0.000284 & 0.000208 & 3911 & 3731 & 1.000612 & 0.088210 & 0.017765 & 0.017775 & 0.059617 & 0.087615 & 0.118110 \\
$\mu_5$ & 0.000524 & 0.000390 & 4036 & 3849 & 1.000759 & 0.157894 & 0.033313 & 0.033325 & 0.106273 & 0.157011 & 0.213798 \\
$\mu_6$ & 0.000108 & 0.000085 & 3918 & 3731 & 1.000117 & 0.028329 & 0.006727 & 0.006747 & 0.018255 & 0.027818 & 0.040354 \\
$\mu_7$ & 0.000226 & 0.000183 & 4245 & 3922 & 0.999731 & 0.042050 & 0.014708 & 0.014786 & 0.021058 & 0.040531 & 0.068871 \\
$\mu_8$ & 0.000420 & 0.000323 & 4106 & 3945 & 0.999803 & 0.080872 & 0.026928 & 0.027046 & 0.041785 & 0.078351 & 0.128391 \\
$\mu_9$ & 0.000329 & 0.000262 & 4063 & 3888 & 0.999703 & 0.063929 & 0.021086 & 0.021141 & 0.032957 & 0.062403 & 0.100715 \\
$\mu_{10}$ & 0.000419 & 0.000345 & 4099 & 3982 & 0.999823 & 0.078085 & 0.027043 & 0.027216 & 0.038962 & 0.075029 & 0.126139 \\
$\mu_{11}$ & 0.000148 & 0.000109 & 3991 & 3468 & 0.999490 & 0.045412 & 0.009324 & 0.009337 & 0.031181 & 0.044923 & 0.061596 \\
$\mu_{12}$ & 0.000380 & 0.000301 & 4212 & 3929 & 1.000031 & 0.060933 & 0.024574 & 0.024692 & 0.026069 & 0.058528 & 0.104872 \\
$\mu_{13}$ & 0.000273 & 0.000204 & 4065 & 4055 & 1.000176 & 0.103408 & 0.017378 & 0.017402 & 0.076986 & 0.102500 & 0.134087 \\
$\mu_{14}$ & 0.000029 & 0.000024 & 4177 & 3775 & 0.999824 & 0.005080 & 0.001858 & 0.001869 & 0.002398 & 0.004877 & 0.008402 \\
$\mu_{15}$ & 0.000048 & 0.000044 & 3449 & 3622 & 1.000405 & 0.004099 & 0.002794 & 0.002849 & 0.000744 & 0.003541 & 0.009496 \\
$\mu_{16}$ & 0.000212 & 0.000150 & 3817 & 3695 & 1.000262 & 0.042417 & 0.013095 & 0.013131 & 0.022861 & 0.041440 & 0.065793 \\
$\mu_{17}$ & 0.000130 & 0.000097 & 4157 & 3735 & 0.999931 & 0.024883 & 0.008341 & 0.008365 & 0.012499 & 0.024254 & 0.039445 \\
$\mu_{18}$ & 0.000346 & 0.000263 & 3828 & 3905 & 0.999887 & 0.062992 & 0.021437 & 0.021507 & 0.030770 & 0.061257 & 0.100322 \\
$\mu_{19}$ & 0.000213 & 0.000169 & 4162 & 3811 & 0.999836 & 0.048502 & 0.013808 & 0.013852 & 0.027500 & 0.047388 & 0.072773 \\
$\mu_{20}$ & 0.000202 & 0.000170 & 4072 & 3948 & 1.000988 & 0.025646 & 0.012941 & 0.013128 & 0.008115 & 0.023435 & 0.049832 \\
$\mu_{21}$ & 0.000125 & 0.000099 & 3944 & 3772 & 1.000535 & 0.027963 & 0.007872 & 0.007892 & 0.016095 & 0.027393 & 0.042215 \\
$\mu_{22}$ & 0.000332 & 0.000261 & 4087 & 3690 & 0.999809 & 0.046654 & 0.021234 & 0.021366 & 0.015990 & 0.044281 & 0.086193 \\
$\mu_{23}$ & 0.000101 & 0.000083 & 3556 & 3993 & 0.999760 & 0.014951 & 0.006110 & 0.006148 & 0.006214 & 0.014266 & 0.025834 \\
$\mu_{24}$ & 0.000095 & 0.000077 & 4017 & 3610 & 1.000540 & 0.015631 & 0.006026 & 0.006038 & 0.006114 & 0.015250 & 0.025990 \\
$\mu_{25}$ & 0.000044 & 0.000035 & 3682 & 3520 & 0.999668 & 0.004384 & 0.002744 & 0.002777 & 0.000709 & 0.003960 & 0.009641 \\
$\mu_{26}$ & 0.000070 & 0.000048 & 3625 & 3973 & 0.999675 & 0.013978 & 0.004191 & 0.004200 & 0.007519 & 0.013702 & 0.021419 \\
$\mu_{27}$ & 0.000054 & 0.000069 & 2964 & 4101 & 1.000721 & 0.003109 & 0.003060 & 0.003187 & 0.000164 & 0.002219 & 0.009282 \\
$\mu_{28}$ & 0.000077 & 0.000061 & 4037 & 3831 & 1.000009 & 0.018455 & 0.004851 & 0.004862 & 0.011230 & 0.018130 & 0.027177 \\
$\mu_{29}$ & 0.000157 & 0.000130 & 3856 & 3904 & 1.000195 & 0.023944 & 0.009611 & 0.009686 & 0.010300 & 0.022738 & 0.041316 \\
$\mu_{30}$ & 0.000123 & 0.000103 & 2666 & 2995 & 1.000239 & 0.008415 & 0.006689 & 0.006858 & 0.000667 & 0.006901 & 0.021580 \\
$\mu_{31}$ & 0.000564 & 0.000475 & 3830 & 3524 & 1.000777 & 0.066053 & 0.035189 & 0.035381 & 0.015547 & 0.062369 & 0.128427 \\
$\mu_{32}$ & 0.000068 & 0.000054 & 4308 & 3911 & 1.000270 & 0.014202 & 0.004484 & 0.004499 & 0.007657 & 0.013831 & 0.022120 \\
$\mu_{33}$ & 0.000265 & 0.000194 & 3920 & 3363 & 1.000598 & 0.043007 & 0.016494 & 0.016494 & 0.015053 & 0.043043 & 0.070446 \\
$\mu_{34}$ & 0.000202 & 0.000143 & 3534 & 3814 & 0.999778 & 0.029122 & 0.012052 & 0.012095 & 0.011111 & 0.028098 & 0.050516 \\
$\mu_{35}$ & 0.000060 & 0.000052 & 3774 & 4021 & 1.000011 & 0.007324 & 0.003718 & 0.003752 & 0.002262 & 0.006817 & 0.014140 \\
$\mu_{36}$ & 0.000148 & 0.000127 & 3854 & 3458 & 0.999992 & 0.019980 & 0.009501 & 0.009581 & 0.006722 & 0.018745 & 0.037228 \\
$\mu_{37}$ & 0.000102 & 0.000107 & 2814 & 3027 & 1.000543 & 0.005834 & 0.005704 & 0.005986 & 0.000329 & 0.004018 & 0.017953 \\
$\mu_{38}$ & 0.000111 & 0.000111 & 3176 & 3720 & 1.000400 & 0.008524 & 0.006594 & 0.006785 & 0.000981 & 0.006925 & 0.020978 \\
$\mu_{39}$ & 0.000055 & 0.000049 & 3146 & 3286 & 1.000248 & 0.005078 & 0.003253 & 0.003296 & 0.000799 & 0.004548 & 0.011119 \\
$\mu_{40}$ & 0.000360 & 0.000355 & 3535 & 3793 & 0.999924 & 0.033277 & 0.022140 & 0.022556 & 0.006336 & 0.028969 & 0.075753 \\
$\mu_{41}$ & 0.000112 & 0.000114 & 3386 & 3603 & 1.001081 & 0.009444 & 0.006698 & 0.006844 & 0.001334 & 0.008040 & 0.022199 \\
$\mu_{42}$ & 0.000054 & 0.000050 & 4010 & 3415 & 0.999542 & 0.006867 & 0.003401 & 0.003453 & 0.002357 & 0.006269 & 0.013169 \\
$\mu_{43}$ & 0.000129 & 0.000118 & 3693 & 3574 & 1.000060 & 0.012296 & 0.008035 & 0.008160 & 0.002001 & 0.010869 & 0.027334 \\
$\mu_{44}$ & 0.000444 & 0.000351 & 3126 & 3596 & 1.001479 & 0.040041 & 0.024731 & 0.025182 & 0.008097 & 0.035299 & 0.087222 \\
$\mu_{45}$ & 0.000228 & 0.000166 & 3203 & 3383 & 0.999907 & 0.026075 & 0.013213 & 0.013327 & 0.007122 & 0.024334 & 0.050056 \\
$\mu_{46}$ & 0.000837 & 0.000940 & 3806 & 3775 & 1.000496 & 0.075711 & 0.052133 & 0.053267 & 0.014471 & 0.064780 & 0.174384 \\
$\mu_{47}$ & 0.000107 & 0.000105 & 4001 & 3778 & 0.999669 & 0.009968 & 0.006813 & 0.006967 & 0.001780 & 0.008512 & 0.023150 \\
$\mu_{48}$ & 0.000047 & 0.000045 & 3488 & 3586 & 1.000251 & 0.003923 & 0.002815 & 0.002857 & 0.000410 & 0.003434 & 0.009043 \\
$\mu_{49}$ & 0.000205 & 0.000201 & 3869 & 3938 & 1.000432 & 0.021565 & 0.012856 & 0.013050 & 0.005152 & 0.019323 & 0.046127 \\
$\mu_{50}$ & 0.000030 & 0.000032 & 3679 & 3764 & 1.000120 & 0.002631 & 0.001859 & 0.001907 & 0.000494 & 0.002208 & 0.006325 \\
$\mu_{51}$ & 0.000266 & 0.000264 & 2927 & 3092 & 1.000511 & 0.019755 & 0.015519 & 0.015973 & 0.001860 & 0.015974 & 0.050568 \\
$\mu_{52}$ & 0.000165 & 0.000214 & 3444 & 3178 & 1.001507 & 0.010043 & 0.009863 & 0.010278 & 0.000478 & 0.007154 & 0.029783 \\
$\mu_{53}$ & 0.000203 & 0.000210 & 3319 & 3446 & 0.999952 & 0.013002 & 0.012235 & 0.012839 & 0.000709 & 0.009111 & 0.037896 \\
$\mu_{54}$ & 0.000220 & 0.000338 & 3537 & 3626 & 1.000493 & 0.013588 & 0.013556 & 0.014174 & 0.000667 & 0.009447 & 0.039194 \\
$\mu_{55}$ & 0.000268 & 0.000403 & 2418 & 3400 & 1.000459 & 0.015773 & 0.015941 & 0.016661 & 0.000780 & 0.010929 & 0.046340 \\
$\mu_{56}$ & 0.000590 & 0.000693 & 3143 & 3321 & 1.000290 & 0.033893 & 0.033136 & 0.034615 & 0.001773 & 0.023883 & 0.100481 \\
$\mu_{57}$ & 0.000493 & 0.000746 & 3386 & 3609 & 1.000301 & 0.029968 & 0.029574 & 0.030746 & 0.001581 & 0.021559 & 0.087785 \\
$\mu_{58}$ & 0.000300 & 0.000399 & 3563 & 3668 & 1.000705 & 0.019008 & 0.018121 & 0.018920 & 0.001006 & 0.013568 & 0.054825 \\
$\mu_{59}$ & 0.000303 & 0.000408 & 3335 & 3721 & 1.000003 & 0.019079 & 0.018303 & 0.019106 & 0.001075 & 0.013597 & 0.055242 \\
$\mu_{60}$ & 0.000475 & 0.000657 & 3158 & 3074 & 1.001896 & 0.029273 & 0.028386 & 0.029537 & 0.001779 & 0.021110 & 0.085679 \\
$\mu_{61}$ & 0.000233 & 0.000278 & 3194 & 3191 & 0.999949 & 0.013998 & 0.013513 & 0.014078 & 0.000805 & 0.010050 & 0.041229 \\
$\mu_{62}$ & 0.000332 & 0.000415 & 3977 & 3933 & 0.999673 & 0.021902 & 0.021103 & 0.021985 & 0.001050 & 0.015737 & 0.064260 \\
$\mu_{63}$ & 0.000336 & 0.000420 & 2905 & 3028 & 1.001776 & 0.021410 & 0.020601 & 0.021471 & 0.001200 & 0.015362 & 0.062279 \\
$\mu_{64}$ & 0.000193 & 0.000260 & 3225 & 3145 & 1.000688 & 0.011919 & 0.011831 & 0.012384 & 0.000623 & 0.008261 & 0.035372 \\
$\mu_{65}$ & 0.000147 & 0.000191 & 2824 & 3206 & 1.000795 & 0.008373 & 0.008437 & 0.008848 & 0.000431 & 0.005709 & 0.025324 \\
$\mu_{66}$ & 0.000212 & 0.000264 & 3537 & 3325 & 1.000886 & 0.013850 & 0.013435 & 0.014152 & 0.000723 & 0.009403 & 0.041281 \\
$\mu_{67}$ & 0.000287 & 0.000348 & 3407 & 3973 & 1.000891 & 0.017780 & 0.017177 & 0.017961 & 0.000992 & 0.012534 & 0.051493 \\
$\mu_{68}$ & 0.001331 & 0.002424 & 3291 & 3430 & 0.999753 & 0.076933 & 0.081426 & 0.084804 & 0.003865 & 0.053237 & 0.238716 \\
$\mu_{69}$ & 0.000226 & 0.000263 & 3032 & 3380 & 1.000435 & 0.013055 & 0.012828 & 0.013457 & 0.000789 & 0.008991 & 0.038689 \\
\bottomrule
\end{longtable}
}
\section{Sensitivity Analysis via Power-Scaling Prior or Likelihoods}\label{app:power_scale_analys}

In these sensitivity analyses, we investigate the impact of the prior and likelihood on the model's parameters by power-scaling them. The results are presented in two tables: one for the power scale analysis of the hyperpriors of the partial pooled model \cref{apptab:power_scale_analysis_partialpooled} and another for the individual-level parameters of the unpooled model \cref{apptab:power_scale_analysis_unpooled}. We exclude power-scale perturbation of the session-level priors, $\Delta \mu$, as the session-level parameters are not of primary interest in this study. A preliminary analysis showed that even with perturbation of $\Delta \mu$ for the partial pooled model, the branching factor $\mu_\alpha$ remained stable.

\begin{longtable}{lrrl}
\label{apptab:power_scale_analysis_partialpooled} \\
\caption{Power scale analysis results for the partial pooled model hyperparameters and individual-level parameters. Each row reports the sensitivity of the posterior distribution for a parameter to power-scaling of the prior and likelihood, respectively. The ``prior'' and ``likelihood'' columns show the cumulative Jensen-Shannon distance between the based and perturbed posterior distribution under power-scaling of the prior or likelihood, respectively. The ``diagnosis'' column indicates whether the parameter is robust (\checkmark) or shows potential sensitivity to the prior or likelihood.}
\\
\toprule
component & prior & likelihood & diagnosis \\
label &  &  &  \\
\midrule
\endfirsthead

\toprule
component & prior & likelihood & diagnosis \\
label &  &  &  \\
\midrule
\endhead

\midrule
\multicolumn{4}{r}{\textit{Continued on next page}} \\
\endfoot

\bottomrule
\endlastfoot
$\mu_\alpha$ & 0.001200 & 0.384590 & \checkmark \\
$\mu_\beta$ & 0.027760 & 0.480850 & \checkmark \\
$\mu_\mu$ & 0.041510 & 0.246800 & \checkmark \\
$\sigma_\beta$ & 0.026360 & 0.512290 & \checkmark \\
$\sigma_\mu$ & 0.032010 & 0.352610 & \checkmark \\
$\sigma_\alpha$ & 0.001290 & 0.464050 & \checkmark \\
$\alpha_0$ & 0.001560 & 0.217080 & \checkmark \\
$\alpha_1$ & 0.001850 & 0.387480 & \checkmark \\
$\alpha_2$ & 0.000670 & 0.101820 & \checkmark \\
$\alpha_3$ & 0.000730 & 0.149280 & \checkmark \\
$\alpha_4$ & 0.001010 & 0.095710 & \checkmark \\
$\alpha_5$ & 0.001410 & 0.346580 & \checkmark \\
$\alpha_6$ & 0.001130 & 0.120970 & \checkmark \\
$\alpha_7$ & 0.001220 & 0.112570 & \checkmark \\
$\alpha_8$ & 0.001100 & 0.173250 & \checkmark \\
$\alpha_9$ & 0.001220 & 0.247070 & \checkmark \\
$\alpha_{10}$ & 0.001150 & 0.148080 & \checkmark \\
$\alpha_{11}$ & 0.000840 & 0.291090 & \checkmark \\
$\alpha_{12}$ & 0.001070 & 0.117290 & \checkmark \\
$\alpha_{13}$ & 0.001190 & 0.532290 & \checkmark \\
$\alpha_{14}$ & 0.000940 & 0.173030 & \checkmark \\
$\alpha_{15}$ & 0.001500 & 0.191710 & \checkmark \\
$\alpha_{16}$ & 0.000710 & 0.245290 & \checkmark \\
$\alpha_{17}$ & 0.001710 & 0.121180 & \checkmark \\
$\alpha_{18}$ & 0.000750 & 0.126220 & \checkmark \\
$\alpha_{19}$ & 0.000910 & 0.282090 & \checkmark \\
$\alpha_{20}$ & 0.000950 & 0.138410 & \checkmark \\
$\alpha_{21}$ & 0.001120 & 0.281770 & \checkmark \\
$\alpha_{22}$ & 0.001390 & 0.120740 & \checkmark \\
$\alpha_{23}$ & 0.001000 & 0.191620 & \checkmark \\
$\alpha_{24}$ & 0.001300 & 0.278990 & \checkmark \\
$\alpha_{25}$ & 0.002420 & 0.102690 & \checkmark \\
$\alpha_{26}$ & 0.000980 & 0.290890 & \checkmark \\
$\alpha_{27}$ & 0.000960 & 0.125080 & \checkmark \\
$\alpha_{28}$ & 0.000980 & 0.210460 & \checkmark \\
$\alpha_{29}$ & 0.001220 & 0.202010 & \checkmark \\
$\alpha_{30}$ & 0.000910 & 0.150630 & \checkmark \\
$\alpha_{31}$ & 0.000590 & 0.219450 & \checkmark \\
$\alpha_{32}$ & 0.000960 & 0.294640 & \checkmark \\
$\alpha_{33}$ & 0.001020 & 0.181210 & \checkmark \\
$\alpha_{34}$ & 0.001520 & 0.147480 & \checkmark \\
$\alpha_{35}$ & 0.001790 & 0.192770 & \checkmark \\
$\alpha_{36}$ & 0.001150 & 0.149300 & \checkmark \\
$\alpha_{37}$ & 0.001470 & 0.195480 & \checkmark \\
$\alpha_{38}$ & 0.000850 & 0.152980 & \checkmark \\
$\alpha_{39}$ & 0.000760 & 0.158220 & \checkmark \\
$\alpha_{40}$ & 0.001050 & 0.194180 & \checkmark \\
$\alpha_{41}$ & 0.000770 & 0.250940 & \checkmark \\
$\alpha_{42}$ & 0.001590 & 0.200740 & \checkmark \\
$\alpha_{43}$ & 0.000860 & 0.170760 & \checkmark \\
$\alpha_{44}$ & 0.000980 & 0.213190 & \checkmark \\
$\alpha_{45}$ & 0.000670 & 0.192530 & \checkmark \\
$\alpha_{46}$ & 0.000970 & 0.157200 & \checkmark \\
$\alpha_{47}$ & 0.001300 & 0.177960 & \checkmark \\
$\alpha_{48}$ & 0.000830 & 0.147370 & \checkmark \\
$\alpha_{49}$ & 0.001080 & 0.190540 & \checkmark \\
$\alpha_{50}$ & 0.001180 & 0.206760 & \checkmark \\
$\alpha_{51}$ & 0.001380 & 0.153350 & \checkmark \\
$\alpha_{52}$ & 0.000910 & 0.221020 & \checkmark \\
$\beta_0$ & 0.001380 & 0.135270 & \checkmark \\
$\beta_1$ & 0.002330 & 0.046710 & \checkmark \\
$\beta_2$ & 0.003080 & 0.200830 & \checkmark \\
$\beta_3$ & 0.003580 & 0.064070 & \checkmark \\
$\beta_4$ & 0.000850 & 0.113690 & \checkmark \\
$\beta_5$ & 0.007450 & 0.404980 & \checkmark \\
$\beta_6$ & 0.000820 & 0.136240 & \checkmark \\
$\beta_7$ & 0.002950 & 0.097240 & \checkmark \\
$\beta_8$ & 0.002890 & 0.177530 & \checkmark \\
$\beta_9$ & 0.002060 & 0.127750 & \checkmark \\
$\beta_{10}$ & 0.003630 & 0.202610 & \checkmark \\
$\beta_{11}$ & 0.001200 & 0.122260 & \checkmark \\
$\beta_{12}$ & 0.003260 & 0.217290 & \checkmark \\
$\beta_{13}$ & 0.008590 & 0.686220 & \checkmark \\
$\beta_{14}$ & 0.003170 & 0.279760 & \checkmark \\
$\beta_{15}$ & 0.003220 & 0.152880 & \checkmark \\
$\beta_{16}$ & 0.004110 & 0.228390 & \checkmark \\
$\beta_{17}$ & 0.001780 & 0.108000 & \checkmark \\
$\beta_{18}$ & 0.001140 & 0.095360 & \checkmark \\
$\beta_{19}$ & 0.002130 & 0.174060 & \checkmark \\
$\beta_{20}$ & 0.001740 & 0.163340 & \checkmark \\
$\beta_{21}$ & 0.000850 & 0.096430 & \checkmark \\
$\beta_{22}$ & 0.002670 & 0.068320 & \checkmark \\
$\beta_{23}$ & 0.000890 & 0.117080 & \checkmark \\
$\beta_{24}$ & 0.006530 & 0.188580 & \checkmark \\
$\beta_{25}$ & 0.009460 & 0.461860 & \checkmark \\
$\beta_{26}$ & 0.003730 & 0.176500 & \checkmark \\
$\beta_{27}$ & 0.004520 & 0.279750 & \checkmark \\
$\beta_{28}$ & 0.004000 & 0.233450 & \checkmark \\
$\beta_{29}$ & 0.002520 & 0.135570 & \checkmark \\
$\beta_{30}$ & 0.009850 & 0.503060 & \checkmark \\
$\beta_{31}$ & 0.004090 & 0.199280 & \checkmark \\
$\beta_{32}$ & 0.001230 & 0.172700 & \checkmark \\
$\beta_{33}$ & 0.006390 & 0.292290 & \checkmark \\
$\beta_{34}$ & 0.009800 & 0.578170 & \checkmark \\
$\beta_{35}$ & 0.005380 & 0.284990 & \checkmark \\
$\beta_{36}$ & 0.004510 & 0.342100 & \checkmark \\
$\beta_{37}$ & 0.011250 & 0.558100 & \checkmark \\
$\beta_{38}$ & 0.006080 & 0.234350 & \checkmark \\
$\beta_{39}$ & 0.006580 & 0.348890 & \checkmark \\
$\beta_{40}$ & 0.004400 & 0.120390 & \checkmark \\
$\beta_{41}$ & 0.003590 & 0.200990 & \checkmark \\
$\beta_{42}$ & 0.004750 & 0.228090 & \checkmark \\
$\beta_{43}$ & 0.008130 & 0.192120 & \checkmark \\
$\beta_{44}$ & 0.007790 & 0.173360 & \checkmark \\
$\beta_{45}$ & 0.008530 & 0.389730 & \checkmark \\
$\beta_{46}$ & 0.008190 & 0.347940 & \checkmark \\
$\beta_{47}$ & 0.021490 & 0.732620 & \checkmark \\
$\beta_{48}$ & 0.018020 & 0.789800 & \checkmark \\
$\beta_{49}$ & 0.011880 & 0.633940 & \checkmark \\
$\beta_{50}$ & 0.015920 & 0.618980 & \checkmark \\
$\beta_{51}$ & 0.009910 & 0.525530 & \checkmark \\
$\beta_{52}$ & 0.009220 & 0.285610 & \checkmark \\
$\mu_0$ & 0.001070 & 0.162590 & \checkmark \\
$\mu_1$ & 0.002130 & 0.345010 & \checkmark \\
$\mu_2$ & 0.002370 & 0.151880 & \checkmark \\
$\mu_3$ & 0.003210 & 0.120760 & \checkmark \\
$\mu_4$ & 0.001510 & 0.080280 & \checkmark \\
$\mu_5$ & 0.004070 & 0.310510 & \checkmark \\
$\mu_6$ & 0.001300 & 0.093270 & \checkmark \\
$\mu_7$ & 0.002130 & 0.164190 & \checkmark \\
$\mu_8$ & 0.001930 & 0.087210 & \checkmark \\
$\mu_9$ & 0.001840 & 0.105780 & \checkmark \\
$\mu_{10}$ & 0.002880 & 0.166110 & \checkmark \\
$\mu_{11}$ & 0.001030 & 0.159870 & \checkmark \\
$\mu_{12}$ & 0.003580 & 0.170270 & \checkmark \\
$\mu_{13}$ & 0.003300 & 0.387700 & \checkmark \\
$\mu_{14}$ & 0.003590 & 0.202720 & \checkmark \\
$\mu_{15}$ & 0.008170 & 0.404400 & \checkmark \\
$\mu_{16}$ & 0.001480 & 0.117640 & \checkmark \\
$\mu_{17}$ & 0.000980 & 0.075830 & \checkmark \\
$\mu_{18}$ & 0.001510 & 0.084420 & \checkmark \\
$\mu_{19}$ & 0.001880 & 0.135580 & \checkmark \\
$\mu_{20}$ & 0.003060 & 0.212720 & \checkmark \\
$\mu_{21}$ & 0.001220 & 0.122070 & \checkmark \\
$\mu_{22}$ & 0.005030 & 0.156370 & \checkmark \\
$\mu_{23}$ & 0.001610 & 0.144920 & \checkmark \\
$\mu_{24}$ & 0.004330 & 0.116400 & \checkmark \\
$\mu_{25}$ & 0.007390 & 0.440750 & \checkmark \\
$\mu_{26}$ & 0.001480 & 0.073900 & \checkmark \\
$\mu_{27}$ & 0.009660 & 0.567520 & \checkmark \\
$\mu_{28}$ & 0.001210 & 0.139600 & \checkmark \\
$\mu_{29}$ & 0.002640 & 0.207710 & \checkmark \\
$\mu_{30}$ & 0.009160 & 0.319100 & \checkmark \\
$\mu_{31}$ & 0.003970 & 0.166070 & \checkmark \\
$\mu_{32}$ & 0.002110 & 0.121440 & \checkmark \\
$\mu_{33}$ & 0.002320 & 0.123760 & \checkmark \\
$\mu_{34}$ & 0.001600 & 0.059360 & \checkmark \\
$\mu_{35}$ & 0.003740 & 0.138330 & \checkmark \\
$\mu_{36}$ & 0.001660 & 0.068780 & \checkmark \\
$\mu_{37}$ & 0.009860 & 0.410560 & \checkmark \\
$\mu_{38}$ & 0.006010 & 0.504310 & \checkmark \\
$\mu_{39}$ & 0.005360 & 0.355350 & \checkmark \\
$\mu_{40}$ & 0.003520 & 0.174180 & \checkmark \\
$\mu_{41}$ & 0.006160 & 0.345340 & \checkmark \\
$\mu_{42}$ & 0.001710 & 0.184750 & \checkmark \\
$\mu_{43}$ & 0.007410 & 0.257960 & \checkmark \\
$\mu_{44}$ & 0.003940 & 0.205030 & \checkmark \\
$\mu_{45}$ & 0.003060 & 0.129150 & \checkmark \\
$\mu_{46}$ & 0.006270 & 0.213570 & \checkmark \\
$\mu_{47}$ & 0.005460 & 0.224500 & \checkmark \\
$\mu_{48}$ & 0.008970 & 0.350320 & \checkmark \\
$\mu_{49}$ & 0.004800 & 0.196330 & \checkmark \\
$\mu_{50}$ & 0.007270 & 0.379510 & \checkmark \\
$\mu_{51}$ & 0.007880 & 0.404040 & \checkmark \\
$\mu_{52}$ & 0.006290 & 0.343940 & \checkmark \\
$\mu_{53}$ & 0.007240 & 0.316480 & \checkmark \\
$\mu_{54}$ & 0.007030 & 0.357190 & \checkmark \\
$\mu_{55}$ & 0.007210 & 0.402440 & \checkmark \\
$\mu_{56}$ & 0.006020 & 0.207320 & \checkmark \\
$\mu_{57}$ & 0.005440 & 0.243050 & \checkmark \\
$\mu_{58}$ & 0.006060 & 0.413630 & \checkmark \\
$\mu_{59}$ & 0.007040 & 0.195290 & \checkmark \\
$\mu_{60}$ & 0.007370 & 0.291900 & \checkmark \\
$\mu_{61}$ & 0.007820 & 0.557700 & \checkmark \\
$\mu_{62}$ & 0.004920 & 0.590890 & \checkmark \\
$\mu_{63}$ & 0.006370 & 0.332520 & \checkmark \\
$\mu_{64}$ & 0.006730 & 0.320250 & \checkmark \\
$\mu_{65}$ & 0.006740 & 0.477950 & \checkmark \\
$\mu_{66}$ & 0.007680 & 0.550830 & \checkmark \\
$\mu_{67}$ & 0.007370 & 0.324150 & \checkmark \\
$\mu_{68}$ & 0.007470 & 0.268960 & \checkmark \\
$\mu_{69}$ & 0.006560 & 0.385100 & \checkmark \\
\bottomrule
\end{longtable}

\begin{longtable}{lrrl}
\label{apptab:power_scale_analysis_unpooled} \\
\caption{Power scale analysis results for the unpooled model hyperparameters and individual-level parameters. Each row reports the sensitivity of the posterior distribution for a parameter to power-scaling of the prior and likelihood, respectively. The ``prior'' and ``likelihood'' columns show the cumulative Jensen-Shannon distance between the based and perturbed posterior distribution under power-scaling of the prior or likelihood, respectively. The ``diagnosis'' column indicates whether the parameter is robust (\checkmark) or shows potential sensitivity to the prior or likelihood.} \\
\toprule
component & prior & likelihood & diagnosis \\
label &  &  &  \\
\midrule
\endfirsthead

\toprule
component & prior & likelihood & diagnosis \\
label &  &  &  \\
\midrule
\endhead

\midrule
\multicolumn{4}{r}{\textit{Continued on next page}} \\
\endfoot

\bottomrule
\endlastfoot
$\alpha_0$ & 0.014470 & 0.080020 & \checkmark \\
$\alpha_1$ & 0.041940 & 0.006640 & \checkmark \\
$\alpha_2$ & 0.006680 & 0.013490 & \checkmark \\
$\alpha_3$ & 0.014310 & 0.014450 & \checkmark \\
$\alpha_4$ & 0.008780 & 0.016600 & \checkmark \\
$\alpha_5$ & 0.042510 & 0.012840 & \checkmark \\
$\alpha_6$ & 0.007750 & 0.014610 & \checkmark \\
$\alpha_7$ & 0.013380 & 0.013910 & \checkmark \\
$\alpha_8$ & 0.006790 & 0.014550 & \checkmark \\
$\alpha_9$ & 0.016340 & 0.013120 & \checkmark \\
$\alpha_{10}$ & 0.012380 & 0.012990 & \checkmark \\
$\alpha_{11}$ & 0.004130 & 0.021650 & \checkmark \\
$\alpha_{12}$ & 0.009990 & 0.011880 & \checkmark \\
$\alpha_{13}$ & 0.065650 & 0.017660 & potential strong prior / weak likelihood \\
$\alpha_{14}$ & 0.010940 & 0.014240 & \checkmark \\
$\alpha_{15}$ & 0.058810 & 0.021880 & potential strong prior / weak likelihood \\
$\alpha_{16}$ & 0.084450 & 0.015310 & potential strong prior / weak likelihood \\
$\alpha_{17}$ & 0.010880 & 0.008830 & \checkmark \\
$\alpha_{18}$ & 0.008700 & 0.019000 & \checkmark \\
$\alpha_{19}$ & 0.014240 & 0.013890 & \checkmark \\
$\alpha_{20}$ & 0.020720 & 0.022240 & \checkmark \\
$\alpha_{21}$ & 0.007890 & 0.009740 & \checkmark \\
$\alpha_{22}$ & 0.017630 & 0.008690 & \checkmark \\
$\alpha_{23}$ & 0.012960 & 0.011370 & \checkmark \\
$\alpha_{24}$ & 0.019430 & 0.013650 & \checkmark \\
$\alpha_{25}$ & 0.046650 & 0.023800 & \checkmark \\
$\alpha_{26}$ & 0.013970 & 0.019130 & \checkmark \\
$\alpha_{27}$ & 0.032480 & 0.017030 & \checkmark \\
$\alpha_{28}$ & 0.009660 & 0.011720 & \checkmark \\
$\alpha_{29}$ & 0.016810 & 0.016530 & \checkmark \\
$\alpha_{30}$ & 0.068550 & 0.015930 & potential strong prior / weak likelihood \\
$\alpha_{31}$ & 0.056110 & 0.012330 & potential strong prior / weak likelihood \\
$\alpha_{32}$ & 0.012100 & 0.013500 & \checkmark \\
$\alpha_{33}$ & 0.068090 & 0.034970 & potential strong prior / weak likelihood \\
$\alpha_{34}$ & 0.161630 & 0.012360 & potential strong prior / weak likelihood \\
$\alpha_{35}$ & 0.026830 & 0.022190 & \checkmark \\
$\alpha_{36}$ & 0.059590 & 0.023580 & potential strong prior / weak likelihood \\
$\alpha_{37}$ & 0.091250 & 0.012050 & potential strong prior / weak likelihood \\
$\alpha_{38}$ & 0.095310 & 0.011110 & potential strong prior / weak likelihood \\
$\alpha_{39}$ & 0.145900 & 0.039310 & potential strong prior / weak likelihood \\
$\alpha_{40}$ & 0.059180 & 0.027300 & potential strong prior / weak likelihood \\
$\alpha_{41}$ & 0.045900 & 0.020490 & \checkmark \\
$\alpha_{42}$ & 0.043980 & 0.013300 & \checkmark \\
$\alpha_{43}$ & 0.100880 & 0.016190 & potential strong prior / weak likelihood \\
$\alpha_{44}$ & 0.069490 & 0.025990 & potential strong prior / weak likelihood \\
$\alpha_{45}$ & 0.110870 & 0.022610 & potential strong prior / weak likelihood \\
$\alpha_{46}$ & 0.145630 & 0.022780 & potential strong prior / weak likelihood \\
$\alpha_{47}$ & 0.109600 & 0.018930 & potential strong prior / weak likelihood \\
$\alpha_{48}$ & 0.121890 & 0.025280 & potential strong prior / weak likelihood \\
$\alpha_{49}$ & 0.067610 & 0.034710 & potential strong prior / weak likelihood \\
$\alpha_{50}$ & 0.152420 & 0.027860 & potential strong prior / weak likelihood \\
$\alpha_{51}$ & 0.120450 & 0.027620 & potential strong prior / weak likelihood \\
$\alpha_{52}$ & 0.126520 & 0.023390 & potential strong prior / weak likelihood \\
$\beta_0$ & 0.019990 & 0.087420 & \checkmark \\
$\beta_1$ & 0.022680 & 0.010020 & \checkmark \\
$\beta_2$ & 0.031400 & 0.010650 & \checkmark \\
$\beta_3$ & 0.028430 & 0.013570 & \checkmark \\
$\beta_4$ & 0.014420 & 0.017490 & \checkmark \\
$\beta_5$ & 0.112750 & 0.017540 & potential strong prior / weak likelihood \\
$\beta_6$ & 0.005290 & 0.013990 & \checkmark \\
$\beta_7$ & 0.023900 & 0.016700 & \checkmark \\
$\beta_8$ & 0.039810 & 0.017040 & \checkmark \\
$\beta_9$ & 0.040250 & 0.009000 & \checkmark \\
$\beta_{10}$ & 0.044180 & 0.026580 & \checkmark \\
$\beta_{11}$ & 0.008320 & 0.011740 & \checkmark \\
$\beta_{12}$ & 0.030710 & 0.013090 & \checkmark \\
$\beta_{13}$ & 0.182550 & 0.019370 & potential strong prior / weak likelihood \\
$\beta_{14}$ & 0.001420 & 0.016630 & \checkmark \\
$\beta_{15}$ & 0.019610 & 0.015220 & \checkmark \\
$\beta_{16}$ & 0.385370 & 0.034900 & potential strong prior / weak likelihood \\
$\beta_{17}$ & 0.003900 & 0.032170 & \checkmark \\
$\beta_{18}$ & 0.025630 & 0.017610 & \checkmark \\
$\beta_{19}$ & 0.035810 & 0.017670 & \checkmark \\
$\beta_{20}$ & 0.039860 & 0.028870 & \checkmark \\
$\beta_{21}$ & 0.009460 & 0.023780 & \checkmark \\
$\beta_{22}$ & 0.040740 & 0.014590 & \checkmark \\
$\beta_{23}$ & 0.006090 & 0.014360 & \checkmark \\
$\beta_{24}$ & 0.013460 & 0.014830 & \checkmark \\
$\beta_{25}$ & 0.016830 & 0.019540 & \checkmark \\
$\beta_{26}$ & 0.030750 & 0.019450 & \checkmark \\
$\beta_{27}$ & 0.002810 & 0.022700 & \checkmark \\
$\beta_{28}$ & 0.049940 & 0.014080 & \checkmark \\
$\beta_{29}$ & 0.047330 & 0.012120 & \checkmark \\
$\beta_{30}$ & 0.064530 & 0.056100 & potential prior-data conflict \\
$\beta_{31}$ & 0.866770 & 0.184320 & potential prior-data conflict \\
$\beta_{32}$ & 0.039830 & 0.014720 & \checkmark \\
$\beta_{33}$ & 0.221890 & 0.035580 & potential strong prior / weak likelihood \\
$\beta_{34}$ & 0.094260 & 0.066840 & potential prior-data conflict \\
$\beta_{35}$ & 0.010720 & 0.015300 & \checkmark \\
$\beta_{36}$ & 0.013190 & 0.041300 & \checkmark \\
$\beta_{37}$ & 0.304790 & 0.077170 & potential prior-data conflict \\
$\beta_{38}$ & 0.025250 & 0.082630 & \checkmark \\
$\beta_{39}$ & 0.078410 & 0.075980 & potential prior-data conflict \\
$\beta_{40}$ & 0.216830 & 0.041700 & potential strong prior / weak likelihood \\
$\beta_{41}$ & 0.143950 & 0.029860 & potential strong prior / weak likelihood \\
$\beta_{42}$ & 0.216480 & 0.081800 & potential prior-data conflict \\
$\beta_{43}$ & 0.266560 & 0.071190 & potential prior-data conflict \\
$\beta_{44}$ & 1.960590 & 0.180970 & potential prior-data conflict \\
$\beta_{45}$ & 0.821850 & 0.134120 & potential prior-data conflict \\
$\beta_{46}$ & 0.595530 & 0.155910 & potential prior-data conflict \\
$\beta_{47}$ & 0.679000 & 0.105940 & potential prior-data conflict \\
$\beta_{48}$ & 0.097710 & 0.079890 & potential prior-data conflict \\
$\beta_{49}$ & 4.976590 & 0.302710 & potential prior-data conflict \\
$\beta_{50}$ & 4.829250 & 0.517090 & potential prior-data conflict \\
$\beta_{51}$ & 2.112620 & 0.130340 & potential prior-data conflict \\
$\beta_{52}$ & 1.230000 & 0.203800 & potential prior-data conflict \\
$\mu_0$ & 0.025390 & 0.080720 & \checkmark \\
$\mu_1$ & 0.050170 & 0.006960 & potential strong prior / weak likelihood \\
$\mu_2$ & 0.037600 & 0.010170 & \checkmark \\
$\mu_3$ & 0.043020 & 0.012500 & \checkmark \\
$\mu_4$ & 0.026810 & 0.016480 & \checkmark \\
$\mu_5$ & 0.076260 & 0.016570 & potential strong prior / weak likelihood \\
$\mu_6$ & 0.006200 & 0.011680 & \checkmark \\
$\mu_7$ & 0.024030 & 0.013540 & \checkmark \\
$\mu_8$ & 0.056860 & 0.008730 & potential strong prior / weak likelihood \\
$\mu_9$ & 0.044550 & 0.012670 & \checkmark \\
$\mu_{10}$ & 0.058670 & 0.012420 & potential strong prior / weak likelihood \\
$\mu_{11}$ & 0.013800 & 0.016420 & \checkmark \\
$\mu_{12}$ & 0.045940 & 0.020900 & \checkmark \\
$\mu_{13}$ & 0.057360 & 0.017110 & potential strong prior / weak likelihood \\
$\mu_{14}$ & 0.000260 & 0.015300 & \checkmark \\
$\mu_{15}$ & 0.003280 & 0.018410 & \checkmark \\
$\mu_{16}$ & 0.123700 & 0.011380 & potential strong prior / weak likelihood \\
$\mu_{17}$ & 0.006600 & 0.014610 & \checkmark \\
$\mu_{18}$ & 0.037350 & 0.010540 & \checkmark \\
$\mu_{19}$ & 0.026280 & 0.026450 & \checkmark \\
$\mu_{20}$ & 0.024940 & 0.026930 & \checkmark \\
$\mu_{21}$ & 0.011730 & 0.016660 & \checkmark \\
$\mu_{22}$ & 0.040920 & 0.011580 & \checkmark \\
$\mu_{23}$ & 0.005300 & 0.021990 & \checkmark \\
$\mu_{24}$ & 0.009980 & 0.018520 & \checkmark \\
$\mu_{25}$ & 0.011300 & 0.015950 & \checkmark \\
$\mu_{26}$ & 0.008220 & 0.010500 & \checkmark \\
$\mu_{27}$ & 0.002260 & 0.018990 & \checkmark \\
$\mu_{28}$ & 0.011140 & 0.017240 & \checkmark \\
$\mu_{29}$ & 0.015390 & 0.012960 & \checkmark \\
$\mu_{30}$ & 0.014330 & 0.023960 & \checkmark \\
$\mu_{31}$ & 0.122920 & 0.023320 & potential strong prior / weak likelihood \\
$\mu_{32}$ & 0.008230 & 0.017570 & \checkmark \\
$\mu_{33}$ & 0.070410 & 0.010910 & potential strong prior / weak likelihood \\
$\mu_{34}$ & 0.012160 & 0.007150 & \checkmark \\
$\mu_{35}$ & 0.003730 & 0.018640 & \checkmark \\
$\mu_{36}$ & 0.010330 & 0.022890 & \checkmark \\
$\mu_{37}$ & 0.064940 & 0.029320 & potential strong prior / weak likelihood \\
$\mu_{38}$ & 0.005720 & 0.023670 & \checkmark \\
$\mu_{39}$ & 0.007810 & 0.015860 & \checkmark \\
$\mu_{40}$ & 0.066270 & 0.021890 & potential strong prior / weak likelihood \\
$\mu_{41}$ & 0.027920 & 0.023480 & \checkmark \\
$\mu_{42}$ & 0.009420 & 0.015930 & \checkmark \\
$\mu_{43}$ & 0.017490 & 0.021810 & \checkmark \\
$\mu_{44}$ & 0.100870 & 0.014940 & potential strong prior / weak likelihood \\
$\mu_{45}$ & 0.048530 & 0.013780 & \checkmark \\
$\mu_{46}$ & 0.142910 & 0.023330 & potential strong prior / weak likelihood \\
$\mu_{47}$ & 0.017900 & 0.030400 & \checkmark \\
$\mu_{48}$ & 0.003820 & 0.022710 & \checkmark \\
$\mu_{49}$ & 0.035420 & 0.012540 & \checkmark \\
$\mu_{50}$ & 0.012310 & 0.023940 & \checkmark \\
$\mu_{51}$ & 0.068750 & 0.023630 & potential strong prior / weak likelihood \\
$\mu_{52}$ & 0.041430 & 0.031130 & \checkmark \\
\bottomrule
\end{longtable}

\section{Posterior Predictive Check}\label{appsec:ppc_rtc}
\begin{figure}[h!]
    \centering
    \begin{subfigure}[b]{0.9\linewidth}
        \centering
        \includegraphics[width=\linewidth]{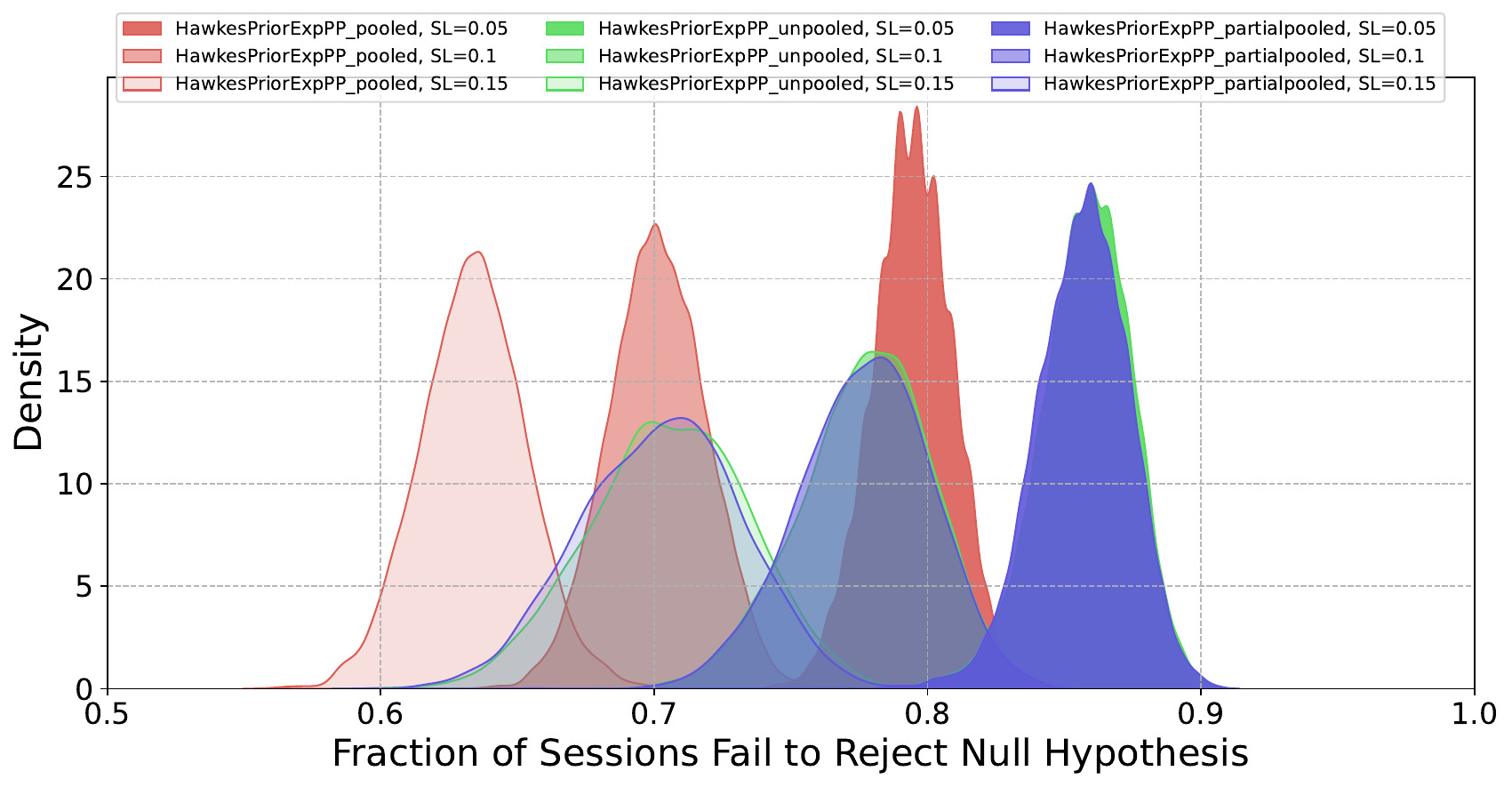}
        \caption{Posterior predictive densities for the \emph{session-level} average number of failed rejections in a Lewis test with Durbin's modification.}
        \label{appfig:ppc_lewis_session}
    \end{subfigure}
    
    \vspace{0.5em} 
    
    \begin{subfigure}[b]{0.9\linewidth}
        \centering
        \includegraphics[width=\linewidth]{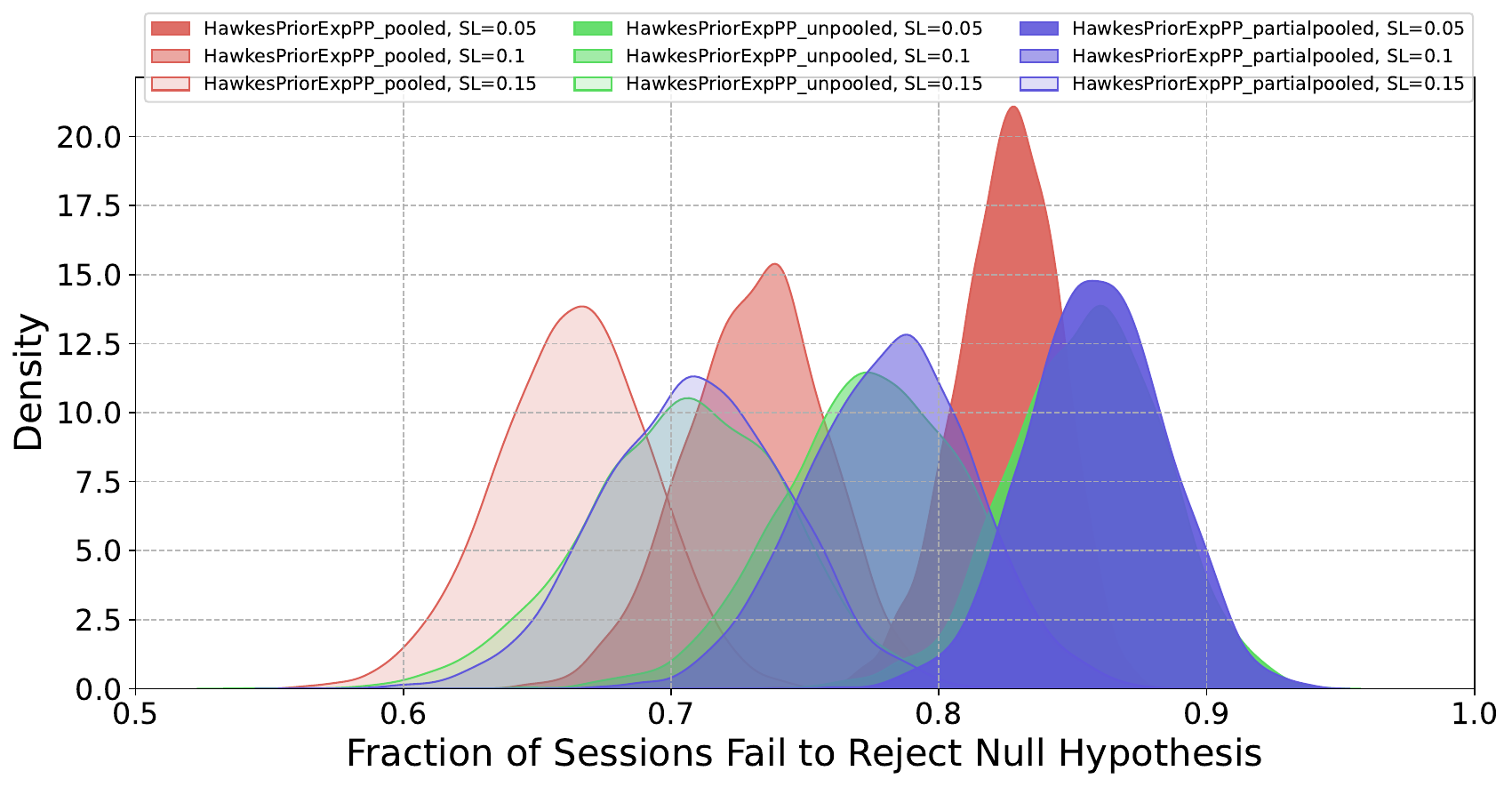}
        \caption{Posterior predictive densities for the \emph{patient-level} average number of failed rejections in a Lewis test with Durbin's modification.}
        \label{appfig:ppc_lewis_patient}
    \end{subfigure}
    
    \caption{Posterior predictive checks for the Lewis test with Durbin's modification, shown at the (a) session level and (b) patient level.}
    \label{appfig:ppc_lewis_combined}
\end{figure}

We perform a Posterior Predictive Check over the Lewis test with Durbin's modification to incorporate the uncertainty over the parameter inference into the calculation of p-value / test statistic \cref{appfig:ppc_lewis_session}. This gives us a distribution over the p-values rather than a point-estimate.
\begin{align*}
\theta^{(m)} & \sim p(\theta \mid \mathcal{S}) \\
S^{(m)}_\text{RTC} &= \{ \Lambda_{\theta^{(m)}}^*(t_1), \Lambda_{\theta^{(m)}}^*(t_2), \ldots, \Lambda_{\theta^{(m)}}^*(t_J) \} \\
\text{p-value}^{(m)} &= \text{Lewis-Test}(S^{(m)}_\text{RTC}) \\
\text{p-values} &= \{\text{p-value}^{(1)},\text{p-value}^{(2)},\ldots, \text{p-value}^{(M)}\}
\end{align*}

\end{appendices}

\end{document}